\documentclass{article}
\usepackage[utf8]{inputenc}
\usepackage{graphics}
\usepackage[sorting = none, backend = biber, style=numeric-comp]{biblatex}
\usepackage{hyperref}
\usepackage[margin=0.8in]{geometry}
\usepackage{todonotes}
\usepackage{array}
\usepackage{overpic}
\usepackage{algorithmic}
\usepackage{tabularx}
\usepackage{amsmath}
\newcolumntype{C}[1]{>{\centering}m{#1}}
\newcolumntype{R}[1]{>{\raggedleft}m{#1}}
\usepackage{float}
\usepackage{authblk}
\usepackage{adjustbox}
\usepackage{tikz}
\usepackage{subcaption}
\usepackage{braket}
\usepackage{amsopn}
\usepackage{amssymb,amsfonts}
\usepackage{caption}
\usepackage{booktabs,multirow}
\usetikzlibrary{quantikz}
\usetikzlibrary{external}

\DeclareMathOperator*{\argmax}{arg\,max}

\let\svthefootnote\thefootnote
\newcommand\freefootnote[1]{%
  \let\thefootnote\relax%
  \footnotetext{#1}%
  \let\thefootnote\svthefootnote%
}

\title{Quantum Volume in Practice: What Users Can Expect from NISQ Devices}
\author[1]{Elijah Pelofske\thanks{epelofske@lanl.gov}}
\author[1]{Andreas Bärtschi}
\author[1]{Stephan Eidenbenz}
\affil[1]{CCS-3 Information Sciences, Los Alamos National Laboratory}

\begin{document}
\date{\vspace{-6ex}}
\maketitle

\begin{abstract}
Quantum volume (QV) has become the de-facto standard benchmark to quantify the capability of Noisy Intermediate-Scale Quantum (NISQ) devices. While QV values are often reported by NISQ providers for their systems, we perform our own series of QV calculations on 24 NISQ devices currently offered by IBM~Q, IonQ, Rigetti, Oxford Quantum Circuits, and Quantinuum (formerly Honeywell). Our approach characterizes the performances that an advanced user of these NISQ devices can expect to achieve with a reasonable amount of optimization, but without white-box access to the device. In particular, we compile QV circuits to standard gate sets of the vendor using compiler optimization routines where available, and we perform experiments across different qubit subsets. We find that running QV tests requires very significant compilation cycles, QV values achieved in our tests typically lag behind officially reported results and also depend significantly on the classical compilation effort invested. 
\end{abstract}

\maketitle

\section{Introduction}
\label{sec:introduction}

\freefootnote{Research presented in this article was supported by the Laboratory Directed Research  and  Development  program  of  Los  Alamos  National Laboratory under project number 20220656ER. LA-UR-22-22058}

Quantum volume (QV)~\cite{QV2019} has been designed as a benchmark measure for Noise Intermediate-Scale Quantum (NISQ) devices. Informally speaking, a NISQ backend that has passed a QV protocol test of $2^m$ will largely correctly execute any quantum circuit on $m$ qubits with up to $m$ random 2-qubit gates on each of those qubits, thus giving a good guideline to users of the device as to what circuit sizes and depths appear reasonable to run on the device. We will give the more formal definition later. In this research, we aim to characterize the QV values of different NISQ backends as it would likely be experienced by regular, albeit somewhat sophisticated users of these systems. 

QV has been defined to include the compilation from abstract circuit representation to the hardware connectivity and native gateset of NISQ devices. This is a necessary part of such a benchmarking definition, however this also means that heavy circuit compiler optimization can be a very large factor impacting the computed QV measure. Our QV testing approach thus starts from a compiler-agnostic perspective. Specifically, all of the initial compiled circuits we send to the different backends are initially compiled using the Qiskit~\cite{Qiskit} transpiler when required, and otherwise directly submitted to the backend. However, once the circuits are sent to the backend, further circuit optimization may occur. A user of NISQ~\cite{Preskill2018quantumcomputingin,nisq-algos} Quantum Processing Units (QPUs) will generally not have the tools, inclination, expertise, or time to perform heavy circuit compilation. Instead, they will use available open source software. This is why we use the Qiskit~\cite{Qiskit} transpiler as an initial basis for comparison (not all systems allow the user to directly handle compilation of the quantum circuits). 

Our main findings are:

\begin{enumerate}
\item Preparation of QV circuits can be remarkably time intense depending on the system and not well standardized across different providers, even though OpenQASM~\cite{cross2017open} is used by several vendors; there is no standardized way across vendors to turn a logical QV circuit, expressed in a standard circuit description language into an optimized, device-specific instruction sequence. 

\item The qubit mapping and routing problem that quantum compilers address is quite difficult~\cite{10.1145/3297858.3304023, nannicini2021optimal}. The compilation toolchain within the software ecosystem and the backend itself greatly impacts circuit execution quality. Therefore, comparisons between QV values should also take into account the compilation method that was used. In particular, we find that using IBM~Q's custom QV compiler for IBM~Q devices indeed improves QV values over a standard Qiskit compilation method.

\item QV values achieved by users typically lag officially reported values, and in some cases the QV values of backends are not reported. The highest QV values we measured for each vendor are IBM at 16 using default transpilation and 32 when using more optimized compilation, Quantinuum (lower bound) at 512, IonQ Harmony at 8, Rigetti at 8, and Oxford Quantum Circuits at $\not=2$. These results are consistent with the backend error rates. 

\item Providing a qubit subset choice for the logical to physical qubit mapping allows us to characterize the regions of the backend (i.e. the qubit and 2-qubit interacting gates) that give the highest fidelity QV results. Not all qubits and connections on a NISQ device are of the same quality; thus even if a device passes a QV test, such success often relies on selecting a good qubit subset, which is not trivial to identify. This fact slightly compromises the original intuitive appeal of the QV measure: passing a QV $2^m$ test does not necessarily imply that the device will generally handle any circuit of depth and width $m$ well because it may not start with a good qubit subset choice.
\end{enumerate}

Overall, we find that hardware vendors have made great progress in the past five years in their device quality. While even advanced users may not quite reach the officially reported QV values, they can expect to just lag a small factor behind. Nevertheless, our findings also point to the need for quantum circuit optimization for any practical application through advancing compilation tools: while we have had the opportunity to run an extensive amounts of QV test runs, most quantum computing users should not have to go through such intense optimization procedures to test their algorithms.

This article is structured as follows. Following a literature review on the current state of quantum volume research, in Section~\ref{sec:methods} we summarize the methods and backends we will test. In particular, we differentiate the tests between \emph{black-box} execution on all backends, and then more customized compilation and execution on the IBM~Q backends, including connected subgraph compilation and more heavily optimized compilation from a toolkit provided by IBM~Q. Next, we present the results of this analysis across all 24 NISQ devices in Section~\ref{sec:results}. We conclude with a discussion in Section~\ref{sec:discussion} about what these results show about the state of NISQ computers and the QV metric. The QV circuits as well as the corresponding measurement counts are publicly available~\cite{pelofske_elijah_2022_6360668}. 

\subsection{Literature review}
\label{sec:introduction_literature_review}

Quantum Volume was proposed as a near-term metric for modern quantum computers that encompasses all aspects of the computational ability of the QPU including connectivity, qubit number, compiler software, and error rates~\cite{QV2019}. Following this, more advanced compiler and routing techniques~\cite{nannicini2021optimal} and Qiskit Pulse optimization~\cite{PRXQuantum.1.020318} allowed measurement of QV 64 on some IBM~Q systems~\cite{Jurcevic_2021}; we use this advanced implementation for some of our experiments.

On Quantinuum (formerly Honeywell) H1 backends, the measured QV has been steadily increasing along with new system upgrades and lower error rates, going from 64~\cite{2021-honywell}, to 1024~\cite{baldwin2021reexamining}, 2048~\cite{S2-QV-published} and finally the maximally possible $4096=2^{12}$~\cite{S2-QV-published-4096} on the 12-qubit \texttt{System Model H1-2} Quantinuum device (this device was previously known as Honeywell S2). 

There are several other NISQ hardware independent benchmarks similar to quantum volume that have been proposed, including mirror circuits~\cite{mirror-circuits-benchmarks}, application oriented benchmarks~\cite{lubinski2021applicationoriented, 2021-application-benchmark}, volumetric benchmarks~\cite{blumekohout2020volumetric} which are generalizations of the QV benchmark, and machine learning motivated metrics such as the proposed q-BAS score~\cite{2019-qBAS}. Using six scalable application oriented benchmarks, numerous devices (across IBM~Q, Rigetti, and IonQ) have been benchmarked~\cite{cornelissen2021scalable}. For another NISQ benchmark, Atos has proposed a metric called \emph{Q-score} as an application relevant benchmark~\cite{Q-score}. 
One of the other key computation metrics that needs to be established for NISQ computers, as it was with classical computers, is a notation of \emph{speed}. To this end, Circuit Layer Operations per Second (CLOPS) has been proposed as a viable NISQ computer speed metric~\cite{wack2021quality}. 

A recent interesting topic is applying error mitigation to the quantum volume protocol in order to demonstrate an increased effective quantum volume~\cite{QV-error-mitigation}.

\section{Methods}
\label{sec:methods}

\newcommand{\sugate}{\gate[2]{\mathrm{SU}(4)}}
\newcommand{\pigate}[2]{\gate[#1]{\pi_{#2}}}
\begin{figure}
    \centering
	\begin{adjustbox}{width=0.8\linewidth}
		\begin{quantikz}[row sep={24pt,between origins},execute at end picture={
					\node at ($(\tikzcdmatrixname-7-2)+(28pt,-15pt)$) {\textcolor{red}{1}};	
					\node at ($(\tikzcdmatrixname-7-4)+(28pt,-15pt)$) {\textcolor{red}{2}};	
					\node at ($(\tikzcdmatrixname-7-10)+(28pt,-15pt)$) {\textcolor{red}{$d$}};	
			}]
			\lstick{$\ket{0}$}\slice{}	& \pigate{7}{1}	& \sugate\slice{}	& \pigate{7}{2}	& \sugate\slice{}	& \qw	& \ldots	& 	& \qw\slice{}	& \pigate{7}{d}	& \sugate\slice{}	& \meter{} 	\\
			\lstick{$\ket{0}$}		& \qw		& \qw			& \qw		& \qw			& \qw	& \ldots	& 	& \qw		& \qw		& \qw			& \meter{} 	\\
			\lstick{$\ket{0}$}		& \qw		& \sugate		& \qw		& \sugate		& \qw	& \ldots	& 	& \qw		& \qw		& \sugate		& \meter{} 	\\
			\lstick{$\ket{0}$}		& \qw		& \qw			& \qw		& \qw			& \qw	& \ldots	& 	& \qw		& \qw		& \qw			& \meter{} 	\\
			\lstick{$\ket{0}$}		& \qw		& \sugate		& \qw		& \sugate		& \qw	& \ldots	& 	& \qw		& \qw		& \sugate		& \meter{} 	\\
			\lstick{$\ket{0}$}		& \qw		& \qw			& \qw		& \qw			& \qw	& \ldots	& 	& \qw		& \qw		& \qw			& \meter{} 	\\
			\lstick{$\ket{0}$}		& \qw		& \qw			& \qw		& \qw			& \qw	& \ldots	& 	& \qw		& \qw		& \qw			& \meter{} 	
		\end{quantikz}
	\end{adjustbox}
	\caption{Quantikz~\cite{quantikz} circuit drawing for a logical Quantum Volume circuit on $m=7$ qubits, where each layer $1,\ldots,d=7$ consists of a random permutation of the qubit labels 
	followed by $m' = 3 =\lfloor \tfrac{7}{2} \rfloor$ random SU(4) gates acting on consecutive pairs of (permuted) qubits.}
    \label{fig:logical_QV_circuit_drawing}
\end{figure}
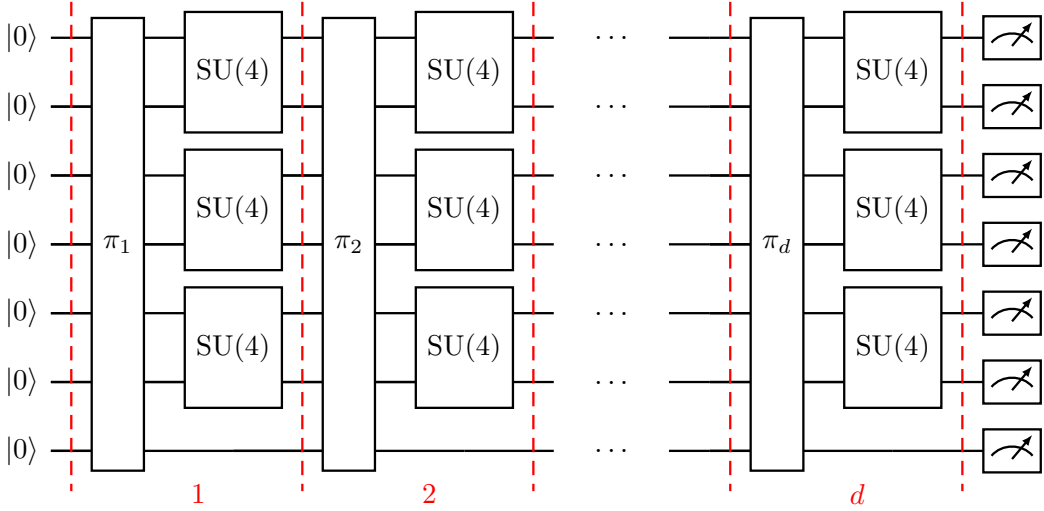

\begin{table*}[h]
    \scriptsize
    \hspace*{-1cm}
    \addtolength{\tabcolsep}{-2pt}
	\centering
	\begin{tabular}{@{}p{2cm}l >{\bfseries}r>{\bfseries}rr rlr lll@{}}
		\toprule
		\	& 		& \multicolumn{3}{c}{Measured QV}		& \multicolumn{3}{c}{QPU} 			& \multicolumn{3}{c}{Mean fidelity}	\\
		\cmidrule(lr){3-5}
		\cmidrule(lr){6-8}
		\cmidrule(lr){9-11}
		Vendor	& Backend name	& Black-box	& Argmax	& Vendor	& \#Qubits	& Type/Topology	& \#Edges	& 2Q gate	& 1Q gate	& SPAM	\\		
		\midrule[\heavyrulewidth]
		Quantinuum & H1-2 & 512* & 512* & 4096 & 12 & All-to-All & 66 & 0.995 & 0.9997 & 0.993\\
		\midrule
		IBM~Q 	& ibmq\_lima & 8 & 8 & 8 & 5 & Falcon r4T & 4 & 0.9898 & 0.9998 & 0.973 \\
		\ 	& ibmq\_belem & 8 & 8 & 16 & 5 & Falcon r4T & 4 & 0.9874 & 0.9998 & 0.9775 \\
		\ 	& ibmq\_quito & 8 & 16 & 16 & 5 & Falcon r4T & 4 & 0.9889 & 0.9998 & 0.9717 \\
		\ 	& ibmq\_jakarta & 8 & 8 & 16 & 7 & Falcon r5.11H & 6 & 0.9896 & 0.9997 & 0.9747 \\
		\ 	& ibmq\_manila & 16 & 16 & 32 & 5 & Falcon r5.11L & 4 & 0.9897 & 0.9997 & 0.9728 \\[1ex]
		\ 	& ibmq\_bogota & 8 & 8 & 32 & 5 & Falcon r4L & 4 & 0.9905 & 0.9998 & 0.9656 \\
		\ 	& ibm\_perth & 8 & 8 & 32 & 7 & Falcon r5.11H & 6 & 0.9781 & 0.9997 & 0.987 \\
		\ 	& ibmq\_casablanca & 8 & 16 & 32 & 7 & Falcon r4H & 6 & 0.9903 & 0.9998 & 0.9805 \\
		\ 	& ibm\_lagos & 8 & 32 & 32 & 7 & Falcon r5.11H & 6 & 0.9924 & 0.9998 & 0.9862 \\
		\ 	& ibmq\_guadalupe & 8 & 32 & 32 & 16 & Falcon r4P & 16 & 0.9892 & 0.9997 & 0.9738 \\[1ex]
		\ 	& ibmq\_sydney & 8 & 16 & 32 & 27 & Falcon r4 & 28 & 0.9487 & 0.9997 & 0.957 \\
		\ 	& ibmq\_toronto & 8 & 16 & 32 & 27 & Falcon r4 & 28 & 0.9787 & 0.9993 & 0.9376 \\
		\ 	& ibmq\_brooklyn & 8 & 32 & 32 & 65 & Hummingbird r2 & 72 & 0.9118 & 0.9995 & 0.9694 \\
		\ 	& ibm\_washington & 8 & 16 & 64 & 127 & Eagle r1 & 142 & 0.9828 & 0.9997 & 0.9737 \\
		\ 	& ibm\_auckland & 8 & 16 & 64 & 27 & Falcon r5.11 & 28 & 0.9536 & 0.9992 & 0.9872 \\[1ex]
		\ 	& ibm\_cairo & 8 & 16 & 64 & 27 & Falcon r5.11 & 28 & 0.9882 & 0.9998 & 0.9807 \\
		\ 	& ibm\_hanoi & 8 & 32 & 64 & 27 & Falcon r5.11 & 28 & 0.9891 & 0.9998 & 0.9778 \\
		\ 	& ibmq\_mumbai & 8 & 16 & 128 & 27 & Falcon r5.1 & 28 & 0.9504 & 0.9994 & 0.972 \\
		\ 	& ibmq\_montreal & 8 & 32 & 128 & 27 & Falcon r4 & 28 & 0.9858 & 0.9996 & 0.9769 \\
		\midrule
		IonQ	& Harmony & 8 & 8 &  & 11 & All-to-All & 55 & 0.96541 & 0.9972 & 0.99709\\
		\midrule
		OQC & Lucy & $< 2$ & $< 2$ & & 8 & LNN ring & 8 & 0.9416 & 0.9991 & 0.9044 \\
		\midrule
		Rigetti	& Aspen-11 & $< 2$ & 4 & 8 & 38 & Octagonal & 43 & 0.9215 & 0.9955 & 0.9678 \\
		\ 	& Aspen-M-1 & $< 2$ & 8 & 8 & 80 & Octagonal & 102 & 0.9113 & 0.9894 & 0.9695 \\
		\bottomrule
	\end{tabular}
	\caption{Table of NISQ QPUs evaluated using the QV protocol. The values in the \emph{Black-box Measured QV} column are what users who execute circuits without tuning can expect, whereas the values in the \emph{Argmax Measured QV} column are the maximum values we were able to validate with significant additional effort as described in Sections~\ref{sec:methods_connected_subg_qv_ibmq}, \ref{sec:methods_rigetti_qubit_subset_enumeration}  and~\ref{sec:methods_ibmq_high_fidelity}.\newline
		Values in the \emph{Vendor Measured QV}, as well as single-qubit \emph{1Q gate}, two qubit \emph{2Q gate} and State Preparation and Measurement (\emph{SPAM}) fidelities are all vendor provided metrics. The mean gate and SPAM fidelities are computed across all operations of the same type available on the device during the whole QV circuit execution duration (which did span several weeks in some cases). The Rigetti device fidelities were computed using the non-simultaneous gate operation vendor provided calibration data. The IBM~Q single qubit error rate averages include the zero error rate \texttt{rz} gate, and not including \texttt{id} gate error rates. The number of edges for each backend was counted simply as the number of connections between qubits (at the time the experiments were executed); this does not count bi-direction gate operations, or multiple different gate operations between adjacent qubits.\newline
	* The QV value for the \texttt{H1-2} device is a lower bound only, not the finally measured QV value; the device has passed all QV experiments up to QV 512 and larger QV circuits are not yet completed.}
	\label{table:NISQ_Devices}
\end{table*}

A QV circuit consists of sequences of random qubit index permutations followed by a layer of random two qubit unitaries from \emph{Special Unitary} matrices of degree 4 (\texttt{SU(4)}). Compiling layers of SU(4) gates with arbitrary connectivity between qubits is especially challenging for compilers~\cite{zulehner2018compiling} in large part due to the hardness of circuit mapping and qubit routing~\cite{https://doi.org/10.4230/lipics.tqc.2019.5} on restricted connectivities. Given in~\cite{QV2019}, a generic square QV circuit with depth $d$ and width $m$ is a sequence of $d$ circuit layers:

\begin{equation}
    U = U^{(d)} \dots U^{(2)} U^{(1)},
    \label{eq:higher-level-QV-circuit-form}
\end{equation}
where each of these layers (e.g. $U^{(d)}$) is of the form
\begin{equation}
    U^{(t)} = U^{(t, m')}_{ \pi_t (2m' -1), \pi_t (2m') } \otimes \dots \otimes U^{(t,1)}_{ \pi_t (1), \pi_t (2) },
    \label{eq:layer-circuit-form}
\end{equation}
indexed by $t$ ranging from $1$ to $d$. Each layer is acting on $m' := \lfloor \frac{m}{2} \rfloor $ pairs of qubits (meaning that if $m$ is odd then one qubit in each layer will be idle). Each layer is created by choosing a uniform random permutation $\pi_t$ of the $m$ qubit indices, and then applying the two qubit unitary gates $U^{(t, i)}_{ \pi_t (2i-1), \pi_t (2i) }$ from \texttt{SU(4)} acting on qubits $\pi_t (2i-1)$ and $\pi_t (2i)$ for all $m'$ pairs of qubits being used in this layer. Figure~\ref{fig:logical_QV_circuit_drawing} is an example diagram of the logical structure of a quantum volume circuit.

For a given QV circuit, the relevant question is how well the quantum device implemented and executed the circuit~\cite{QV2019}. To this end the QV protocol uses the \emph{heavy output generation problem}~\cite{10.5555/3135595.3135617}. Given a QV circuit $U$, it will have an ideal bitstring output distribution of 

\begin{equation}
    p_U (x) = |\braket{x|U|0}|^2
    \label{eq:U_prob}
\end{equation}

where $x$ is a bitstring output with length equal to $m$. The central idea of the heavy output generation problem is to partition all possible observable bitstrings into two equal-sized sets; one of which is has a lower ideal output probability, and one of which has a higher ideal output probability (i.e. the set of heavy outputs). The quantum hardware implementation of $U$ is then considered successful if more than $\frac{2}{3}$ of the measured output bitstrings fall into the heavy output set. More formally, given the full ideal probability distribution $p_U(x)$, we sort each probability such that $p_1 \leq p_2 \dots \leq p_{2^m}$. Then we can partition according the median of the probabilities $p_{\mathrm{median}}$ in order to get the heavy output set of bitstrings for $U$:

\begin{equation}
    H_U = \{ x \in \{0, 1\}^m \text{~such that~} p_U(x) > p_{\mathrm{median}}\}
    \label{eq:H_U}
\end{equation}

Thus the \emph{heavy output probability} (HOP) for a quantum circuit $U$ implemented on a backend is defined as the number of heavy bitstrings found in the distribution (i.e. the number of elements in $H_U$) out of the total number of samples taken on the backend. This measurement is then repeated for multiple QV circuits (say $k$ distinct QV circuits, each with their own random permutations and random seeds) in order to to determine if the quantum backend in question can reliably sample heavy output probability distributions with probability greater than $\frac{2}{3}$. In the limit of the number of QV circuits (and for large circuit size $m$ and depth $d$ values) the expected mean HOP approaches $\frac{1 + \ln(2)}{2} \approx 0.85$~\cite{10.5555/3135595.3135617, QV2019}. For $k$ distinct QV circuits, each circuit $QV_i$ has a measured heavy output probability denoted as $\mathit{HOP}(QV_i)$. Treating the $\mathit{HOP}$ outcome as a binomial distribution (i.e. either the backend passes $\frac{2}{3}$ or it fails to pass $\frac{2}{3}$), over many circuits this can be approximated as a normal distribution. Using this approximation we can compute confidence intervals on the resulting distribution: Equations~\eqref{eq:HOP-mean} and~\eqref{eq:sigma} show the formula for computing the mean $\mathit{HOP}$, then the standard deviation of the distribution, and then Equations~\eqref{eq:z-value} and~\eqref{eq:z-conf} show the formula for computing the $99\%$ confidence interval. These statistical tests are important because they show when a particular distribution of \emph{heavy output probabilities} is above the $\frac{2}{3}$ threshold with a high confidence level. 

\begin{figure*}[h!]
    \centering
    \begin{overpic}[width=0.85\textwidth]{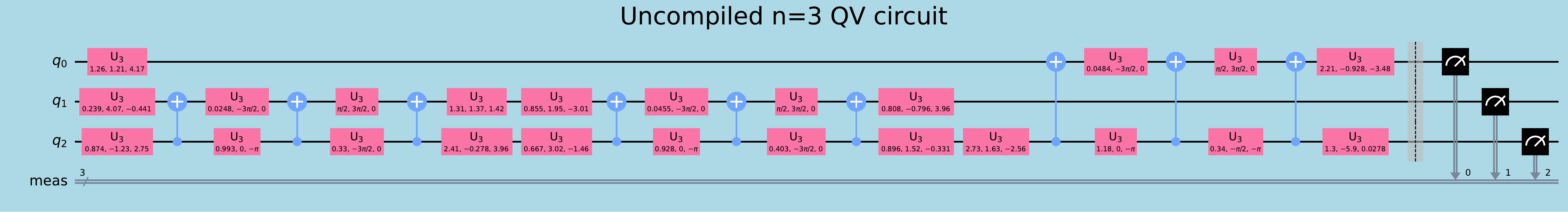}
    \put(101,6.2){\textbf{(1)}}
    \end{overpic}
    \begin{overpic}[width=0.85\textwidth]{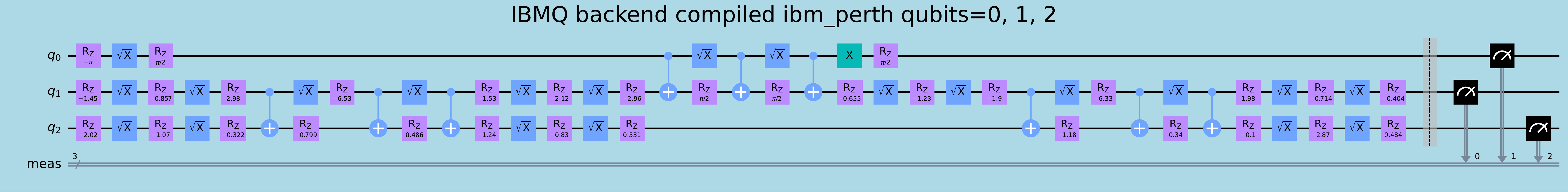}
    \put(101,5.8){\textbf{(2)}}
    \end{overpic}
    \begin{overpic}[width=0.85\textwidth]{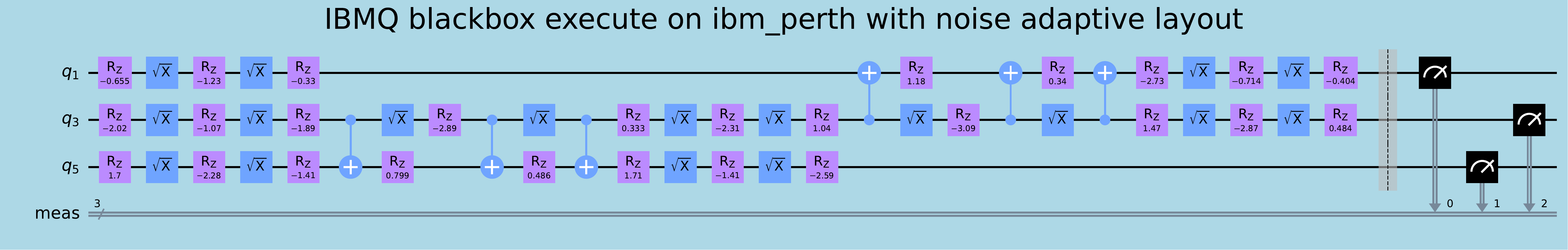}
    \put(101,7.5){\textbf{(3)}}
    \end{overpic}
    \begin{overpic}[width=0.85\textwidth]{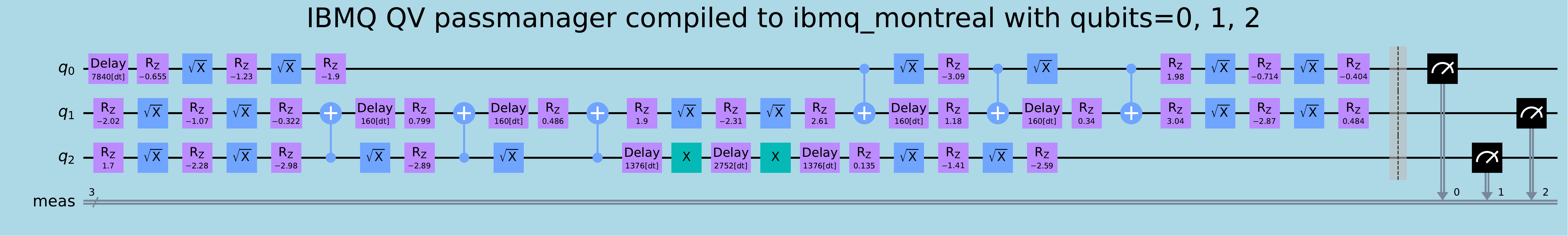}
    \put(101,7.5){\textbf{(4)}}
    \end{overpic}
    \begin{overpic}[width=0.85\textwidth]{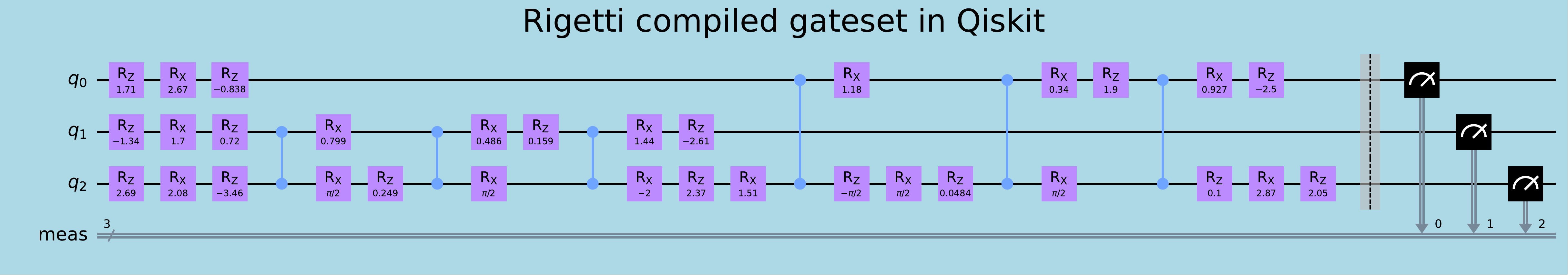}
    \put(101,8.1){\textbf{(5)}}
    \end{overpic}
    \begin{overpic}[width=0.85\textwidth]{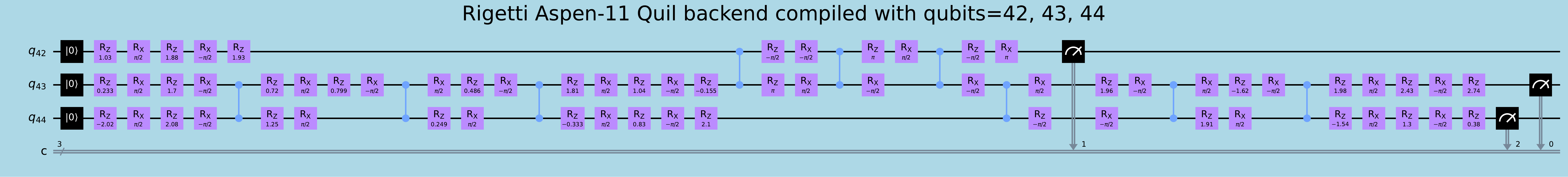}
    \put(101,5.4){\textbf{(6)}}
    \end{overpic}
    \begin{overpic}[width=0.85\textwidth]{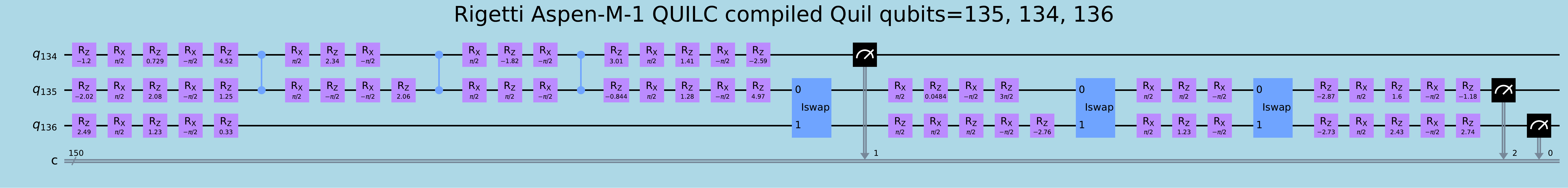}
    \put(101,5.4){\textbf{(7)}}
    \end{overpic}
    \begin{overpic}[width=0.85\textwidth]{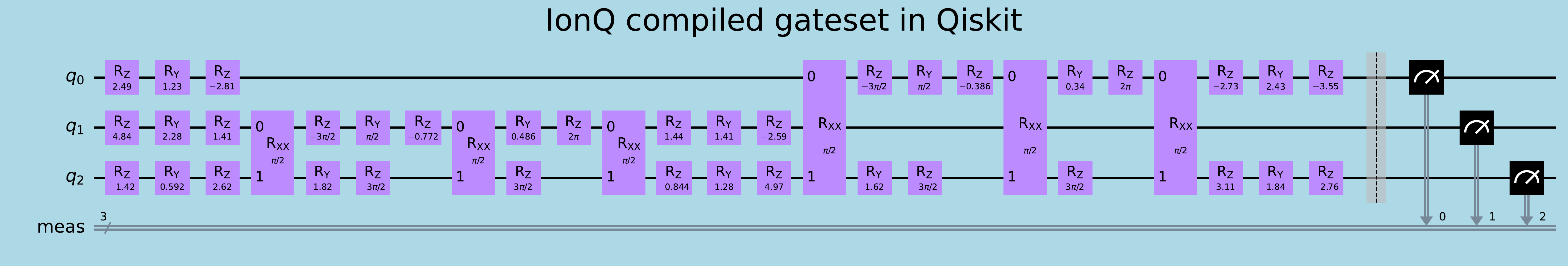}
    \put(101,8.1){\textbf{(8)}}
    \end{overpic}
    \begin{overpic}[width=0.85\textwidth]{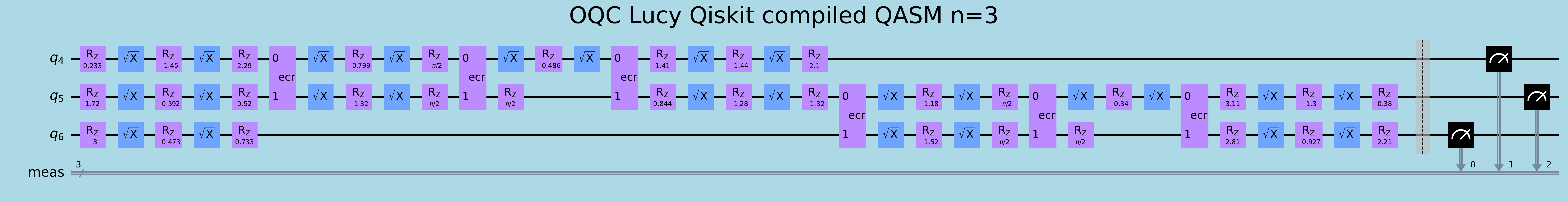}
    \put(101,6.3){\textbf{(9)}}
    \end{overpic}
    \begin{overpic}[width=0.85\textwidth]{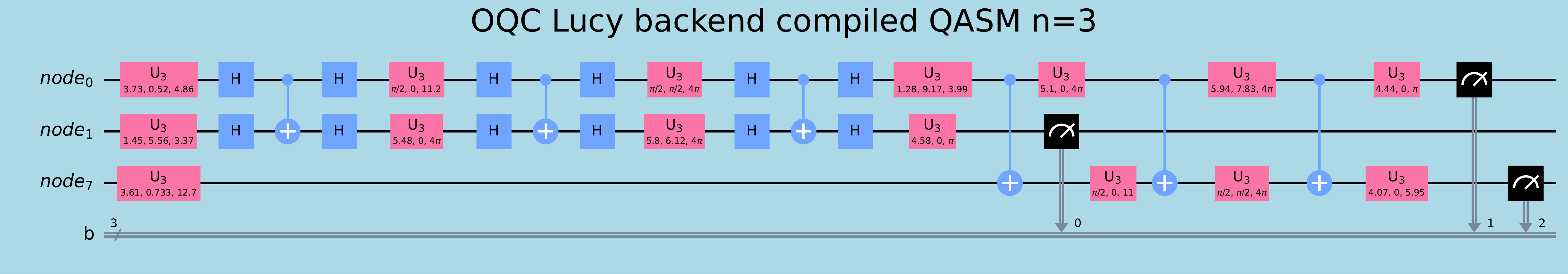}
    \put(101,8.2){\textbf{(10)}}
    \end{overpic}
    \caption{Impact of compilers: starting from the original QV circuit (top) defined in Qiskit using \texttt{u3} and \texttt{cx} gates we compile to backends with different connectivities, gatesets, and software. We use a simple $QV=8$ ($n=3$) circuit for illustration purposes. 
    Note that the backend compiled circuits for IonQ and Quantinuum are not shown; these backends do not currently support returning the backend executed circuits to users. }
    \label{fig:QV_circuit_drawings}
\end{figure*}

\begin{equation}
    \text{mean} = \frac{\sum_i^k HOP(QV_i)}{k}
    \label{eq:HOP-mean}
\end{equation}

\begin{equation}
    \sigma = \text{mean} \cdot \sqrt{\frac{(1-\text{mean})}{k}}
    \label{eq:sigma}
\end{equation}

\begin{equation}
    z = \frac{(\text{mean} - \frac{2}{3})}{\sigma}
    \label{eq:z-value}
\end{equation}

\begin{equation}
    Z_{\mathit{conf}} = 0.5 \cdot (1 + \mathrm{erf}(\frac{z}{\sqrt{2}}))
    \label{eq:z-conf}
\end{equation}

In order to determine if a device (or qubit subset of the device) passed the QV protocol, we use the following criteria: 
\begin{enumerate}
    \item The mean of the heavy output probability (HOP) states is above $\frac{2}{3}$ (see Equation~\eqref{eq:HOP-mean}). 
    \item $2 \sigma$ below the mean of the HOP state probability is also above $\frac{2}{3}$ (see Equation~\eqref{eq:sigma}). 
    \item The distribution is above $\frac{2}{3}$ with a $0.99$ z-confidence interval (see Equations~\eqref{eq:z-value} and~\eqref{eq:z-conf}).
\end{enumerate}
These requirements increase in strictness; $2 \sigma$ corresponds to a z-confidence of $0.977$ (which has a corresponding to z value of $2$), 
compared to the 0.99 z-confidence. 
The QV metric was originally defined on a quantum computer with at least $m$ qubits~\cite{QV2019} and $d(m)$ denoting the largest $d$ such that criteria 1) and 2) are satisfied:
\begin{equation}
    \log_2 \mathrm{QV} = \argmax_m \min(m, d(m))
\end{equation}
When trying to find the quantum volume, we can restrict ourselves to tests with $d=m$.
For simplicity, in the remainder of the article we will denote $n = \log_{2} \mathit{QV}$, and $n=m=d$. 

\begin{figure*}[h!]
    \centering
    \includegraphics[width=1\textwidth]{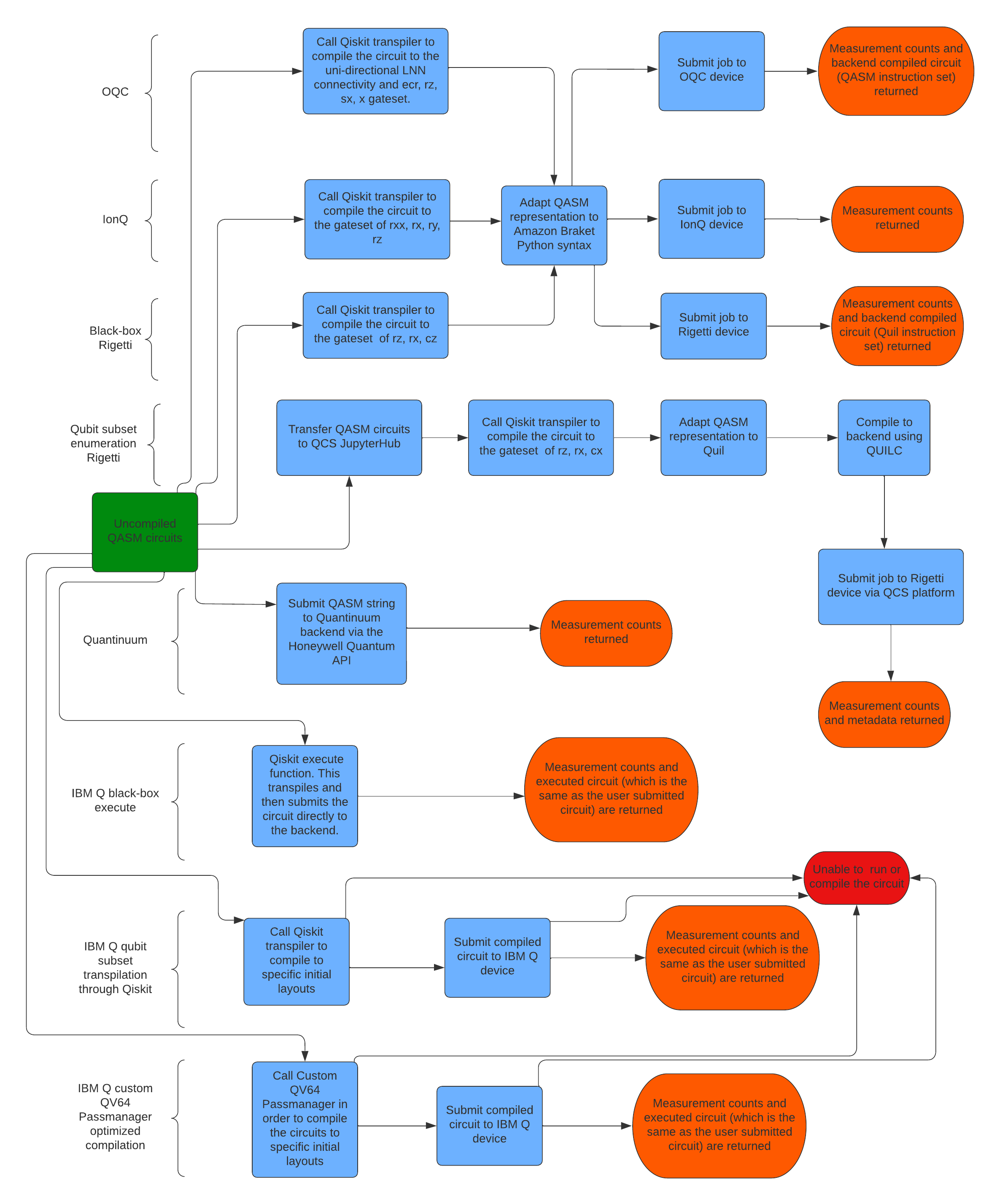}
    \caption{High level compilation procedure diagram}
    \label{fig:flowchart}
\end{figure*}

\begin{figure*}[h!]
    \centering
    \includegraphics[width=0.49\textwidth]{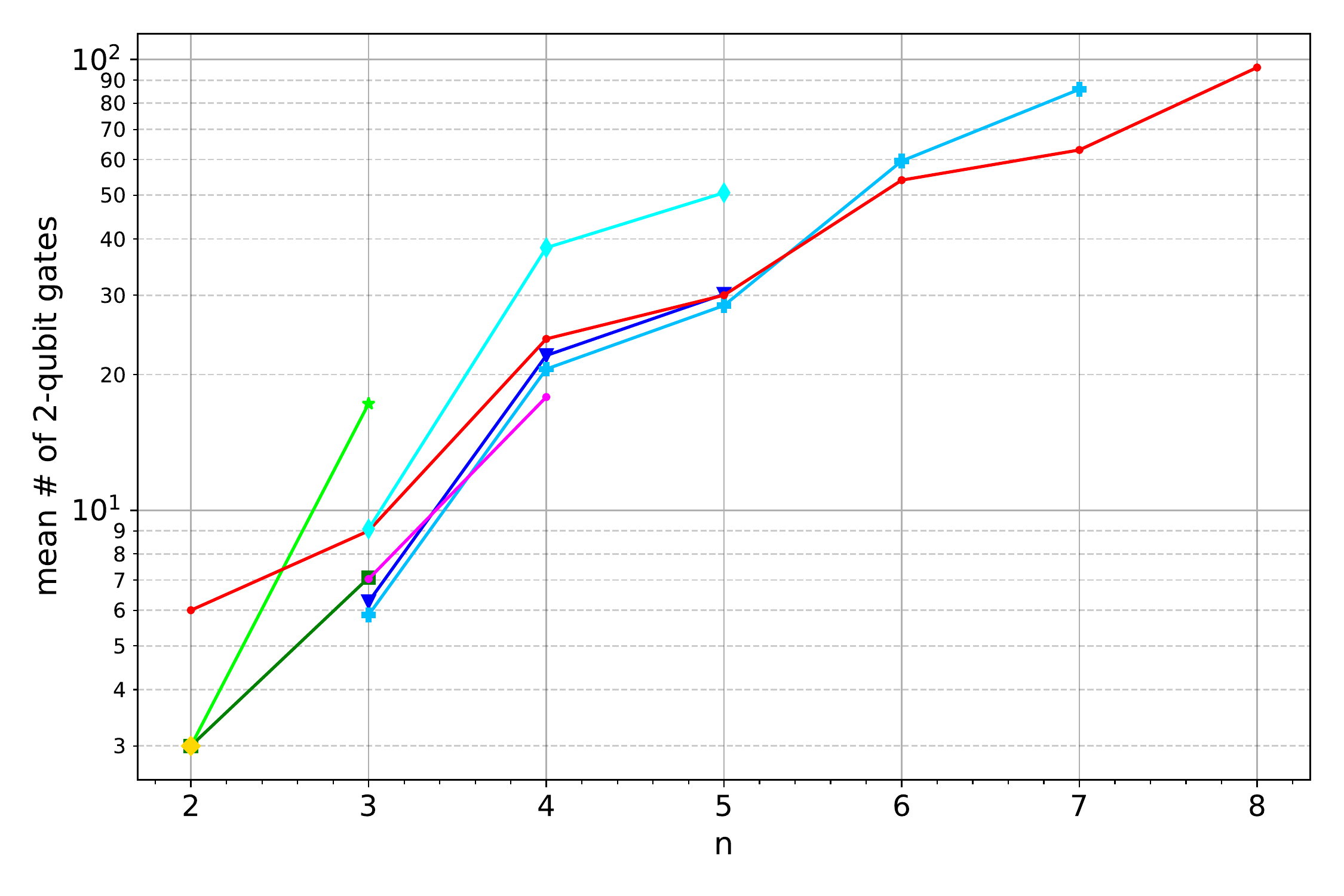}\hfill%
    \includegraphics[width=0.49\textwidth]{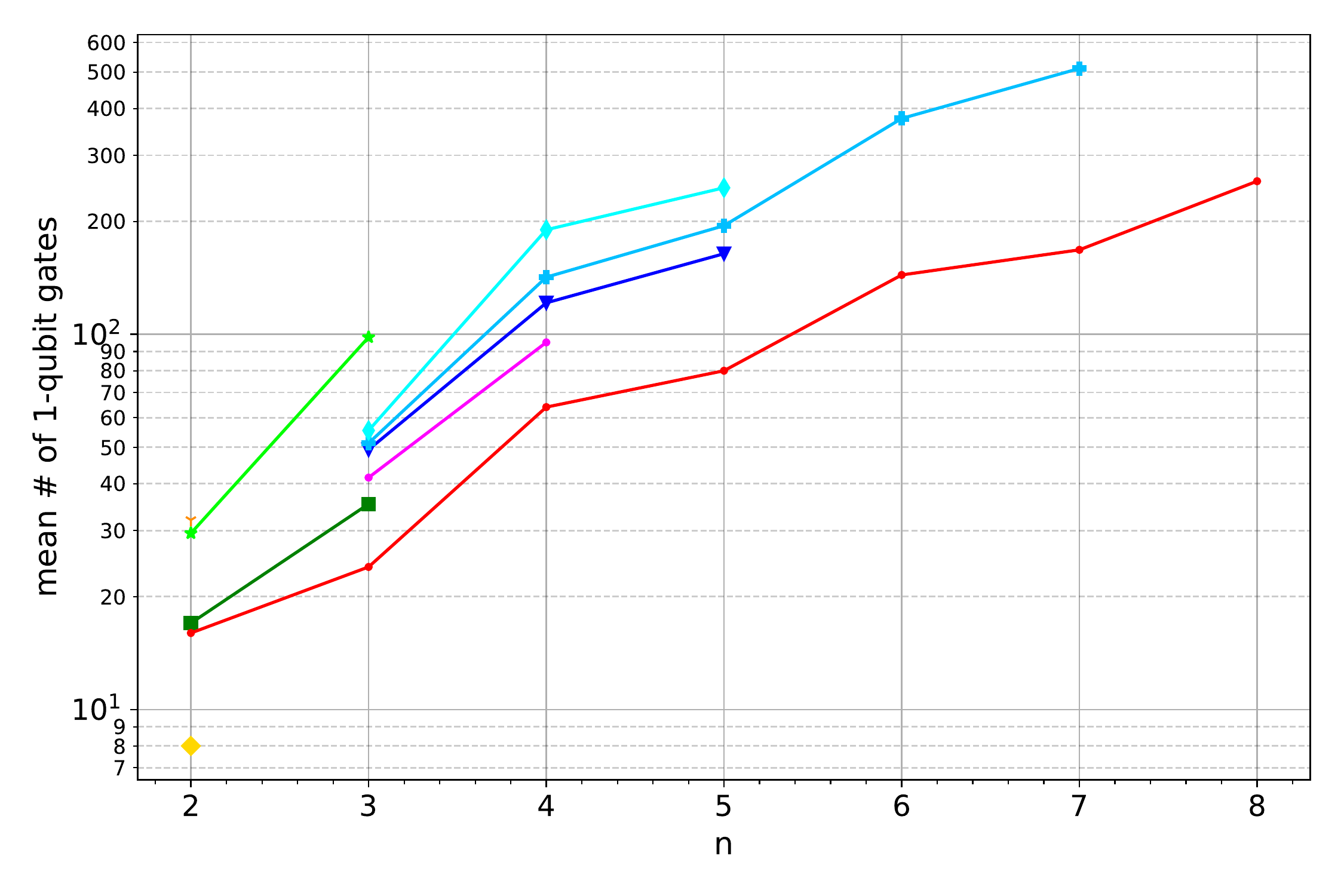}\\
    \includegraphics[width=0.49\textwidth]{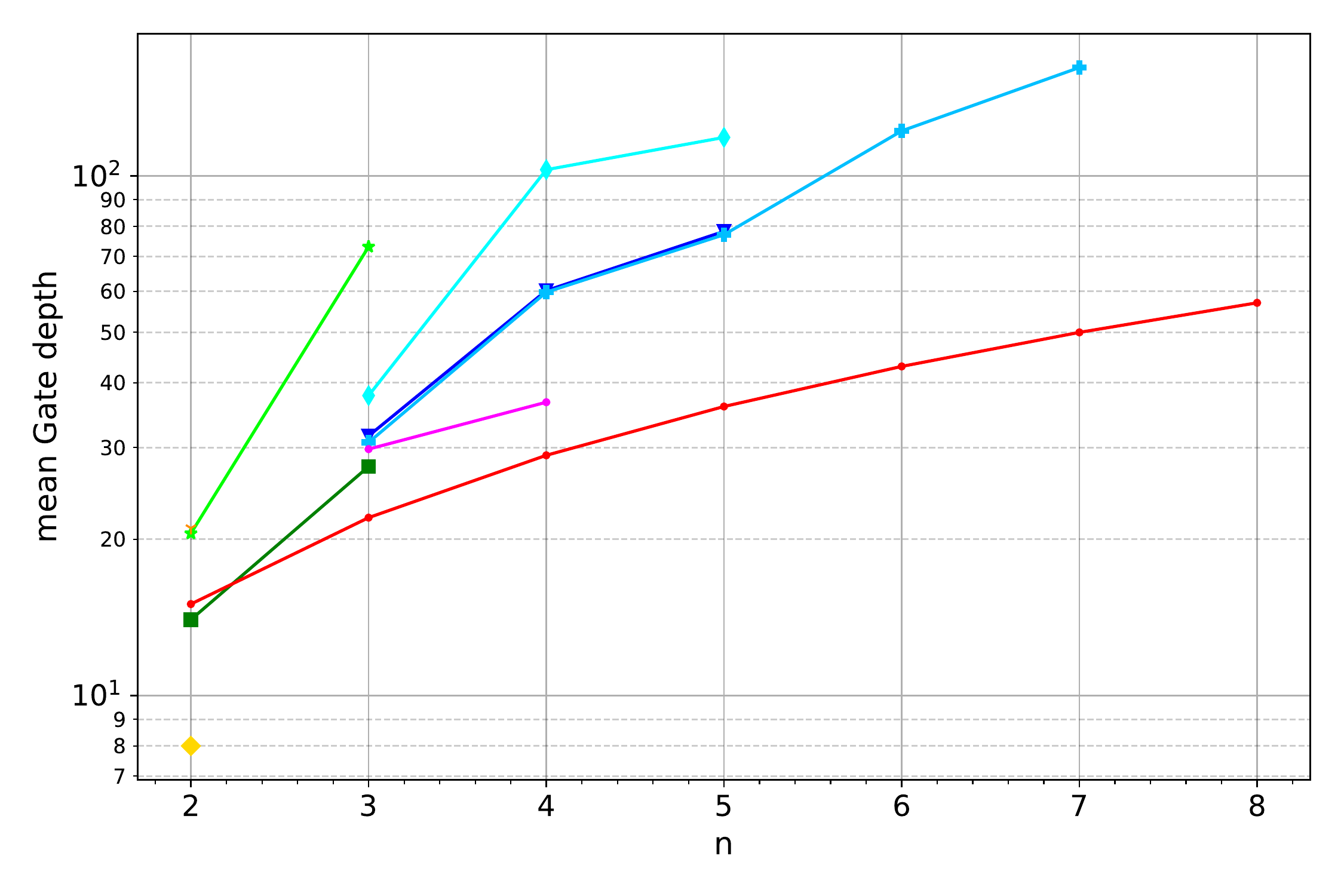}\hfill%
    \includegraphics[width=0.42\textwidth]{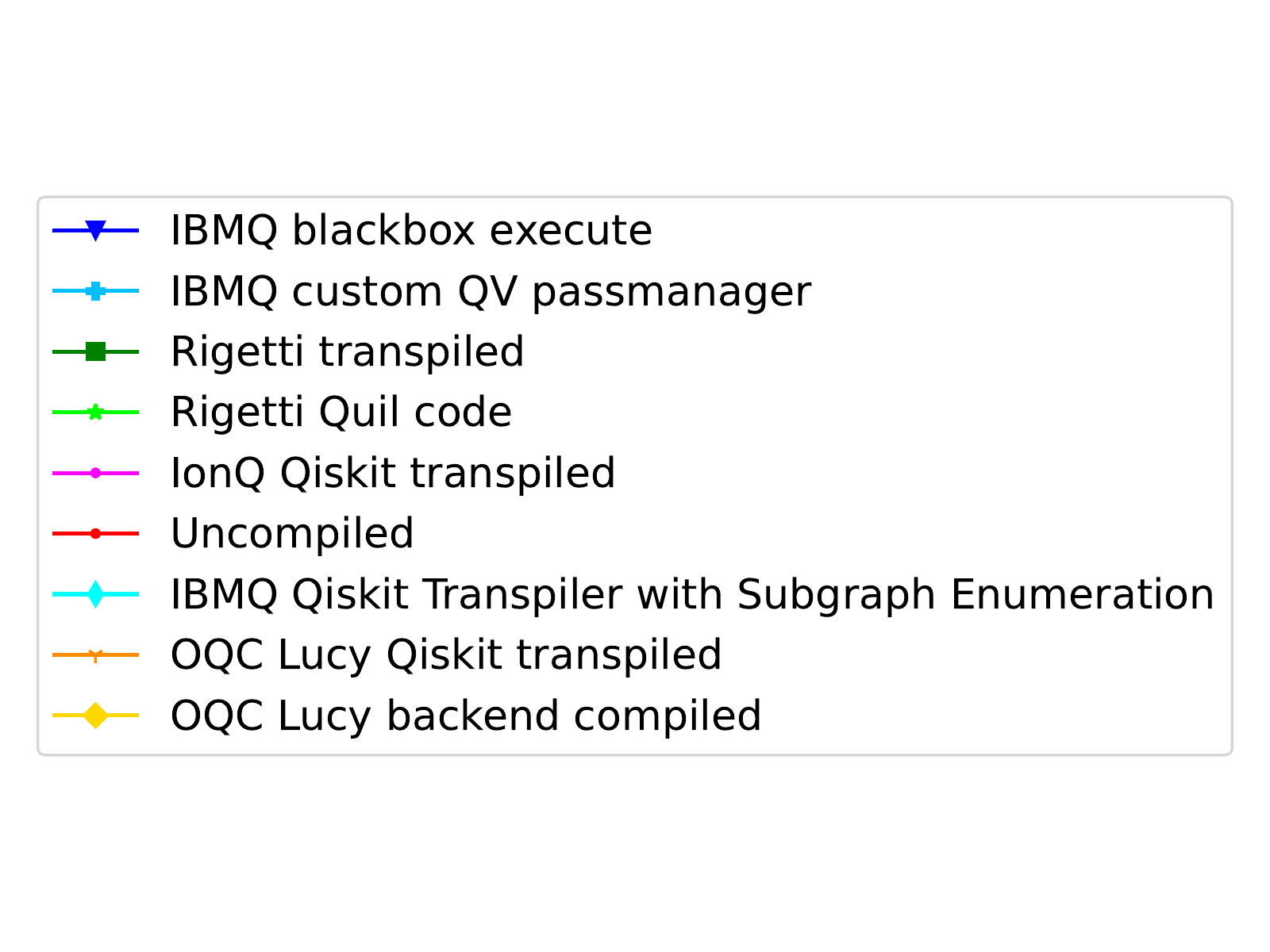}
    \caption{QV circuit operations summary figures. Average single qubit operations (upper right), two qubit operations (upper left), SPAM operations (bottom left), and the legend (bottom right).
    The 2-qubit gate counts were computed as follows; Qiskit transpiled code to the Rigetti gateset: \texttt{cz}, Rigetti Quil backend black box compiled code: \texttt{cz, XY} (the CPHASE gate was never introduced into the circuits by the QUILC compiler during blackbox execution), IonQ Qiskit transpiled code: \texttt{RXX}, IBM~Q backend compiled code: \texttt{cx}, uncompiled code: \texttt{cx}, Qiskit compiled to OQC Lucy gateset: \texttt{ECR}, OQC Lucy backend compiled circuit: \texttt{cx}. 
    The 1-qubit gate counts were computed as follows; Qiskit transpiled code to the Rigetti gateset: \texttt{rx, rz}, Rigetti Quil backend compiled code: \texttt{rx, rz}, IonQ Qiskit transpiled code: \texttt{rz, rx, ry}, IBM~Q backend compiled code: \texttt{rz, x, sx}, uncompiled code: \texttt{U3}, Qiskit compiled to OQC Lucy gateset: \texttt{sx, x, rz}, OQC Lucy backend compiled circuit: \texttt{U3}. See the Appendix~\ref{sec:Appendix_gate_def} for details on the gate definitions. Note that the \emph{delay} gates in the pulse optimized IBM~Q circuits are \emph{not} counted in the gate depth or single qubit gate counts. Measurement operations are also not counted in the single qubit counts. All plots have a log y-axis scale. }
    \label{fig:circuit_statistics}
\end{figure*}

By this definition, the best QV value found on a backend will be the QV value of that backend. In particular this means that distinctions between different qubit subsets have not been specifically reported when applying the QV protocol. One of the methods we investigate is distinguishing how the QV circuits perform on specific qubit subsets, as opposed to simply taking the best performing value found on the device (see Section~\ref{sec:methods_connected_subg_qv_ibmq}). However, this is not possible on all devices because some systems do not currently support specifying a qubit subset to compile the logical circuit to. 

In order to standardize the circuits used on all backends, 1,000 QV circuits per QV circuit size are generated using the Qiskit quantum volume method~\cite{Qiskit-QV-method}. Figure~\ref{fig:circuit_statistics} shows the gate counts for the raw uncompiled circuits. 

Table~\ref{table:NISQ_Devices} summarizes the details and published hardware metrics of the 24 NISQ backends we test. Table~\ref{table:NISQ_Devices} also summarizes the QV values we found on each of the 24 backends. These values are differentiated between \texttt{1.} the QV value found when using the black-box execution method (little or no qubit mapping or basis gate conversions), and \texttt{2.} the best QV value found across all circuits executed on that backend that were compiled using more time intensive compilation procedures (these heavier compilation procedures are specific to IBM~Q and Rigetti backends, see Sections~\ref{sec:methods_connected_subg_qv_ibmq}, \ref{sec:methods_rigetti_qubit_subset_enumeration} and~\ref{sec:methods_ibmq_high_fidelity}). The procedure on all backends is to start at a small QV circuit size (e.g. $n=3$), and then iterate to either larger circuit sizes or smaller circuit size depending on the results of the initial test. Once the device clearly fails to pass at a given $n$, the procedure is terminated. However, this procedure has not fully completed for the \texttt{H1-2} Quantinuum backend due to usage limitations, so far passing experiments up to $n=9$. Therefore, its reported value of QV 512 is a lower bound on the true (black-box) quantum volume of the device.

\begin{figure*}[h]
    \centering
    \includegraphics[width=0.49\textwidth]{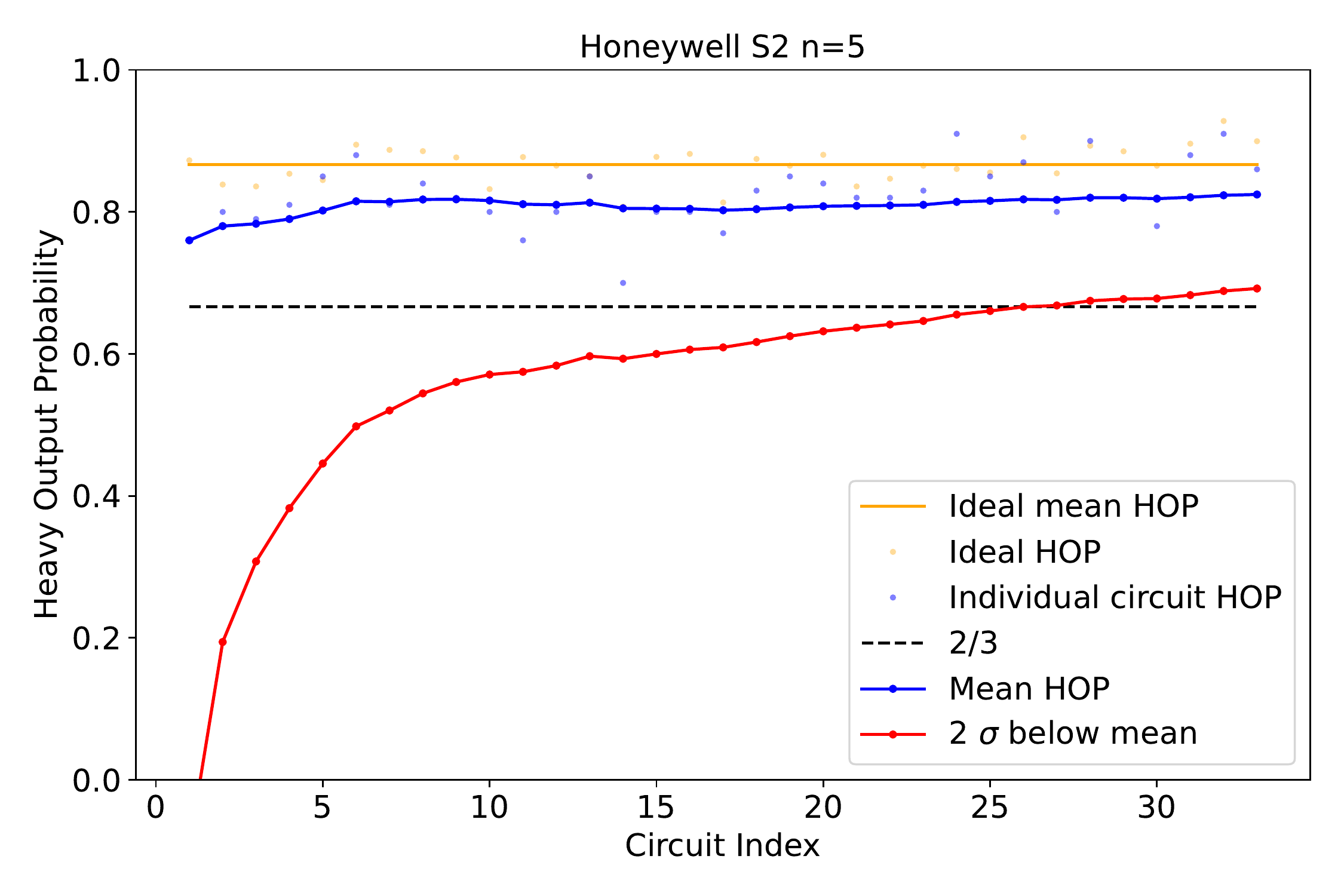}\hfill%
    \includegraphics[width=0.49\textwidth]{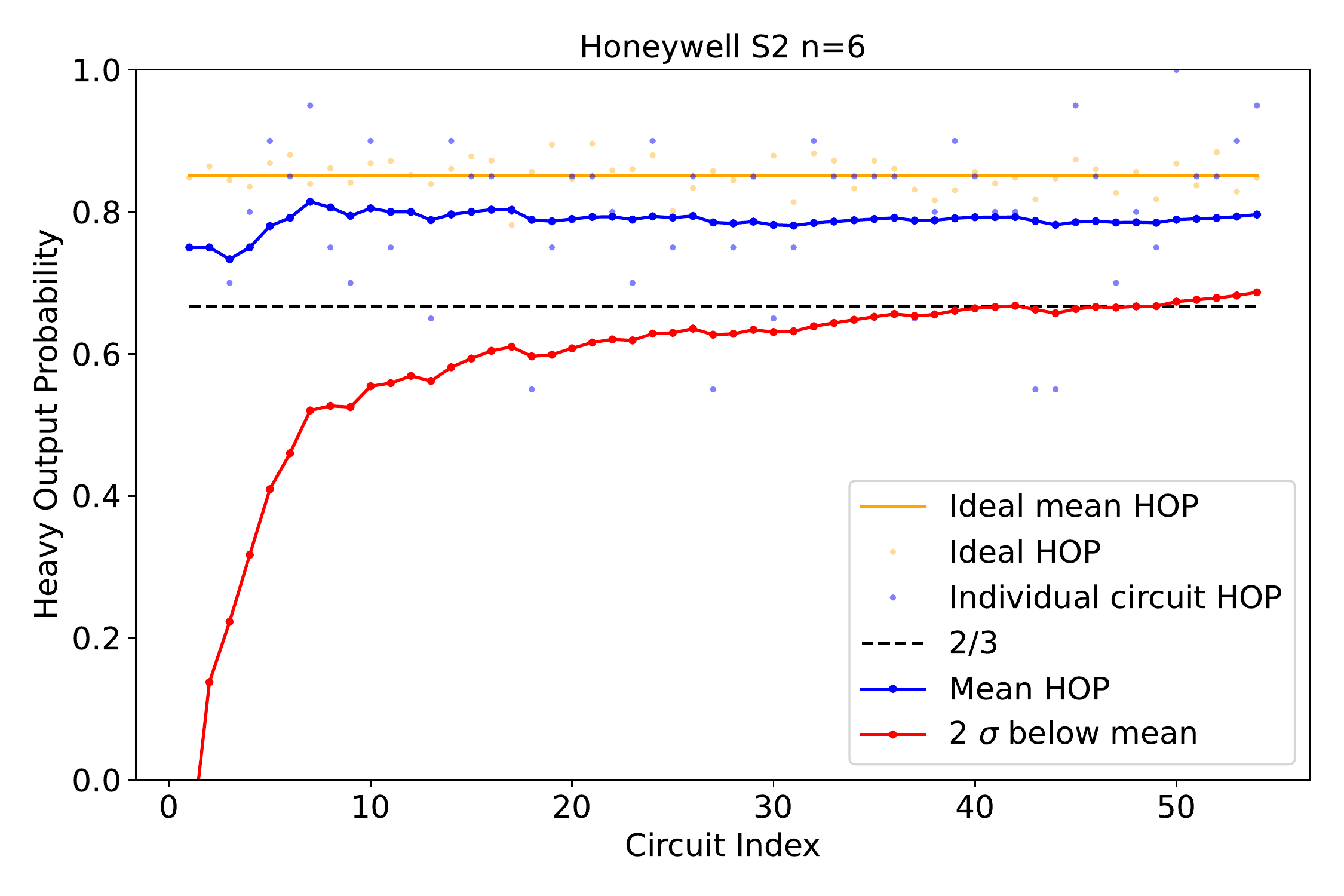}\\
    \includegraphics[width=0.49\textwidth]{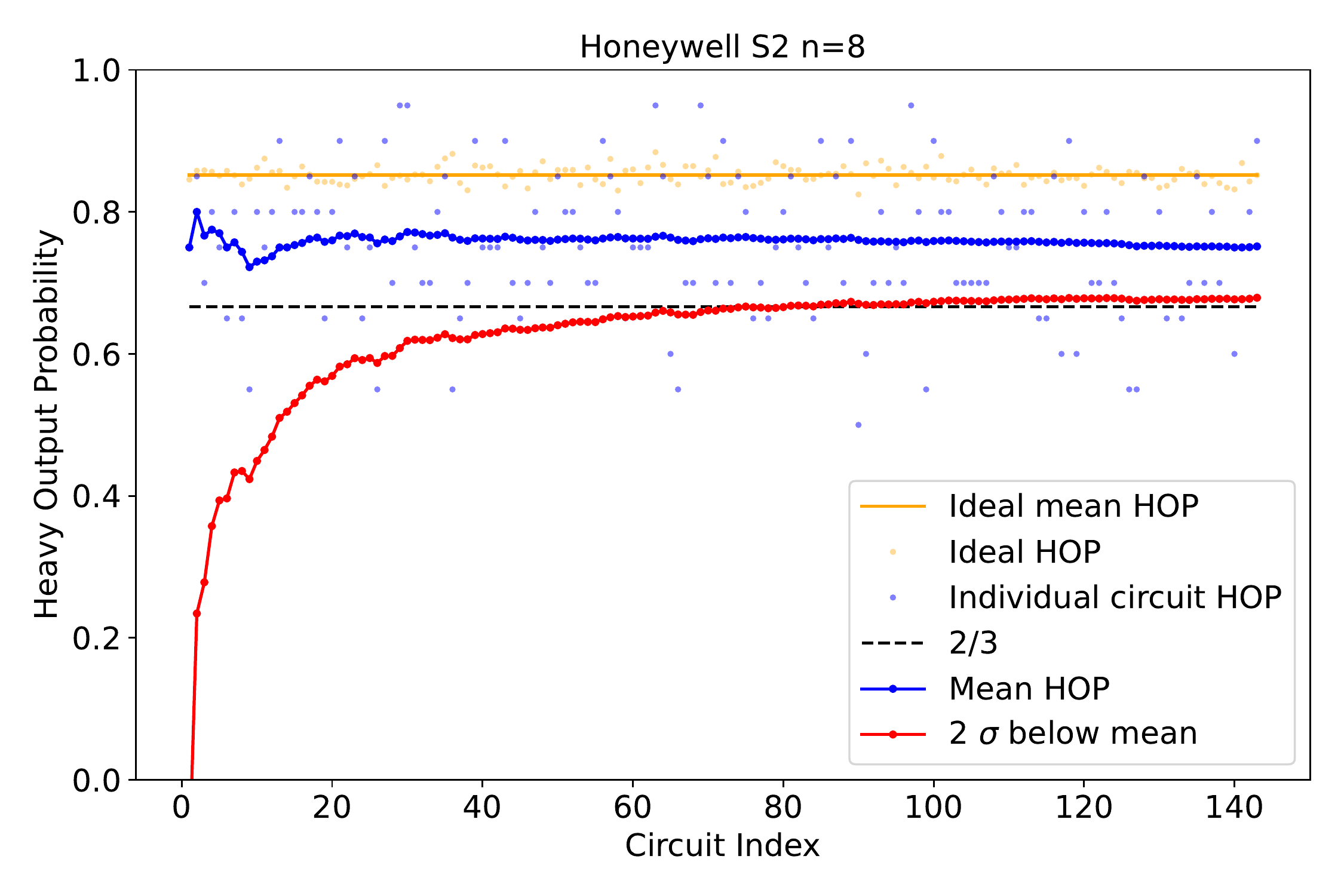}\hfill%
    \includegraphics[width=0.49\textwidth]{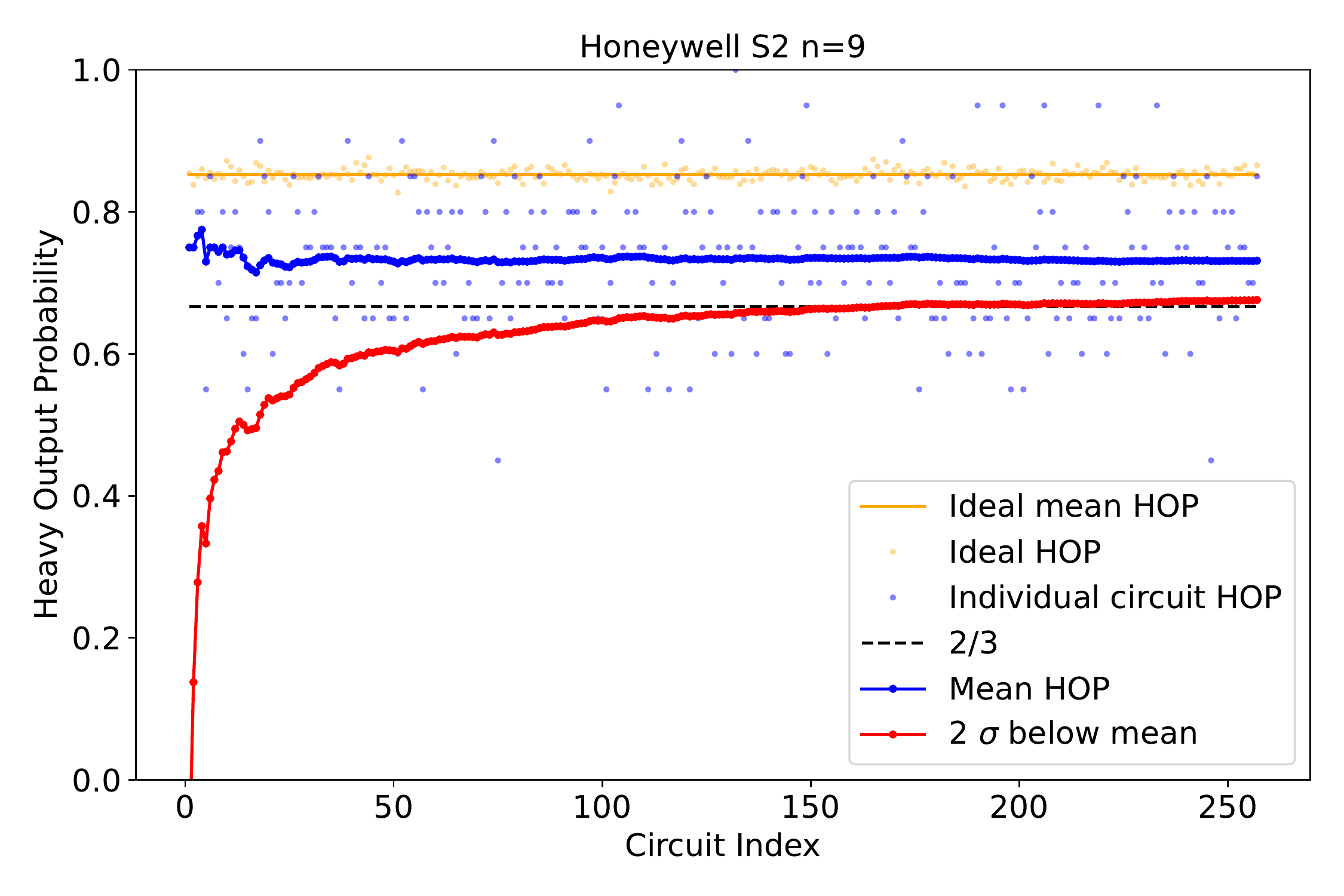}
    \caption{Heavy Output Probabilities as a function of circuit index for the Quantinuum (formerly Honeywell) \texttt{H1-2} device at $n=5$ (top left), $n=6$ (top right), $n=8$ (bottom left), $n=9$ (bottom right). Mean HOP and $2 \sigma$ below the mean are both computed cumulatively using the historical data up to point $i$ along the x-axis. }
    \label{fig:Quantinuum}
\end{figure*}

Next we will detail the full QV circuit compilation procedure, a high level diagram of which is shown in Figure~\ref{fig:flowchart}. Across all backends (except the Quantinuum backend) we use the Qiskit transpiler method~\cite{Qiskit} to take the quantum volume circuits and compile them to a specific gateset and a specific qubit layout. The gateset supplied for compilation corresponds to the native gateset that is supported by the backend to the best of our knowledge. No other information (e.g. gate execution times, error rates, etc) is supplied to the transpiler. The software versions we use are \texttt{qiskit=0.33.1}, \texttt{qiskit-terra=0.19.1}, and \texttt{amazon-braket-sdk=1.9.5}, up to \texttt{amazon-braket-sdk=1.21.0}. 

When submitting jobs through the Amazon Braket SDK (this includes the backend providers IonQ, Rigetti, and OQC), there is a small but important difference between the Amazon Braket SDK and Qiskit; Amazon Braket does not have a measure gate. Instead, the measure gates are applied implicitly to qubits which had gates applied. This becomes important in the $n=3$ QV circuit case, where it is possible to create a QV circuit where all $3$ layers are only acting on 2 qubits throughout the entire QV circuit; resulting in an idle qubit. In the Qiskit implementation, this qubit was still measured (even though no gate operations had been applied to it). Therefore, to maintain consistency with Qiskit and to not encounter non-contiguous qubit indexing errors, in the Amazon Braket implementation a single identity gate was applied to the idle qubit. 

The IBMQ backends allow \emph{jobs} to be submitted to the backend, where a job is a collection of circuits. However, there are constraints on the number of circuits (and the number of samples) that can be used per job. For all IBMQ experiments, we submit 4 jobs each composed of $250$ QV circuits in order to execute each sequence of $1,000$ QV circuits. Thus, while these jobs are submitted in sequence, there is possibly a time delay in between these $4$ blocks of circuits. 

Figures in this article were generated using Qiskit~\cite{Qiskit} and Matplotlib~\cite{Hunter:2007, thomas_a_caswell_2021_5194481}.

\subsection{Black-box quantum volume}
\label{sec:methods_black_box_qv}
The simplest method we use is to simply submit the uncompiled circuits to the specified backend (and let the backend or system handle compilation). However, how this method is implemented varies depending on the backend. In some cases, directly submitting the uncompiled circuits to the backend is not possible because the gateset is incompatible or the software is incompatible, therefore requiring custom code to be developed to handle this conversion (for example in the case of the backends provided through Amazon Braket). In other cases the system allows for very quick and direct submission of the uncompiled circuits (for example in the case of Quantinuum). We give details in the following subsections for each hardware vendor.

\subsubsection{Quantinuum}
\label{sec:methods_black_box_qv_Quantinuum}
The black-box method to access the \texttt{H1-2} Quantinuum backend is simple: The Honeywell API allows users to submit OpenQASM~\cite{cross2017open} code to each of these backends with no initial compilation needed. 

In accordance with the black-box execution approach, we do not compile or optimize these circuits at all before submitting them to the backend via the provided Quantinuum Python API. The API allows the optional \texttt{no-opt} compiler flag to be specified, which for the black-box approach we set to the default (which is False), which allows the backend to perform compiler optimizations. The current published QV value of the \texttt{H1-2} system is 4096~\cite{S2-QV-published-4096} (increased from the previous 2048~\cite{S2-QV-published}).

Our access to the Quantinuum backend was granted through Oak Ridge National Laboratories OLCF program.

\begin{figure*}[h]
    \centering
    \includegraphics[width=0.49\textwidth]{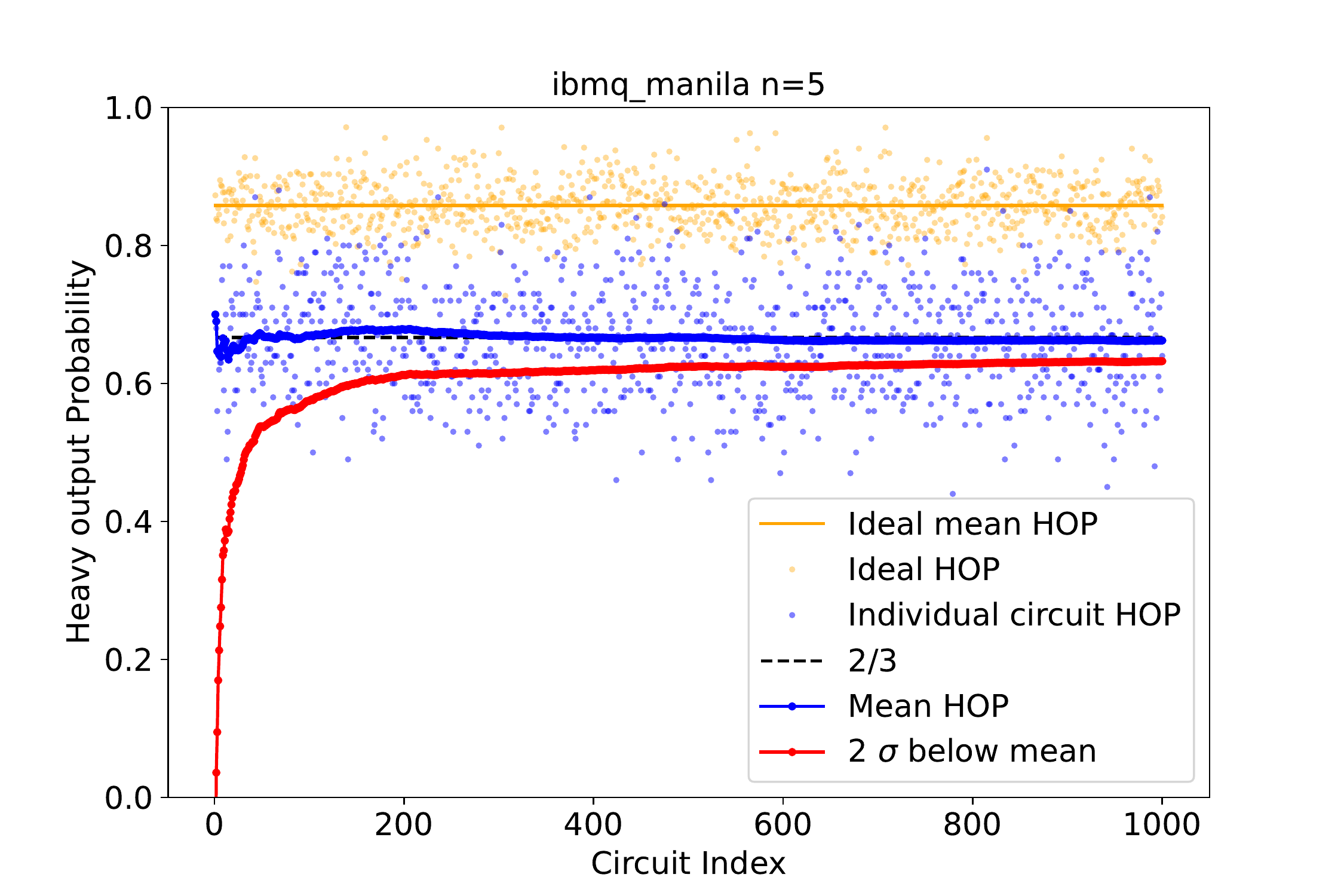}\hfill%
    \includegraphics[width=0.49\textwidth]{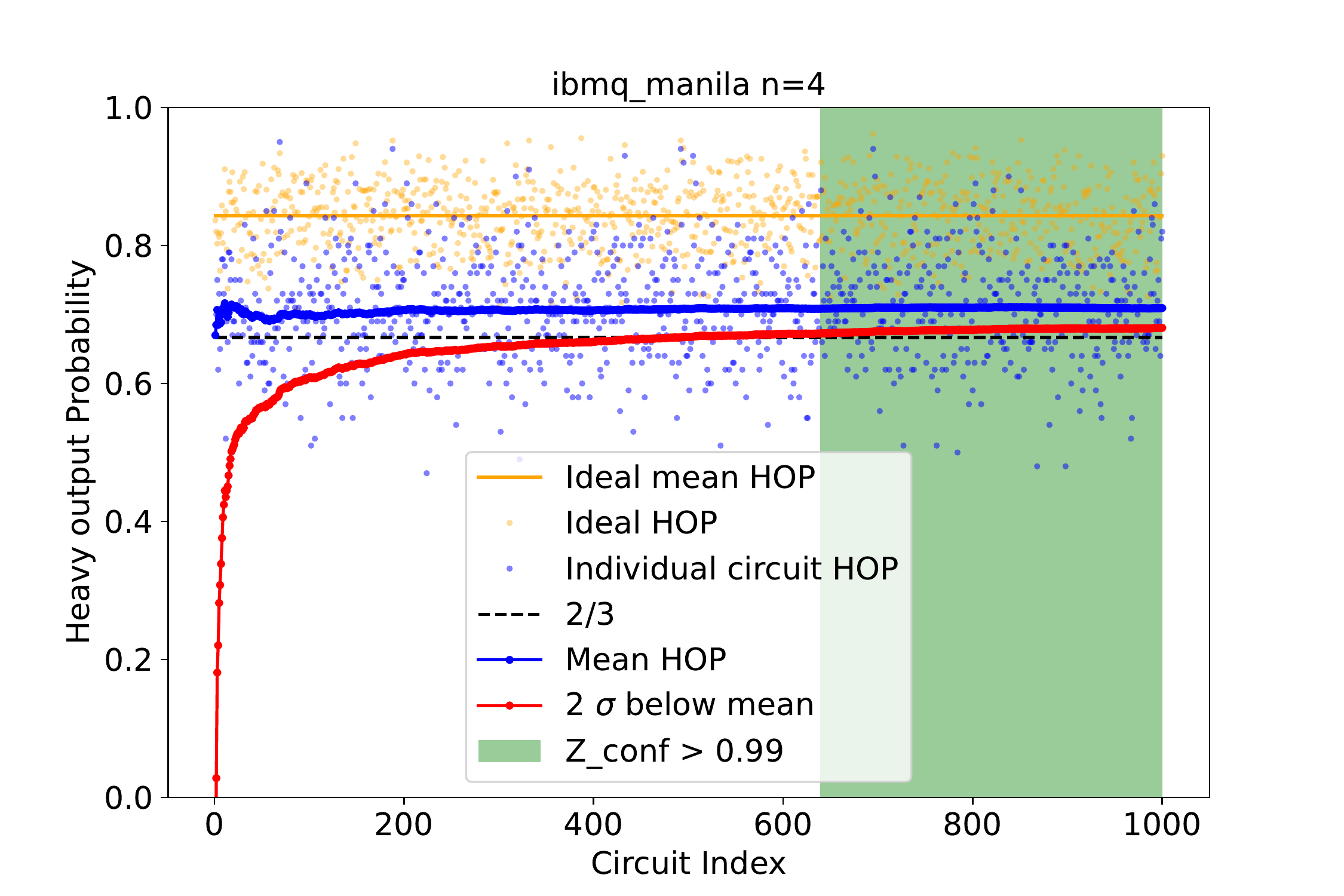}
    \caption{Heavy Output Probabilities (HOP) as a function of circuit index for the \texttt{ibmq\_manila} device using black-box circuit execution for $n=4$ (right) and $n=5$ (left).}
    \label{fig:ibmq_blackbox}
\end{figure*}

\subsubsection{Black-box IBM~Q}
\label{sec:methods_black_box_qv_ibmq}
For the IBM-Q backends, the black-box method we use is to call the Qiskit \texttt{execute} method~\cite{ibm-quantum, Qiskit, execute-method} using the flags \texttt{optimization\_level=3} and \texttt{layout\_method=noise\_adaptive}. Although not specified by the user, \texttt{sx, rz, cx, x} are the IBM~Q basis gates. This method compiles the original QV circuits onto that backend. Note that the Qiskit transpiler does not always successfully compile a group of circuits; the transpiler exits after unsuccessfully compiling for 1,000 iterations. Therefore, for a group of circuits we run the \texttt{execute} method until it successfully compiles. Overall, this makes the method not real-time efficient.

\subsubsection{IonQ}
\label{sec:methods_black_box_ionq}
The 11 qubit IonQ Harmony backend was accessed through Amazon Braket. In order to submit jobs through this service, the circuits need to be specified using the supported gates. In order to help the compilation, we compile the QV circuits to the IonQ gateset. The compiler we use in this stage, as with the other backends, is the Qiskit~\cite{Qiskit} transpiler. Here the Qiskit gateset we use for compilation is \texttt{rxx, ry, rz, rx}. Once converted to QASM, we convert the Qiskit gateset to a supported Amazon Braket gateset; \texttt{xx, ry, rz, yx} (with \texttt{XX} being a native two qubit gate supported by the IonQ backend~\cite{Grzesiak2020}). These gates can then be converted to Amazon Braket SDK code, and submitted to the IonQ backend. However, the circuit that is compiled and run on the backend is not visible to the user. 

IonQ has not published the quantum volume of the 11 qubit trapped ion quantum computer Harmony, available through Amazon Braket~\cite{lubinski2021applicationoriented}. However, there has been application oriented benchmarking of the 11 qubit device~\cite{Wright2019}.

\subsubsection{Oxford Quantum Circuits (OQC)}
\label{sec:methods_black_box_OQC}
The Oxford Quantum Circuits backend \texttt{Lucy}~\cite{coaxmon} can be accessed through Amazon Braket. Using the Qiskit transpiler the QV circuits were compiled using the uni-directional Linear-Nearest-Neighbors (LNN) ring gate connectivity, optimization level 3, and basis gates \texttt{rz}, \texttt{sx}, \texttt{x}, and \texttt{ECR} (these are the basis gates for the OQC Lucy backend). Then these circuits are converted into Amazon Braket syntax and submitted to the backend. As with the other Amazon Braket available backends, we allow the compilation software stacks of Amazon Braket and OQC handle all aspects of compilation and qubit assignment. One of the options the user has when submitted circuits to the OQC Lucy backend via Amazon Braket is to specify whether to turn off all circuit optimizations (this is accomplished in the Amazon Braket case by creating the circuit with a \emph{verbatim\_box} option), or to leave on all circuit optimizations. By default we leave on all circuit optimizations which allows the OQC compiler to select which qubits to use and to optimize the circuit (Figure~\ref{fig:QV_circuit_drawings} circuit \textbf{(10)} shows what one of these optimized circuits looks like, having been returned from the backend).

\begin{figure*}[h]
    \centering
    \includegraphics[width=0.49\textwidth]{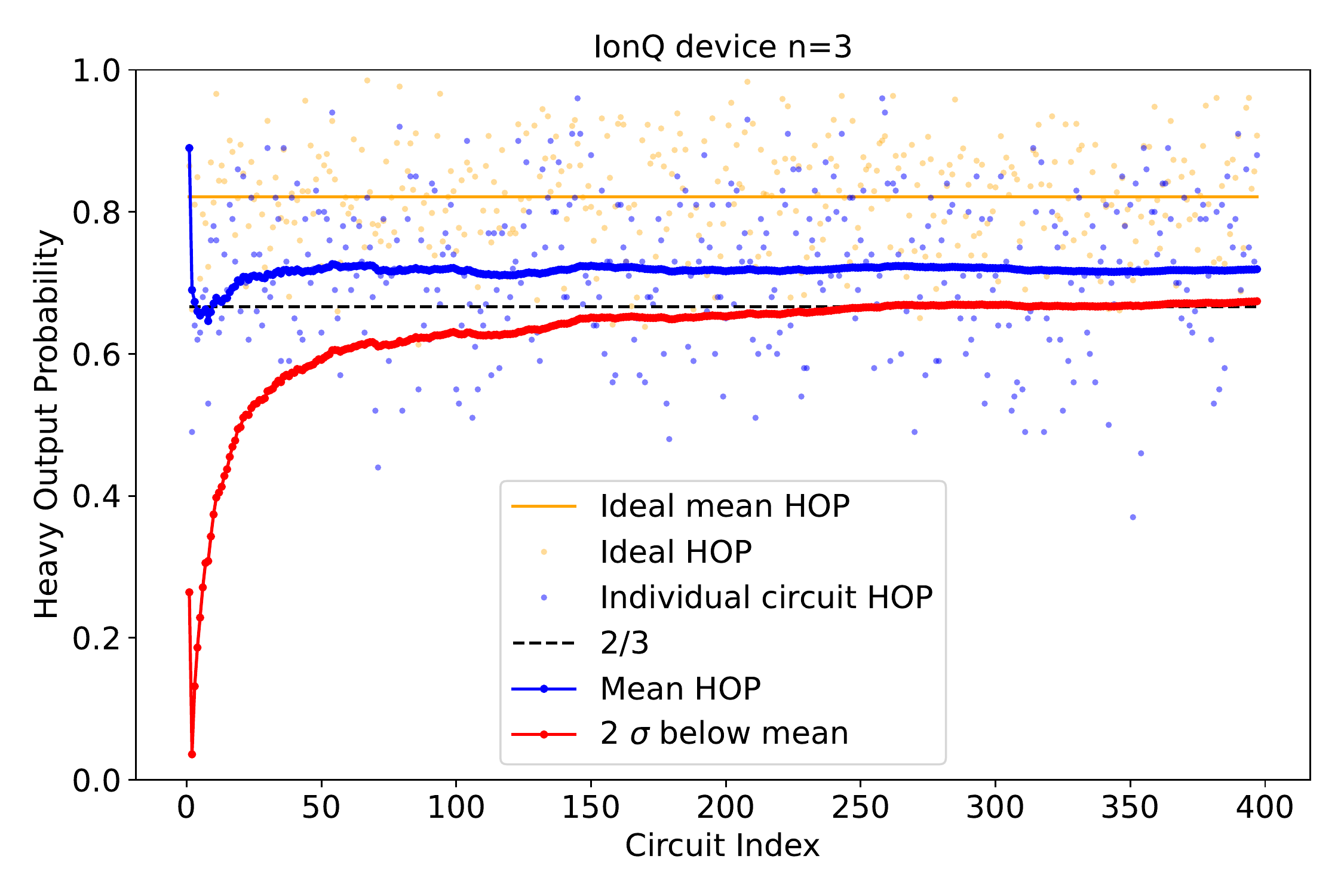}\hfill%
    \includegraphics[width=0.49\textwidth]{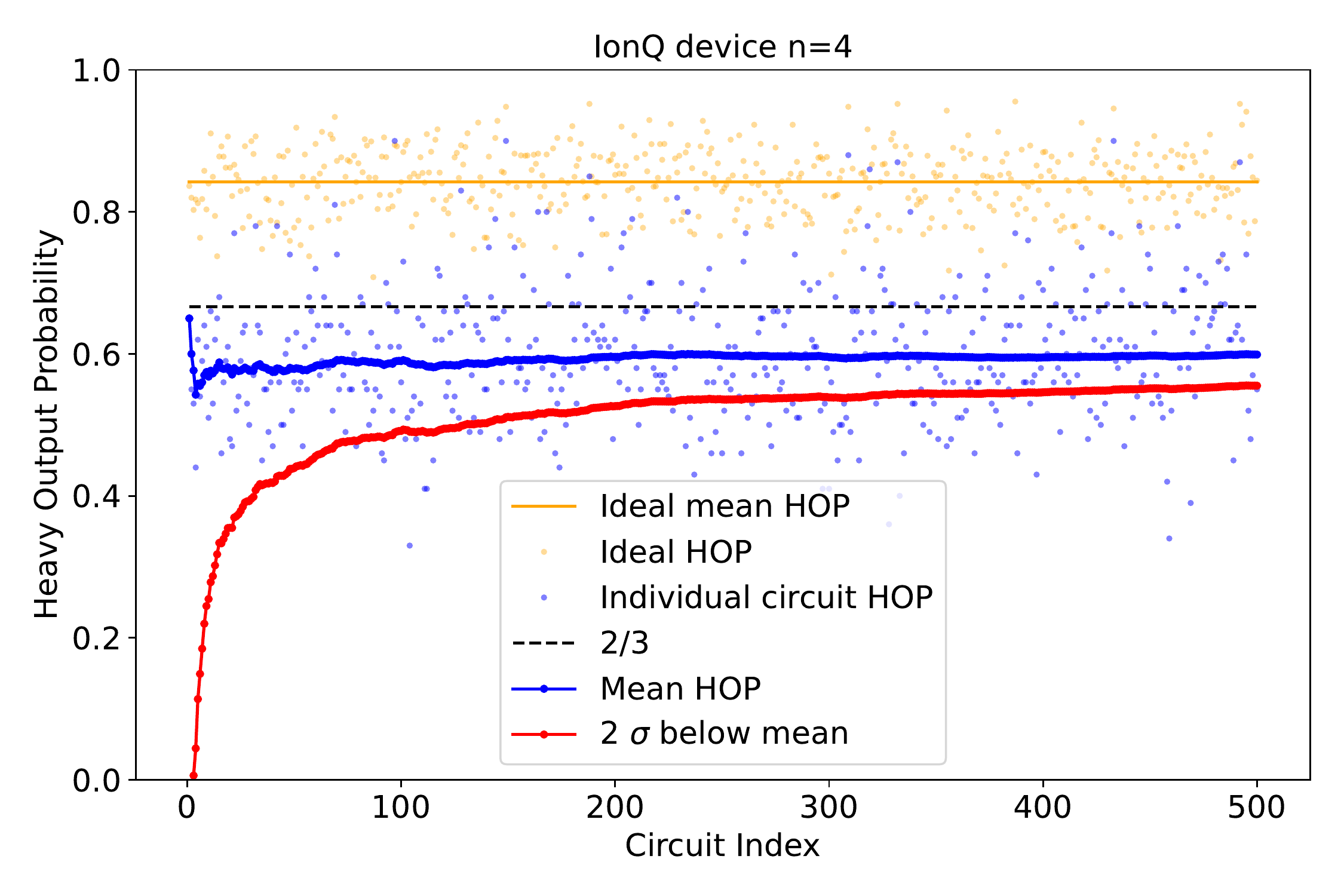}
    \caption{IonQ Harmony HOP plots for $n=3$ (left) and $n=4$ (right): IonQ Harmony passes the n=3 QV protocol at about 380 circuits, but is still far from passing at n=4 after 500 circuits.}
    \label{fig:IonQ}
\end{figure*}

\subsubsection{Black-box Rigetti}
\label{sec:methods_black_box_rigetti}

The route we took for the blackbox compilation on Rigetti devices was by accessing the \texttt{Aspen-11} and \texttt{Aspen-M-1} devices through Amazon Braket.The Qiskit gateset we use for compilation is \texttt{cz, rz, rx}. Once compiled to QASM, we convert the Qiskit gateset to a supported Amazon Braket gateset; \texttt{cz, rz, rx}. Note that \texttt{XY} is also a native gate of the Rigetti devices~\cite{Abrams2020}, however it is not currently a supported gate for the Qiskit transpiler. Additionally \texttt{CPHASE} is another native gate for the Rigetti backends, however it was not used by the Qiskit to implement the QV circuits in this compilation route. These gates (\texttt{CPHASE, XY, cz, rz, rx}) form the native gateset of the \texttt{Aspen-11} and \texttt{Aspen-M-1} devices~\cite{Karalekas_2020}. Once each circuit is represented in the Amazon Braket compatible gateset, it is submitted to the Rigetti backends. From there, the Rigetti QUILC compiler~\cite{smith2020opensource} compiles the supplied circuit to the backend connectivity based on the latest calibration data of the backend. The resulting compiled circuit is sent back to the user as Quil code, allowing us to analyze the circuit that was executed on the backend~\cite{smith2016practical, robert_s_smith_2020_3677541, Karalekas_2020, appleby_2020_3762258}. In other words, for the black-box approach we let the compilation stacks of Amazon Braket and Rigetti handle compilation and qubit assignment. 

The quantum volume of the \texttt{Aspen-11} and the \texttt{Aspen-M-1} have both been measured to be $8$~\cite{rigetti-clops}, and previous Rigetti Aspen devices (specifically Aspen-4) have had a measured quantum volume of $8$~\cite{Karalekas_2020} as well. 

\subsection{Rigetti Qubit Subset Enumeration}
\label{sec:methods_rigetti_qubit_subset_enumeration}
For a more detailed analysis of Rigetti Aspen-11 and Aspen-M-1 devices, we used the Rigetti Cloud Computing Services (QCS) platform in order to perform qubit subset enumeration for the QV circuits. Figure~\ref{fig:flowchart} shows the data pipeline for this compilation; starting at the uncompiled QV QASM circuits we employ the Qiskit transpiler to convert the circuits into a connectivity matching a qubit subset of either Aspen-11 or Aspen-M-1 using the gateset of rz, rx and cx. Next this representation was converted into the Quil instruction set~\cite{smith2016practical} and then the Rigetti QUILC compiler~\cite{smith2020opensource} was called in order to compile the Quil program into a device compatible form. Lastly the circuit was submitted to the device to be executed (using 100 samples). 

One important distinction between this method and the black-box Rigetti approach described in Section~\ref{sec:methods_black_box_rigetti} is the use of \emph{Active Reset}~\cite{Karalekas_2020}. Active Reset decreases the amount of computation time used on the device by speeding up the process of qubit reset; the cost is a decrease in qubit fidelity. For the Aspen-11 and Aspen-M-1, the Active Reset fidelities, based on vendor provided calibration data on the QCS platform are typically approximately 99\%. The Amazon Braket execution process (i.e. the blackbox approach of Section~\ref{sec:methods_black_box_rigetti}) always uses Active Reset. By contrast, in the case of the qubit subset enumeration done on the QCS platform, we do not use the Active Reset capability. This difference can be directly seen in Figure~\ref{fig:QV_circuit_drawings}; circuit \textbf{(6)} was submitted through Amazon Braket and it specifies Active Reset for the three qubits at the start of the circuit, whereas circuit \textbf{(7)} was compiled using QUILC for the purpose of the qubit subset enumeration procedure, and no active reset is used in that circuit.

\subsection{Qiskit Transpiler with Qubit Subset Enumeration: IBM~Q}
\label{sec:methods_connected_subg_qv_ibmq}

In order to more thoroughly test QV values achievable, we specify which groups of qubits to compile the QV circuits to, instead of letting the compiler (local or backend) handle this. This is specified using the \emph{initial layout} parameter in the Qiskit transpiler. 

The Qiskit gateset we use for compilation is \texttt{sx, rz, cx, x}. We compile each of the 1,000 QV circuits onto each of the connected subgraphs of each of the IBM~Q backends. We learned that this compilation process is quite time intensive, and requires HPC multiprocessing. We spent approximately 100,000 CPU hours (including the custom QV passmanager compilation time) to compile the $n=3$ through $n=7$ circuits onto 19 different IBM~Q backends (see Table~\ref{table:ibmq_connected_subgraph_results}). The main reason for the enormous amount of compile time required is that the compiler often reaches a \emph{maximum 1,000 iteration error}, thus possibly requiring many attempts to compile a given circuit onto a connected subgraph of the hardware. Although this is an allowed component of the QV protocol, it is not time efficient for users; the black-box compilation (Section~\ref{sec:methods_black_box_qv_ibmq}) is closer to what a typical user would implement. 

The arguments we use for the transpiler method are the \texttt{coupling\_map} of the backend, \texttt{optimization\_level=3}, \texttt{initial\_layout} of the connected subgraph, and \texttt{basis\_gates=x, sx, cx, rz}. Occasionally, the Qiskit transpiler will use some neighboring qubits outside of this connected subgraph in an attempt to improve circuit fidelity. Based on the compilation described in~\cite{QV2019}, this is an accepted part of the definition for QV. One of the causes of the transpiler reaching a \emph{maximum 1,000 iteration error} is because of a qubit subset choice that is poorly chosen. For this procedure, we do not care about the order of the qubit subset choice (because the order of the qubits used in the circuit can be remapped for different connectivities); therefore we also randomly shuffle the qubit subset while attempting to compile each circuit.

\subsection{Custom Compilation using QV Passmanager: IBM~Q}
\label{sec:methods_ibmq_high_fidelity}
Lastly we use the custom QV compiler techniques introduced in~\cite{nannicini2021optimal, Jurcevic_2021, PRXQuantum.1.020318} in order to compare how heavier compilation affects the measured QV compared to using the standard Qiskit transpiler. The software used in these experiments is published on the \emph{qiskit-tutorials} Github~\cite{qv64-source1-github, qv64-source2}. Specifically we implement the same qubit subset compilation of Section~\ref{sec:methods_connected_subg_qv_ibmq}, except we now use the custom QV compiler~\cite{qv64-source1-github, qv64-source2}. In particular, we attempt to compile a sorted qubit subset, and then several random permutations of the qubit subset. However, some circuits fail to compile to some qubit subsets using the custom QV compiler. The passmanger flags are all set to the default, in particular these are flags we specify: \texttt{basis\_gates}, \texttt{coupling\_map}, \texttt{qubit\_subset}, \texttt{backend\_props=backend.properties()}, \texttt{instruction\_durations} (provided from the backend), \texttt{synthesis\_fidelity=.99}, \texttt{pulse\_optimize=True}. 

This method requires even more computation time than the compilation with qubit subset enumeration from Section~\ref{sec:methods_connected_subg_qv_ibmq}. The routing and qubit assignment optimization~\cite{nannicini2021optimal, Jurcevic_2021} is done with the CPLEX solver~\cite{cplexv12}; meaning that this method requires a CPLEX license to compile these circuits. For all compilation we set the \texttt{BIPMapping} (i.e. the CPLEX optimization of routing and qubit assignment) timeout to $5000$ seconds~\cite{BIP-mapping}. This custom transpilation also uses Qiskit Pulse~\cite{Alexander_2020, PRXQuantum.1.020318} optimization to increase circuit fidelity. The use of Qiskit Pulse additionally requires precise timing of the gate instructions, which is enforced by delay gates in the circuits. An example of the usage of these delay gates can be seen in Figure~\ref{fig:QV_circuit_drawings} circuit \textbf{(4)}. 

Additionally, we cut off compilation of all of the circuits after a few days using HPC resources; in some cases there were circuits (and qubit subset choices) that were not attempted to be compiled because of the time constraint we set. Therefore, either because of compiler errors preventing compilation, or because of the time constraint we imposed, not all circuits for all qubit subsets could be compiled across the backends we tested using this compilation method. We also restricted these compilations to a subset of the available IBM~Q backends, as opposed to Section~\ref{sec:methods_connected_subg_qv_ibmq} where we compiled circuits for all available IBM~Q backends. 

\begin{figure}[h]
    \centering
    \includegraphics[width=0.49\textwidth]{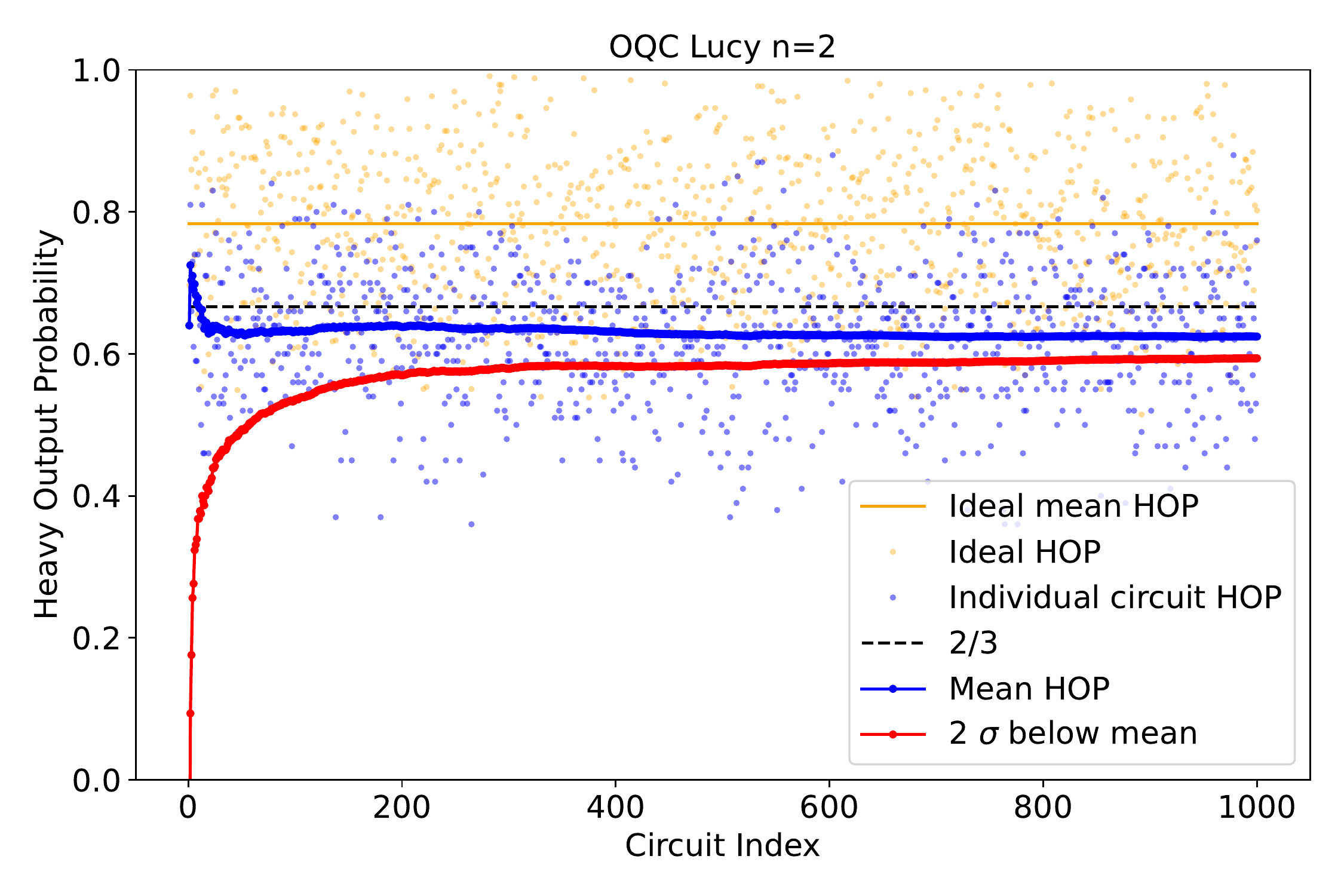}
    \caption{OQC Lucy backend HOP plot for $n=2$ show that the mean HOP is consistently below $\frac{2}{3}$}
    \label{fig:OQC_Lucy}
\end{figure}

\begin{figure}[h!]
    \centering
    \includegraphics[width=0.49\textwidth]{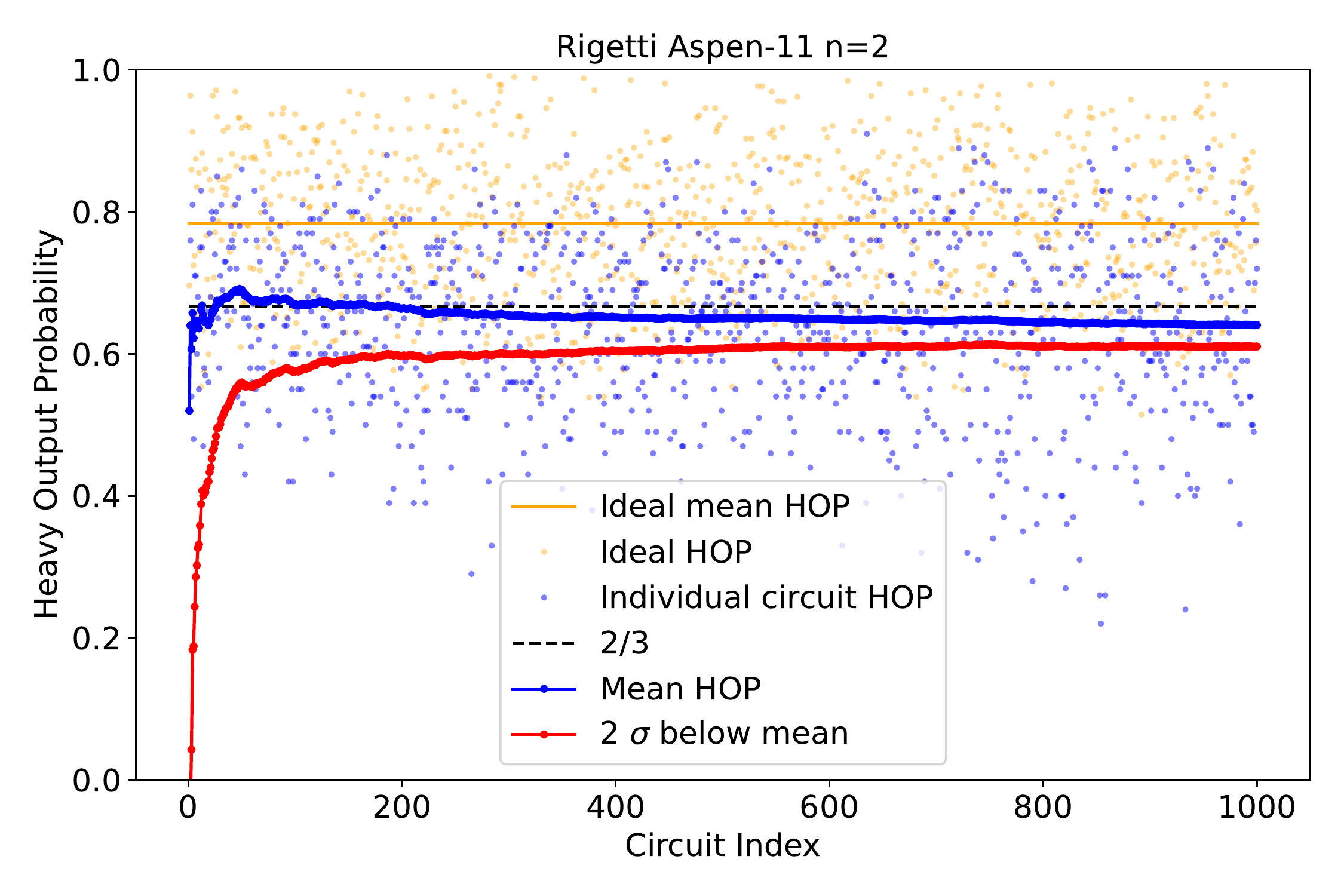}
    \includegraphics[width=0.49\textwidth]{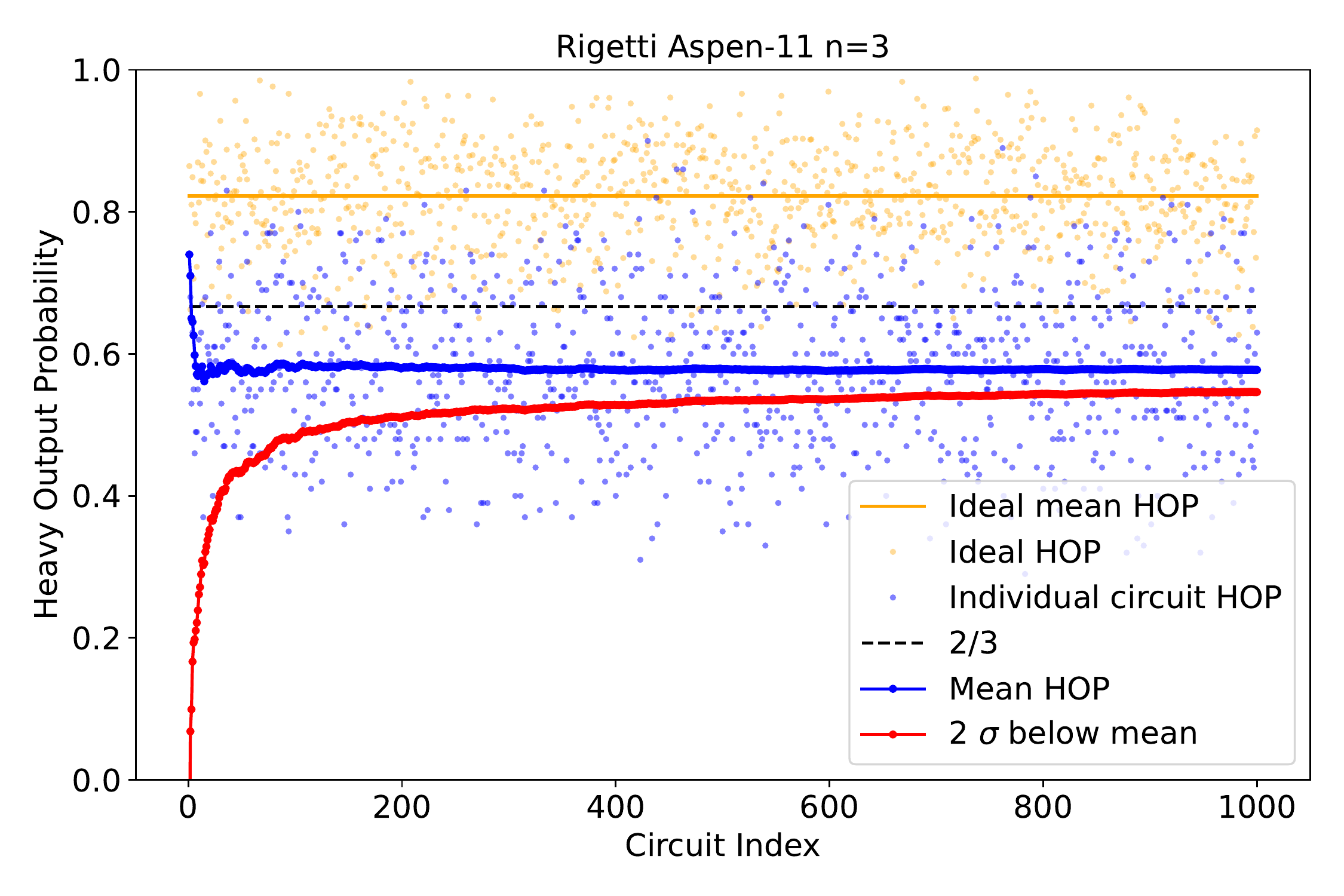}
    \includegraphics[width=0.49\textwidth]{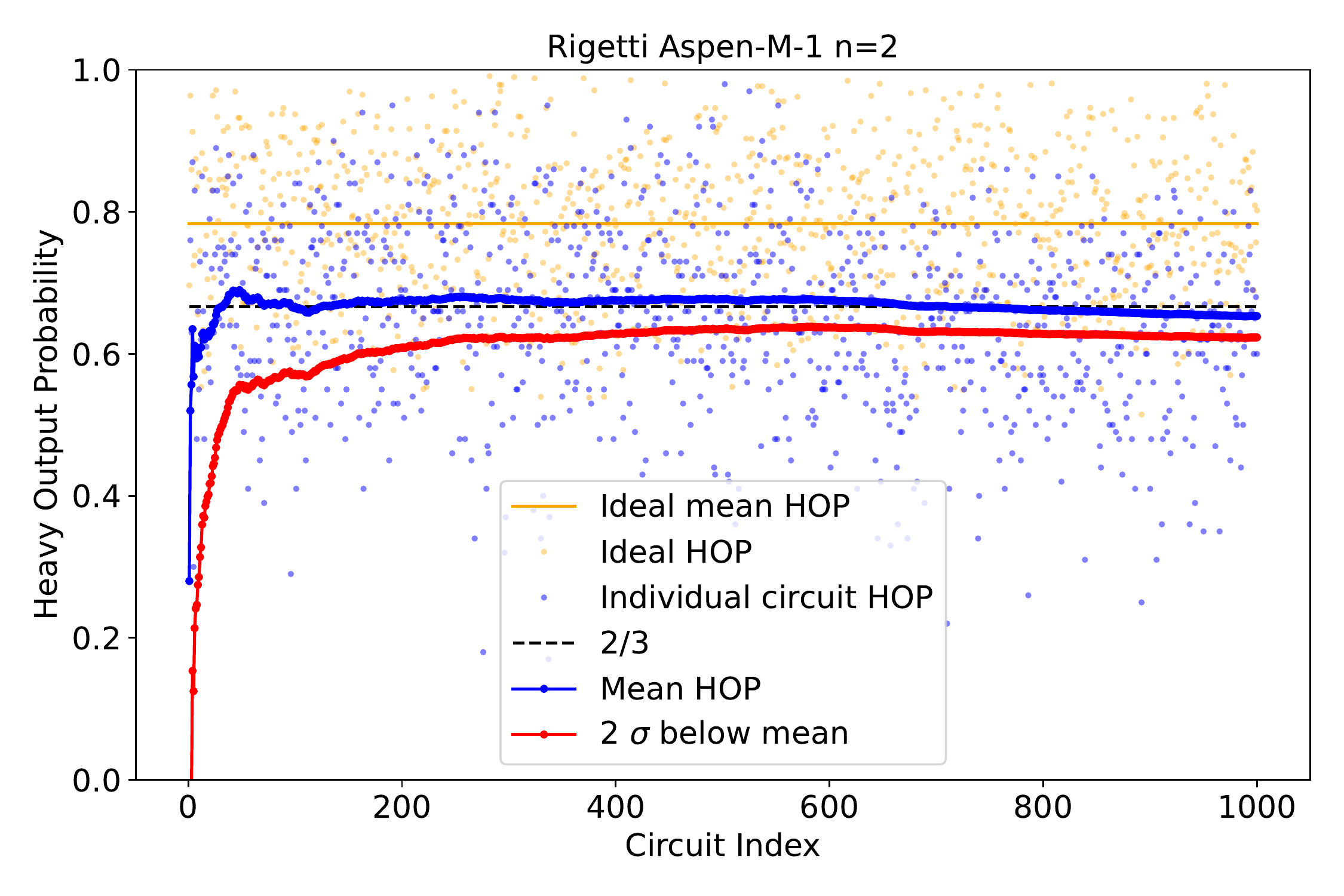}
    \caption{Black-box execution of Rigetti \texttt{Aspen-11} HOP distribution for $n=2$ (top left) and $n=3$ (top right), and \texttt{Aspen-M-1} HOP distribution for $n=2$ (bottom). We see that at $n=2$ the mean HOP is very close to $\frac{2}{3}$, at times passing $\frac{2}{3}$ for a smaller number of circuits, but the $2 \sigma$ value consistently remains below $\frac{2}{3}$. }
    \label{fig:rigetti}
\end{figure}

\section{Results}
\label{sec:results}
We compile and execute the QV circuits on the backends listed in Table~\ref{table:NISQ_Devices}. First we show results for black-box compilation and execution, which is the most general and widely available method across the backends (for example, not all backends allow full specification of which qubits to use in the circuit). Next we show results for IBM~Q backends when we enumerate compilation across the qubit subsets (of size $n$) of a backend. This allows us to characterize the QV protocol results across the entire chip of the IBM~Q backends. Lastly, we evaluate the \texttt{custom IBM~Q QV passmanager} for circuit compilation~\cite{qv64-source1-github, qv64-source2, Jurcevic_2021} on a restricted set of IBM~Q backends and connectivities on those backends. Unless otherwise noted, all experiments used 100 samples for each circuit execution. 

\begin{figure*}[h!]
    \centering
    \includegraphics[width=0.49\textwidth]{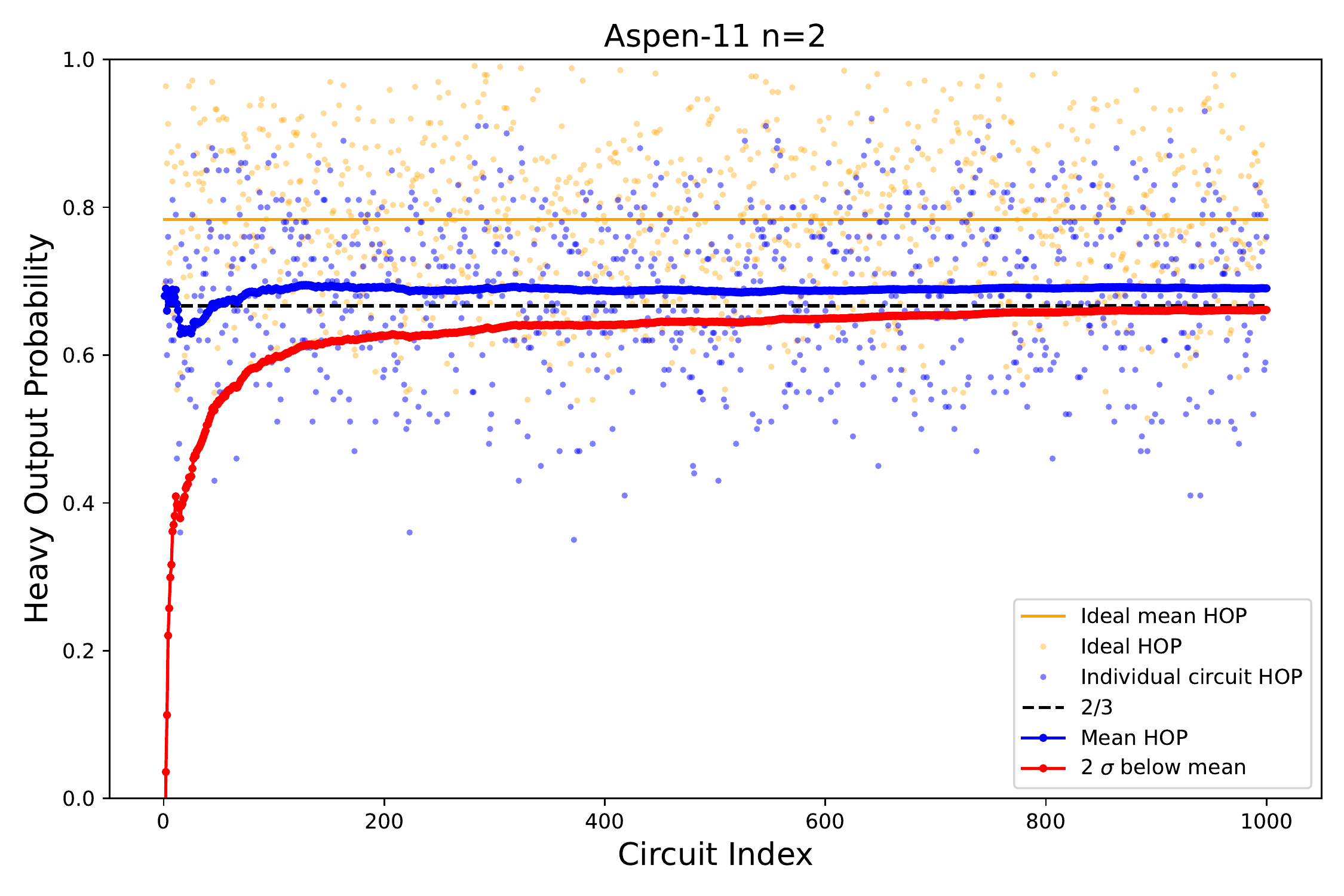}\hfill%
    \includegraphics[width=0.49\textwidth]{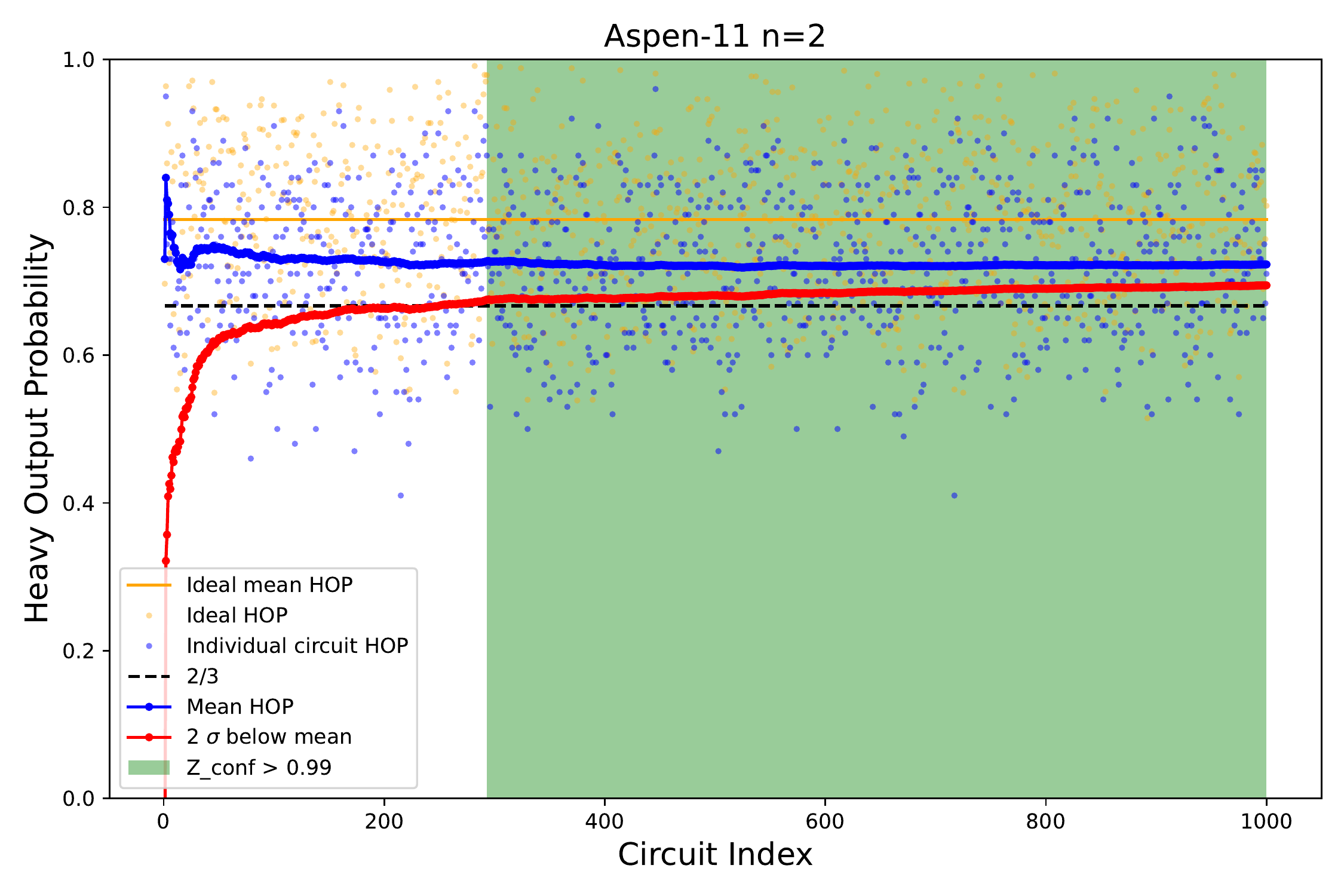}\\
    \includegraphics[width=0.49\textwidth]{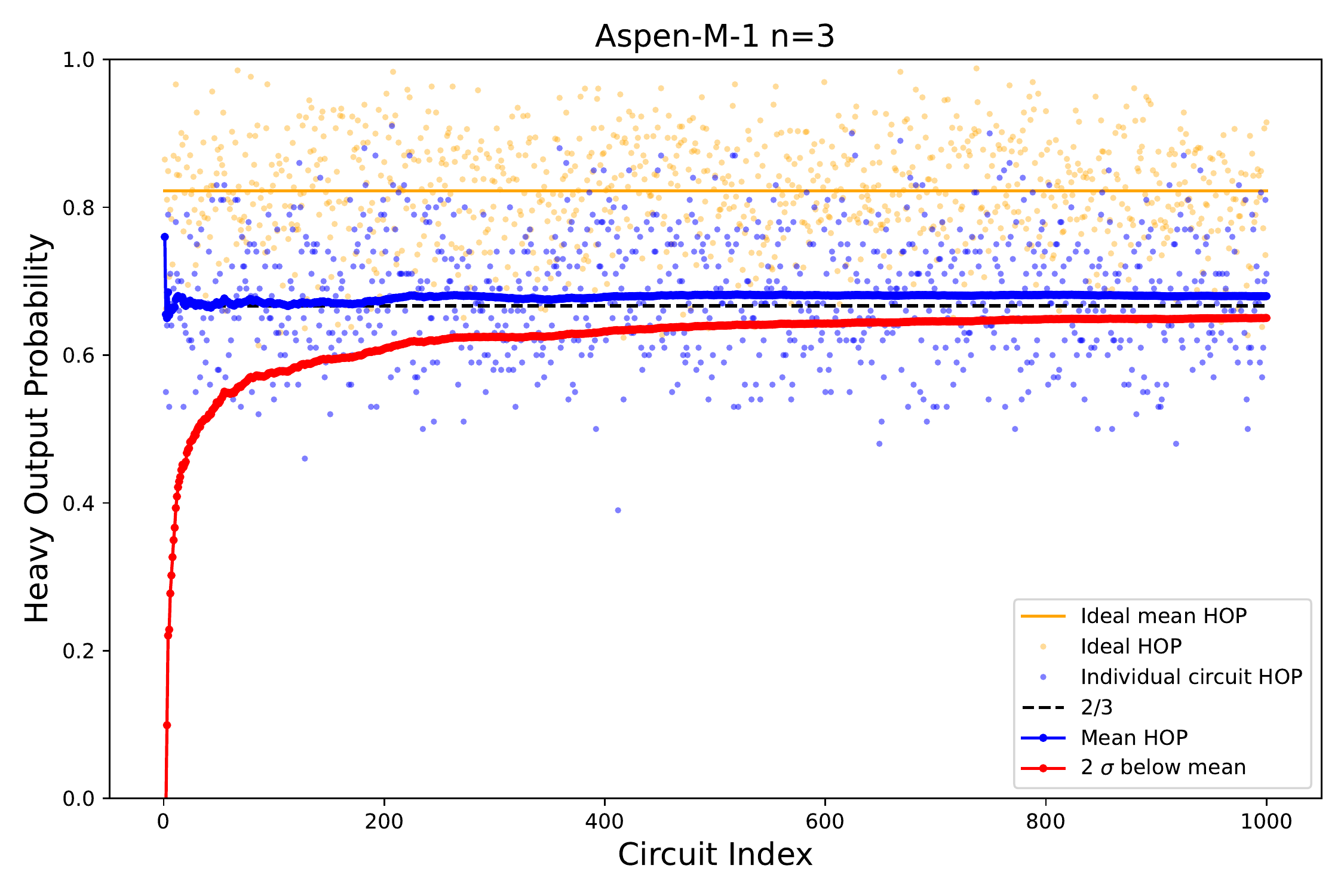}\hfill%
    \includegraphics[width=0.49\textwidth]{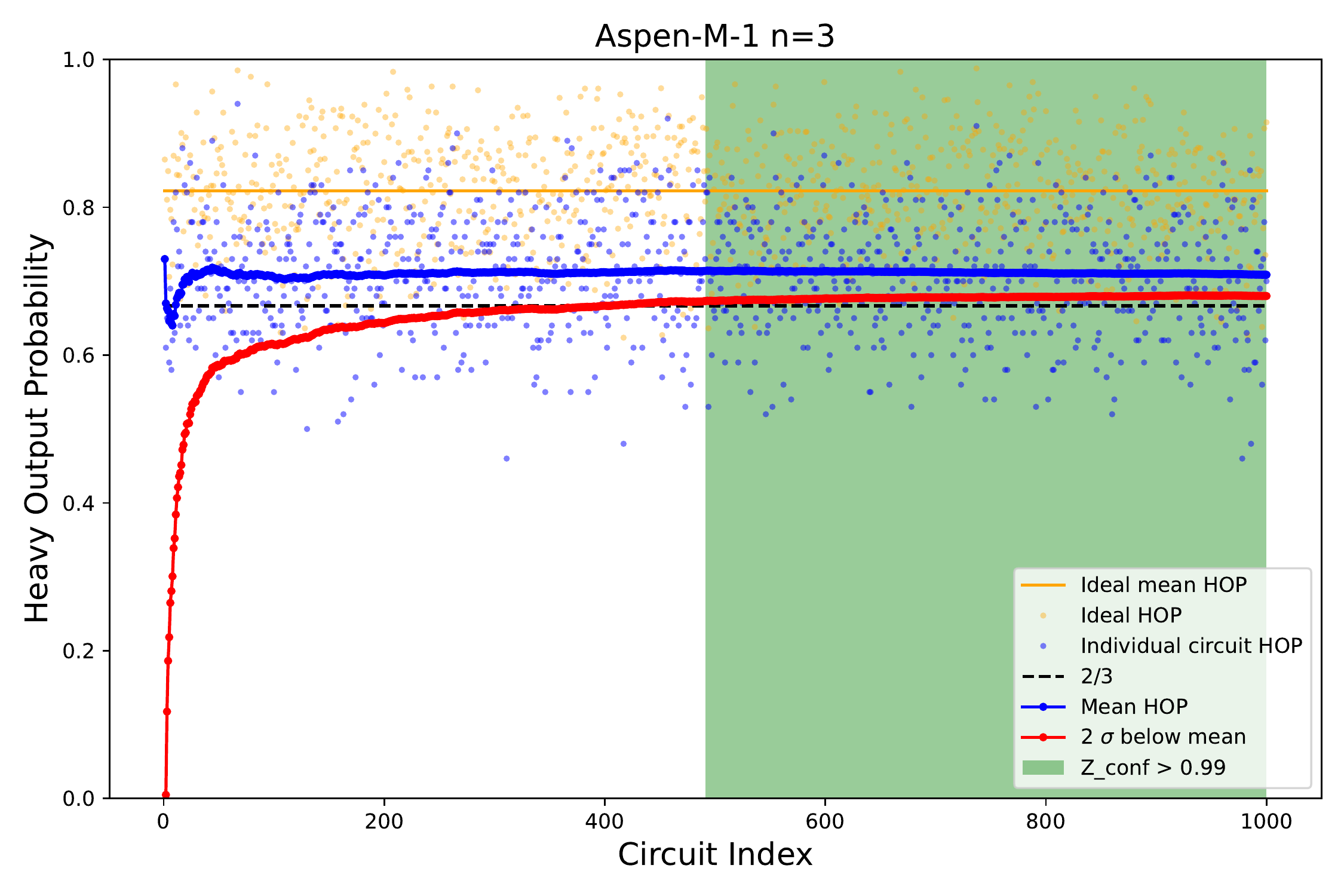}
    \caption{Cumulative HOP plots for Rigetti qubit subset enumeration procedure. Top left: $n=2$ QV circuits run on qubits 14 and 15 of the Aspen-11 device did not pass a mean HOP above $\frac{2}{3}$. Top right: $n=2$ on qubits 26 and 27 of the Aspen-11 device passed the QV test after roughly 300 circuits. Bottom left: $n=3$ QV circuits on qubits 146, 131 and 132 of Aspen-M-1 did not pass $\frac{2}{3}$, whereas $n=3$ QV circuits ran on qubits 112, 113 and 114 passed the test after roughly 500 circuits. }
    \label{fig:rigetti_qubit_subset}
\end{figure*}

Figure~\ref{fig:QV_circuit_drawings} show the differences in circuit compilation starting from the original un-compiled circuit to the compiled circuits that are submitted to the different devices. In order from the top to bottom circuit drawings in Figure~\ref{fig:QV_circuit_drawings}:

\noindent
\textbf{(1)} uncompiled QV circuit; all subsequent circuits use this description to compile to the other gatesets or connectivities. 

\noindent
\textbf{(2)} Qiskit transpiled circuit on qubits 0, 1, 2 on ibm\_perth with optimization level $3$. Corresponding methodology described in Section~\ref{sec:methods_connected_subg_qv_ibmq}. 

\noindent
\textbf{(3)} Transpiled using the black-box \texttt{execute} method on ibm\_perth. Corresponding methodology described in Section~\ref{sec:methods_black_box_qv_ibmq}. 

\noindent
\textbf{(4)} Compiled using the custom IBM~Q QV passmanager with pulse level optimization. Corresponding methodology described in Section~\ref{sec:methods_ibmq_high_fidelity}. 

\noindent
\textbf{(5)} Qiskit compiled circuit for the Rigetti Aspen-11 backend that was submitted to Amazon Braket for the blackbox Rigetti method. Corresponding methodology described in Section~\ref{sec:methods_black_box_rigetti}. 

\noindent
\textbf{(6)} Compiled Quil circuit that was executed on the Rigetti Aspen-11 backend having been submitted through Amazon Braket. Corresponding methodology described in Section~\ref{sec:methods_black_box_rigetti}. The active reset instructions on all three qubits can be seen at the start of the circuit. 

\noindent
\textbf{(7)} QUILC compiled circuit on Aspen-M-1 for the qubit subset enumeration method. This circuit used native XY two qubit gates, but in this Qiskit circuit drawing we use iSWAP gate in place of the XY($\pi$) gates since those two are equivalent. Corresponding methodology described in Section~\ref{sec:methods_rigetti_qubit_subset_enumeration}. 

\noindent
\textbf{(8)} Qiskit compiled circuit that was submitted to the IonQ Harmony backend (having been converted into Amazon Braket python code). Corresponding methodology described in Section~\ref{sec:methods_black_box_ionq}. 

\noindent
\textbf{(9)} Qiskit compiled circuit to the Lucy OQC gateset and uni-directional connectivity that was submitted to the backend via Amazon Braket. Corresponding methodology described in Section~\ref{sec:methods_black_box_OQC}. 

\noindent
\textbf{(10)} OQC Lucy backend compiled circuit that was returned with the job metadata. Corresponding methodology described in Section~\ref{sec:methods_black_box_OQC}. 

In Figure~\ref{fig:QV_circuit_drawings}, the difference in structure that results from the same logical $n=3$ circuit is perhaps surprising, even though some of the diversity can be explained with the different native gate sets that the compilers aim to optimize to. Interestingly, when comparing circuit \textbf{(6)} and circuit \textbf{(5)} in Figure~\ref{fig:QV_circuit_drawings}, we see that although the supplied QV circuit used $6$ two-qubit cz gates, the software stack that compiled the circuits chose to use a total of $9$ cz gates in the final Quil compiled circuit, which is sub-optimal. Note that although the OQC Lucy backend native gateset uses ECR as the two qubit gate, the returned QASM code that was compiled for the backend (circuit \textbf{(10)}) used CNOT gates for two qubit interactions.

\begin{figure*}[h]
    \centering
    \includegraphics[width=0.49\textwidth]{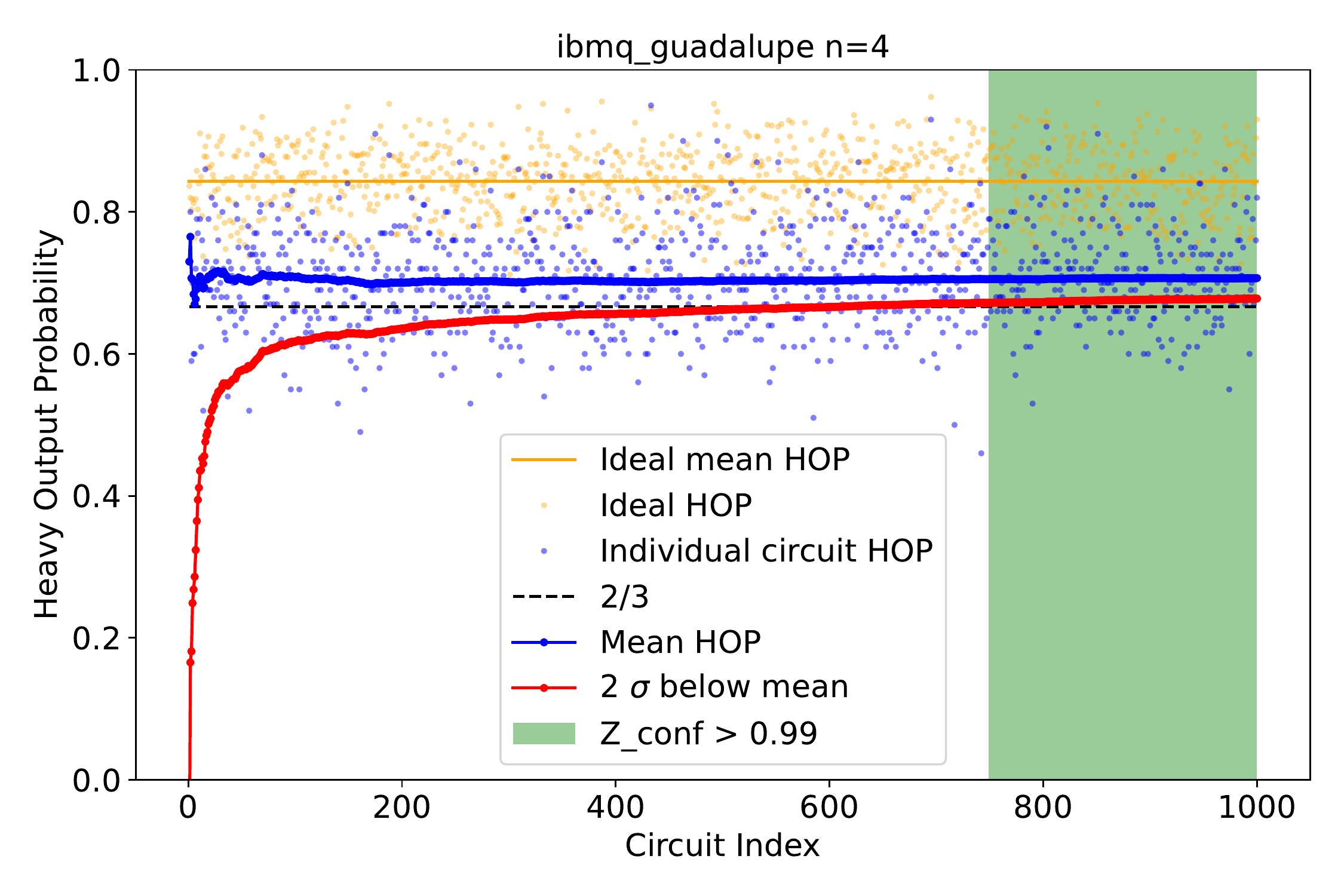}
    \includegraphics[width=0.49\textwidth]{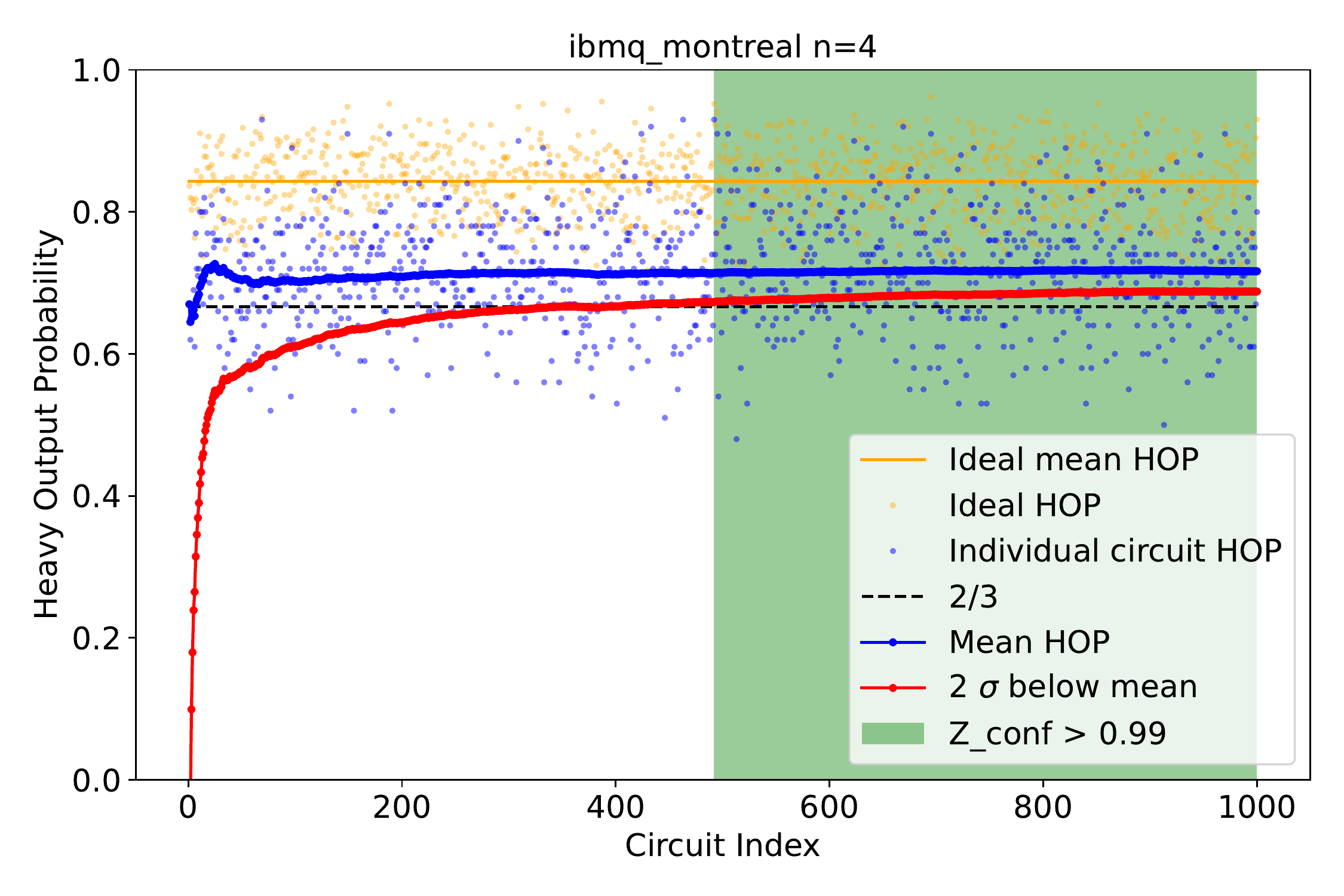}
    \caption{IBM~Q cumulative HOP on qubits \texttt{8, 9, 11, 14} of \texttt{ibmq\_guadalupe} (left) and qubits \texttt{16, 19, 20, 14} of \texttt{ibmq\_montreal} (right) at $n=4$. The shaded green regions show where the HOP distribution becomes statistically significant above $\frac{2}{3}$ with z-confidence $> 0.99$}
    \label{fig:IBMQ_connected_subgraph}
\end{figure*}

Figure~\ref{fig:circuit_statistics} show the average circuit statistics in terms of gate depth, one qubit gate counts, and two qubit gate counts, across the different device compiled circuits. Across different QV circuit sizes the \texttt{custom QV passmanager} compilation reduces the CNOT count on average compared to the two other IBM~Q compilation procedures. 

In order to visualize how the QV protocol progresses as we execute each circuit on the given QPU, we show cumulative Heavy Output Probability (HOP) figures where the x-axis is the circuit index, and the y-axis is the Heavy Output Probability (HOP). In these figures, we plot the ideal HOP distribution, the individual circuit measured HOP values, the cumulative mean of the HOP values (up to index $i$ in the plot), $2 \sigma$ below the cumulative HOP mean, and lastly we color shade the region with z-confidence $> 0.99$ if more circuits are executed past that confidence level (this can be seen in Figures~\ref{fig:ibmq_blackbox},~\ref{fig:IBMQ_connected_subgraph}, and~\ref{fig:heavy-compilation-ibmq}). We encourage close attention to the x-axis whenever comparing HOP plots: while we usually plot up to the full 1,000 circuits, we cut off earlier when it is clear that the test has been passed. 

In all HOP figures we plot both the mean HOP (solid orange horizontal line), as well as the individual ideal HOP values for each circuit (shown as high transparency orange points). Note that the 1,000 QV circuits have smaller ideal HOP at $n=2$ and $n=3$ compared to larger values of $n$. This can be seen in the ideal distributions of Figure~\ref{fig:rigetti} for $n=2$ compared to $n=5$ or greater plots (for example Figure~\ref{fig:Quantinuum}. This is to be expected for smaller QV circuit sizes~\cite{baldwin2021reexamining}, even though in the limit the ideal HOP distribution approaches $\frac{1+ln(2)}{2}$.

\subsection{Black-box Quantinuum}
\label{sec:results_Quantinuum}
As described in Section~\ref{sec:methods_black_box_qv_Quantinuum}, the QV circuits were submitted directly to the backend as the uncompiled QASM file (the circuit statistics on the uncompiled QV circuits can be see in Figure~\ref{fig:circuit_statistics}) which is entirely comprised of CNOT and U3 gates (see Figure~\ref{fig:QV_circuit_drawings}). No user side circuit optimization, basis gate conversions, or transpilation was performed on these circuits. 

Figure~\ref{fig:Quantinuum} shows that the \texttt{H1-2} device passes the QV test for circuit sizes up to $n=8$. In order to save resources, as with IonQ, execution was terminated once the QV protocol criteria were met. Due to usage constraints, larger circuit sizes are still being tested. Therefore, for the \texttt{H1-2} backend we can only provide a lower bound ($n=9$) on the QV value of the QPU. 
The $n=5$ experiments used $100$ shots for each circuit. The $n=6, 8, 9$ experiments used $20$ shots for each circuit. As a comparison, $n=8$ was reached after only 140 circuits, which is considerably shorter than in particular most IBM took to reach some of their best results.

\begin{figure*}[h]
    \centering
    \includegraphics[width=0.32\textwidth]{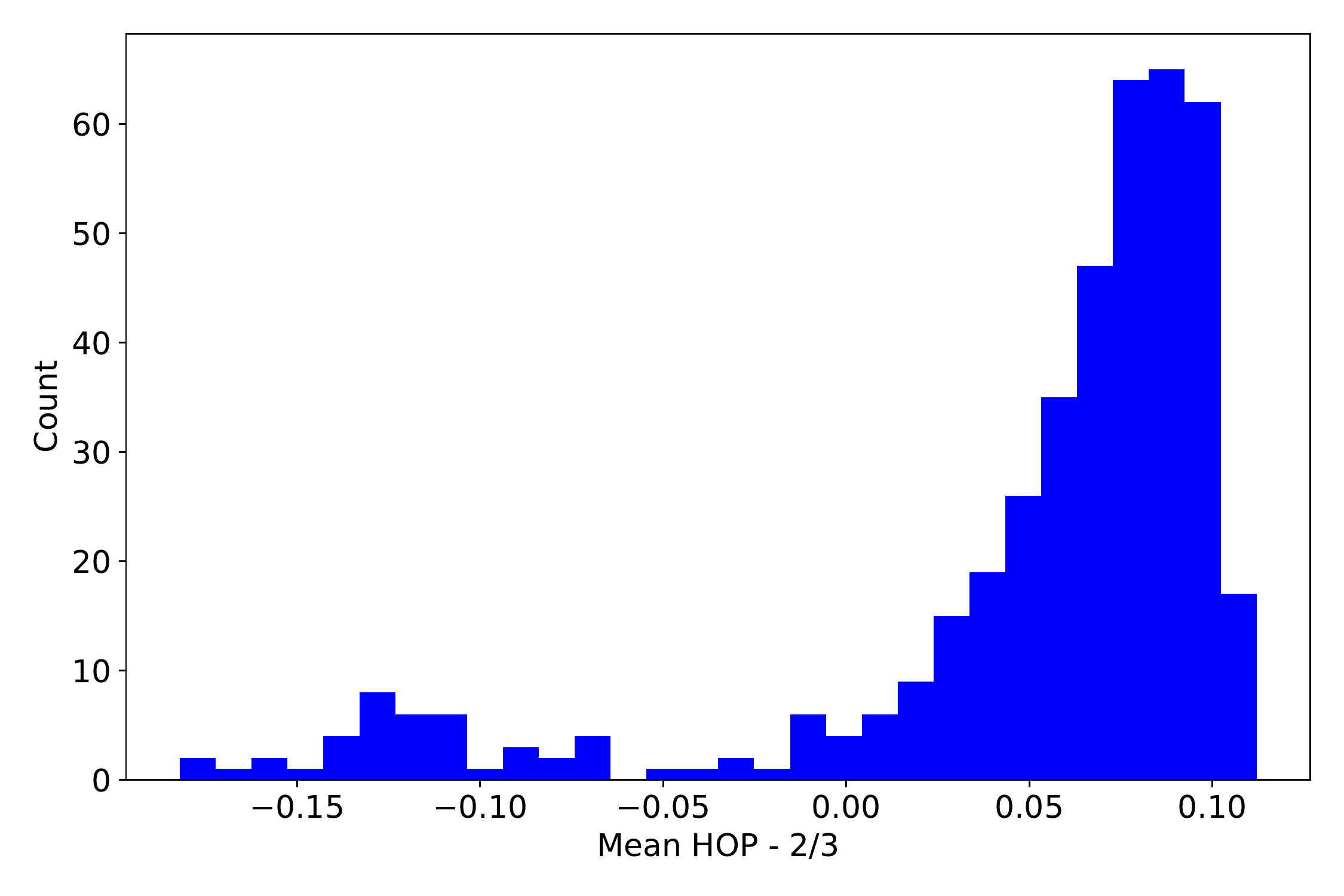}\hfill%
    \includegraphics[width=0.32\textwidth]{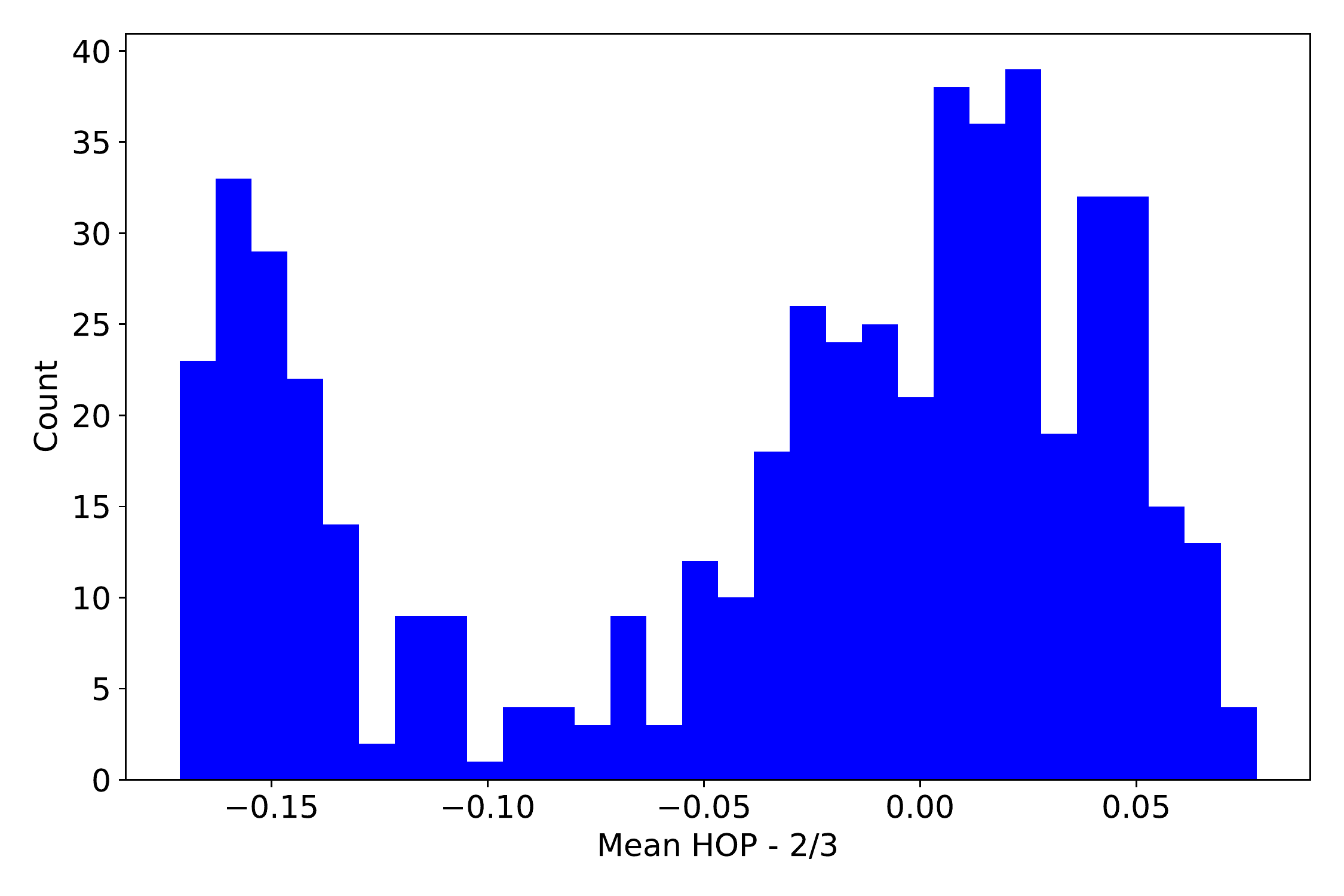}\hfill%
    \includegraphics[width=0.32\textwidth]{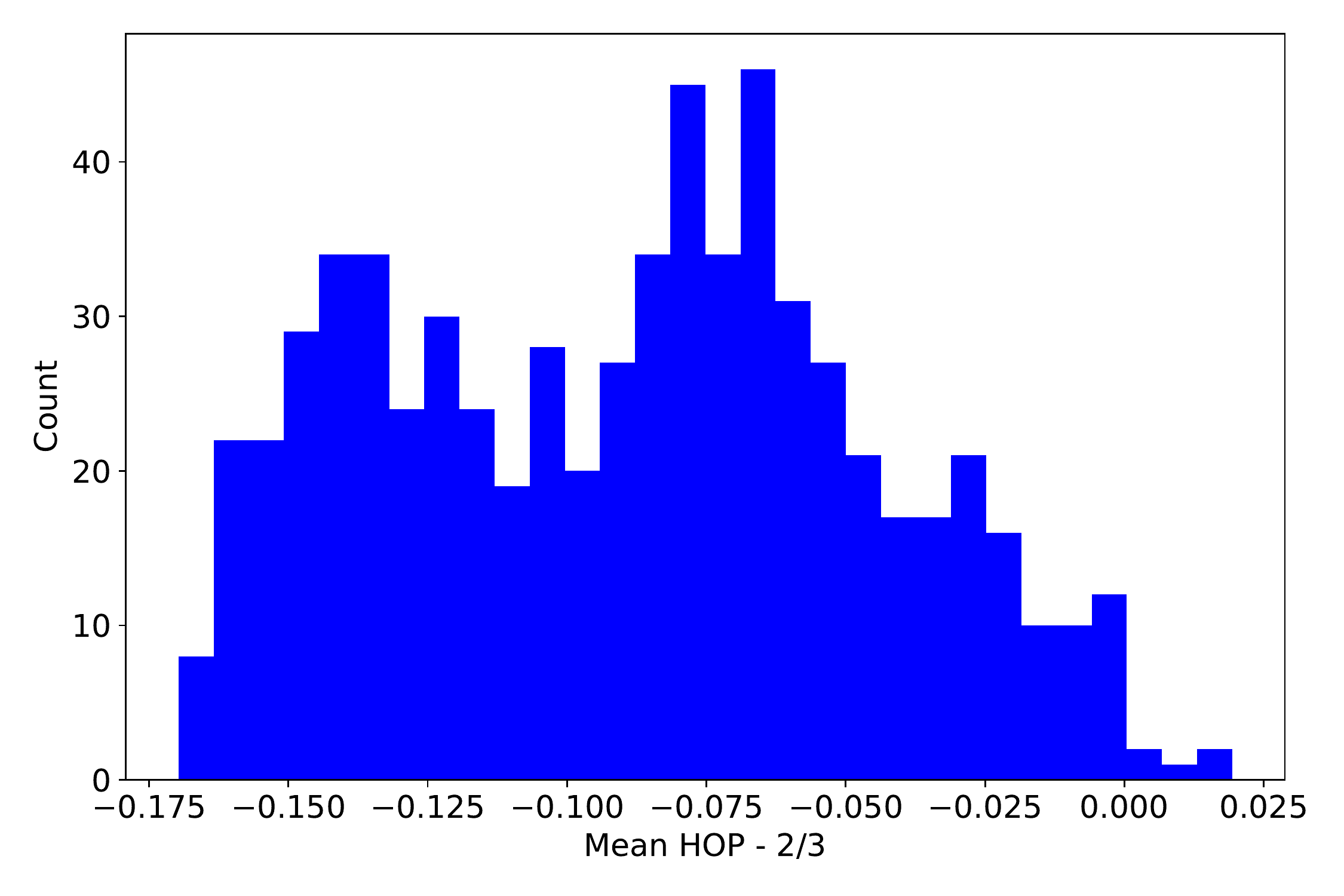}%
    \caption{Distribution of mean HOP minus $2/3$ across all IBM~Q backends and connected subgraphs (see the full QV results from this datatset in Table~\ref{table:ibmq_connected_subgraph_results}) for $n=3$ (left), $n=4$ not including \texttt{ibm\_washington} (middle), $n=5$ (right) when using the Qiskit transpiler in order to compile the raw circuits onto each backend and qubit subset choice. }
    \label{fig:histogram_HOP_diff_IBMQ}
\end{figure*}

\subsection{Black-box IBM~Q}
\label{sec:results_ibmq}
The QV results were nearly identical across all IBM~Q backends when using the black-box \texttt{execute} method. That is, every backend passed the $n=3$ QV test, and failed $n=4$ and $n=5$ with the exception of a particular qubit subset of \texttt{ibmq\_manila} for $n=4$. Table~\ref{table:NISQ_Devices} summarizes these results. Figure~\ref{fig:ibmq_blackbox} shows the results where \texttt{ibmq\_manila} passed for $n=4$ sized QV circuits, but failed to pass at $n=5$. 

\subsection{Black-box IonQ}
\label{sec:results_ionq}
Figure~\ref{fig:IonQ} shows that the 11 qubit IonQ backend passes the QV test at $n=3$ and fails to pass at $n=4$. For $n=3$, execution was stopped once the results passed the z-confidence threshold of $0.99$. Because the mean HOP for $n=4$ was definitively lower than $\frac{2}{3}$, we stopped execution at 500 circuits. 

\subsection{Black-box Oxford Quantum Circuits (OQC)}
\label{sec:results_OQC}
Figure~\ref{fig:OQC_Lucy} shows the HOP distribution from executing the 1,000 QV circuits at $n=2$ on the OQC Lucy backend. This plot shows that the mean HOP is consistently below $\frac{2}{3}$. In the compilation of these circuits, the OQC compiler applied optimizations to the circuit(s) that were submitted through Amazon Braket. 

\subsection{Black-box Rigetti}
\label{sec:results_rigetti}

Figure~\ref{fig:rigetti} shows that using black-box compilation and job submission, the Aspen-11 device fails to pass the QV test at $n=3$ and $n=2$. Note that $n=2$ was tested, unlike the other backends, since the $n=3$ test failed. We additionally tested the Aspen-M-1 backend for $n=2$, which also failed to pass the QV protocol. 

\subsection{Rigetti Qubit Subset Enumerations}
\label{sec:results_rigetti_qubit_subset_enumeration}

Here we summarize the results from executing the qubit subset enumeration QV protocol on Aspen-M-1 and Aspen-11 on the QCS platform~\cite{Karalekas_2020}. Not all qubit subsets of Aspen-11 and Aspen-M-1 were tested. For Aspen-11 at $n=2$, 12 out of the possible 48 connected qubit subsets were tested, 8 of which passed with a 99\% z-confidence level. For Aspen-11 at $n=3$, 51 out of the possible 72 connected qubit subsets were compiled to and executed, none of which passed with high confidence. For Aspen-M-1 at $n=3$ 26 out of the possible 184 connected qubit subsets were tested, one of which passed with a 99\% z-confidence level. Figure~\ref{fig:rigetti_qubit_subset} shows four specific examples of cumulative HOP distributions where $n=2$ both passed and failed on Aspen-11, and where $n=3$ both passed and failed on Aspen-M-1.

\subsection{Qiskit Transpiler with Qubit Subset Enumeration: IBM~Q}
\label{sec:results_ibmq_connected_subg}
Table~\ref{table:ibmq_connected_subgraph_results} shows the qubit subset enumeration results for each of the IBM~Q backends when the QV protocol is applied using the Qiskit transpiler with no modifications to the transpilation procedure; only compilation flag level 3 (which is the highest flag supported by this method), the connectivity graph, and the required basis gates are provided as additional transpiler arguments. Because of the size of the backend, only some of the qubit subsets of \texttt{ibm\_washington} at $n=4$ were tested. Running  $n=3, 5$ on \texttt{ibm\_washington} would have also required significant additional QPU time. 

Figure~\ref{fig:IBMQ_connected_subgraph} shows two HOP plots for two different IBM~Q backends at $n=4$. Importantly, this compilation procedure resulted in the highest QV value found across all tested IBM~Q backends was $n=4$ ($QV=16$). 

\begin{figure*}[h]
    \centering
    \includegraphics[width=0.24\textwidth]{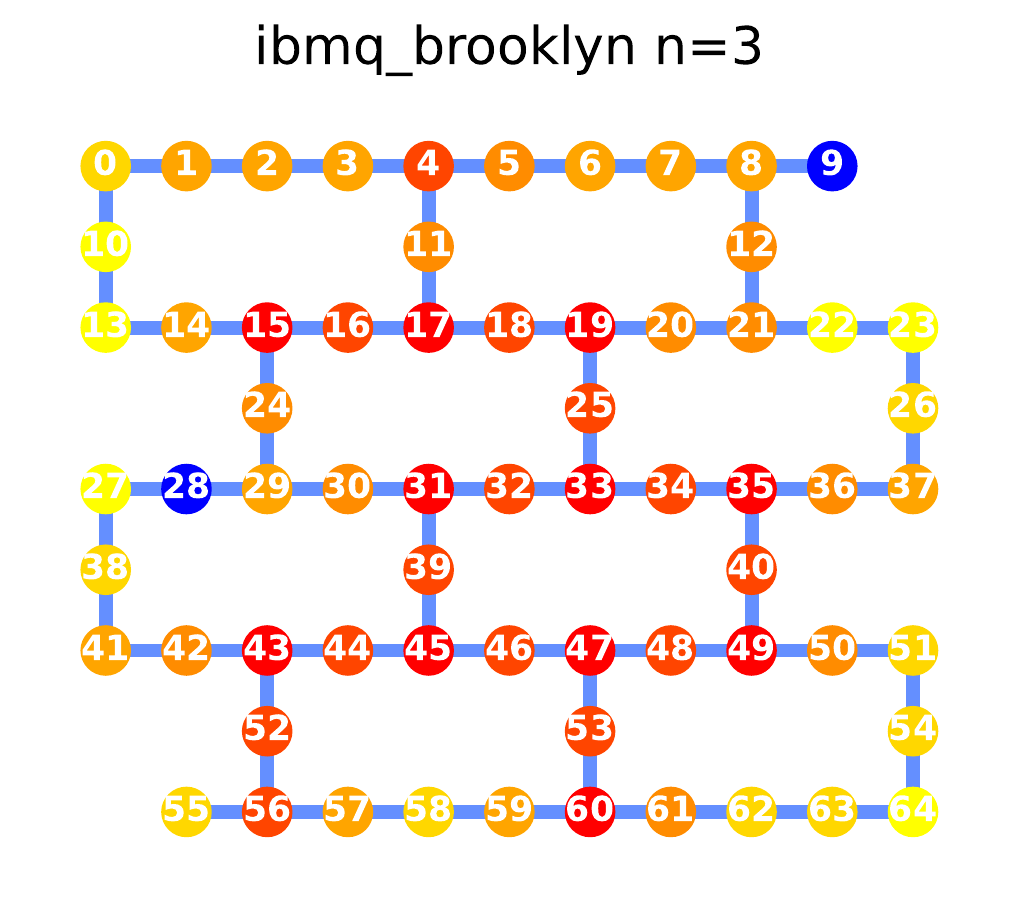}
    \includegraphics[width=0.24\textwidth]{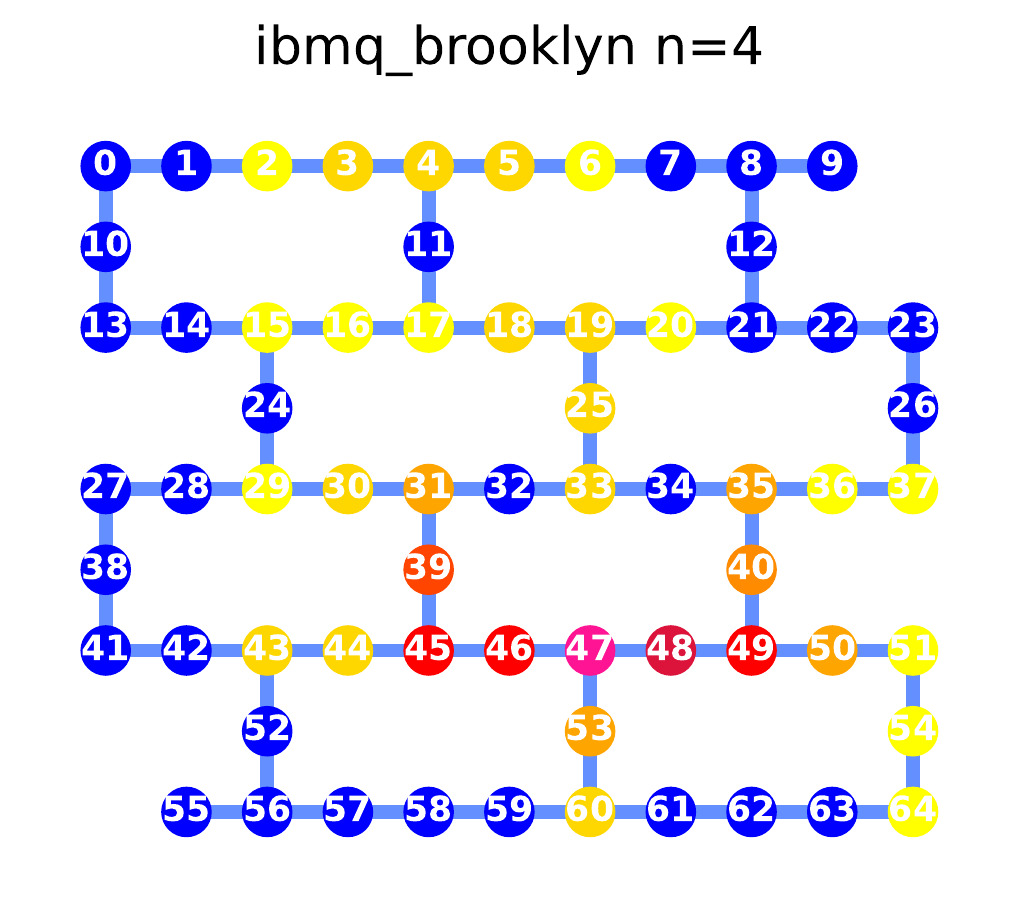}\\
    \includegraphics[width=0.24\textwidth]{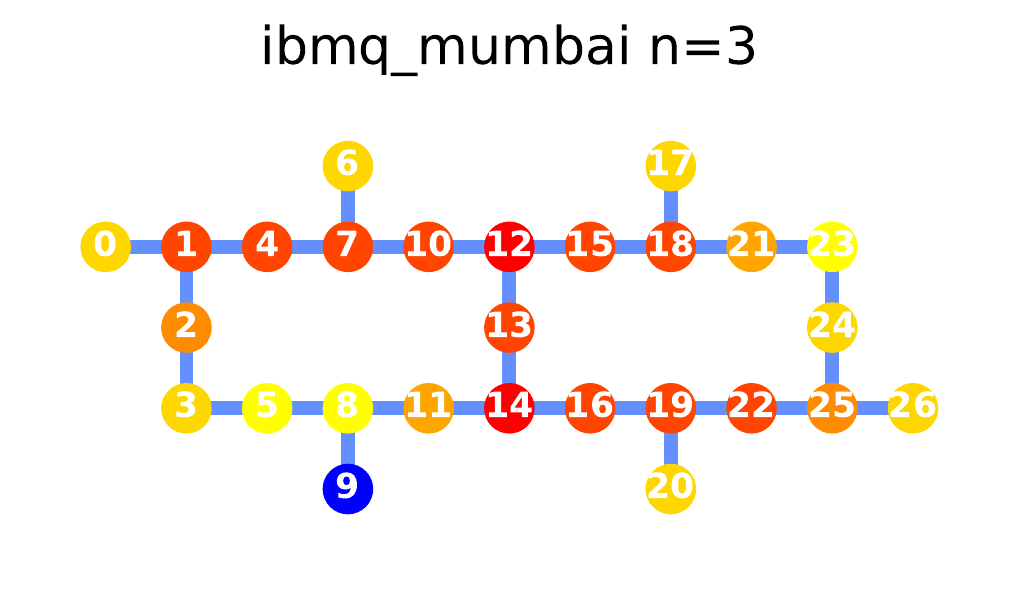}
    \includegraphics[width=0.24\textwidth]{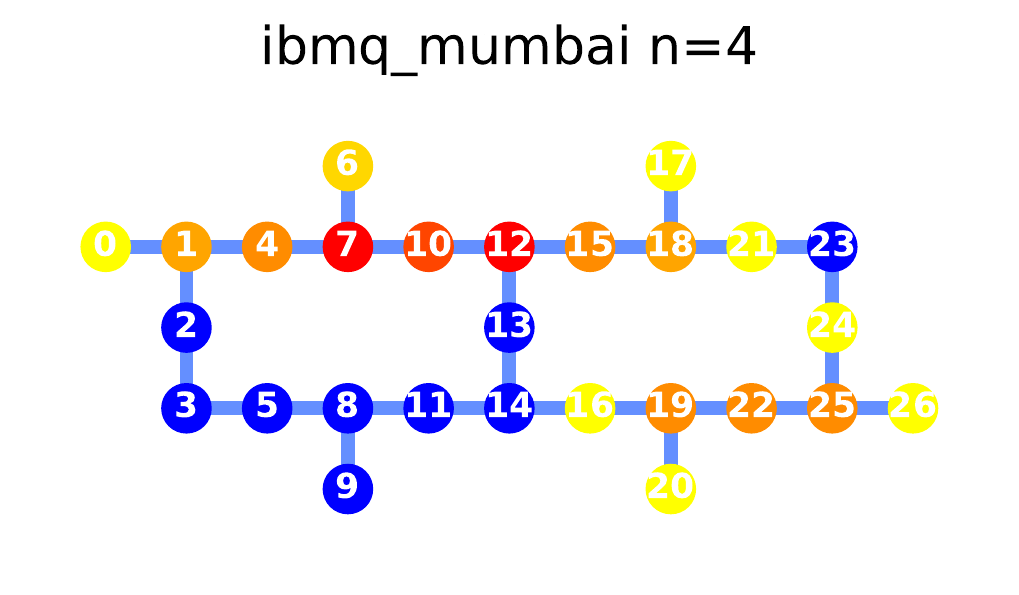}
    \includegraphics[width=0.24\textwidth]{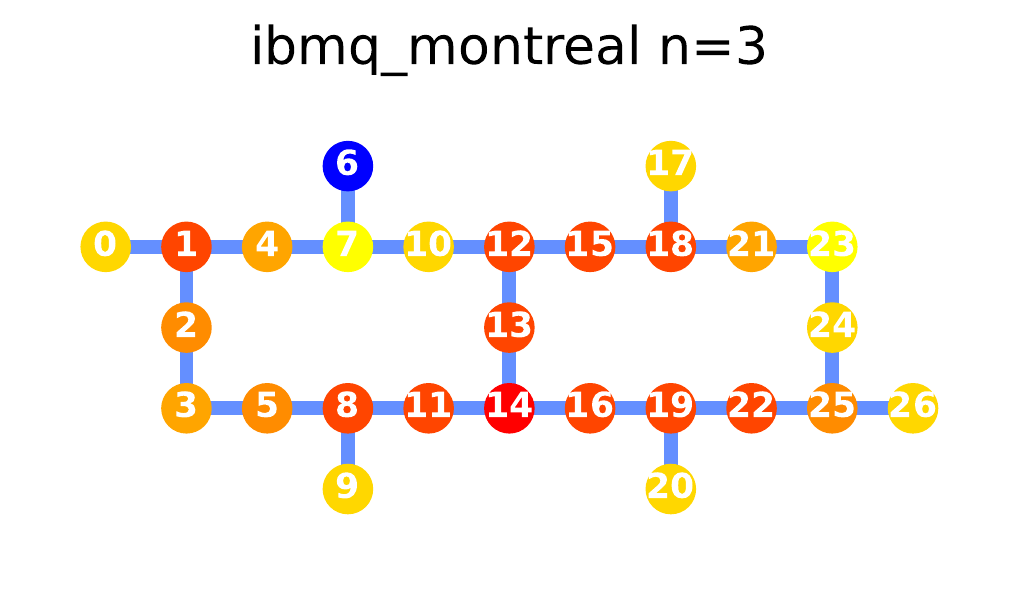}
    \includegraphics[width=0.24\textwidth]{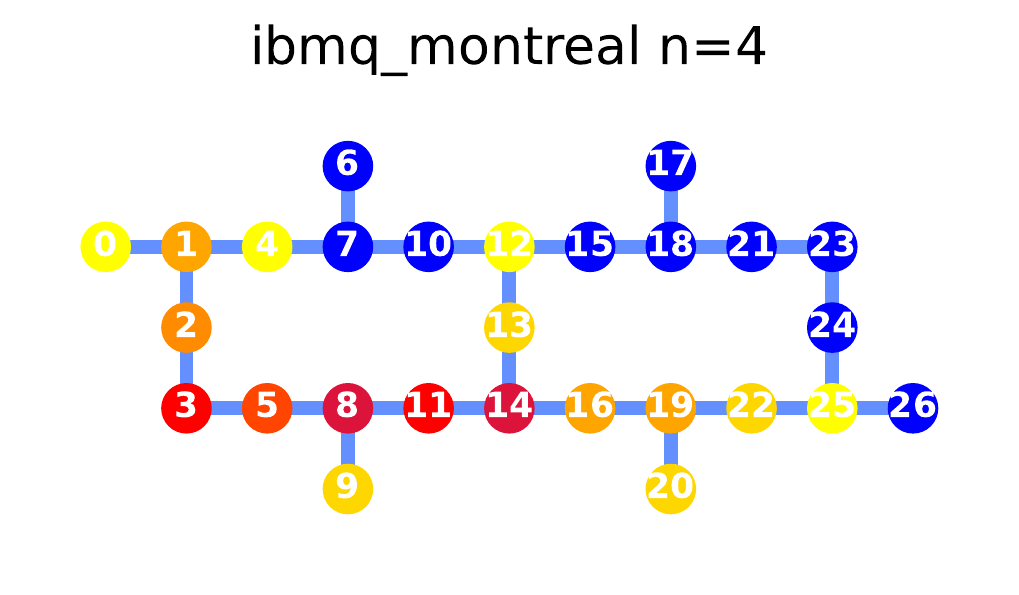}\\
    \includegraphics[width=0.24\textwidth]{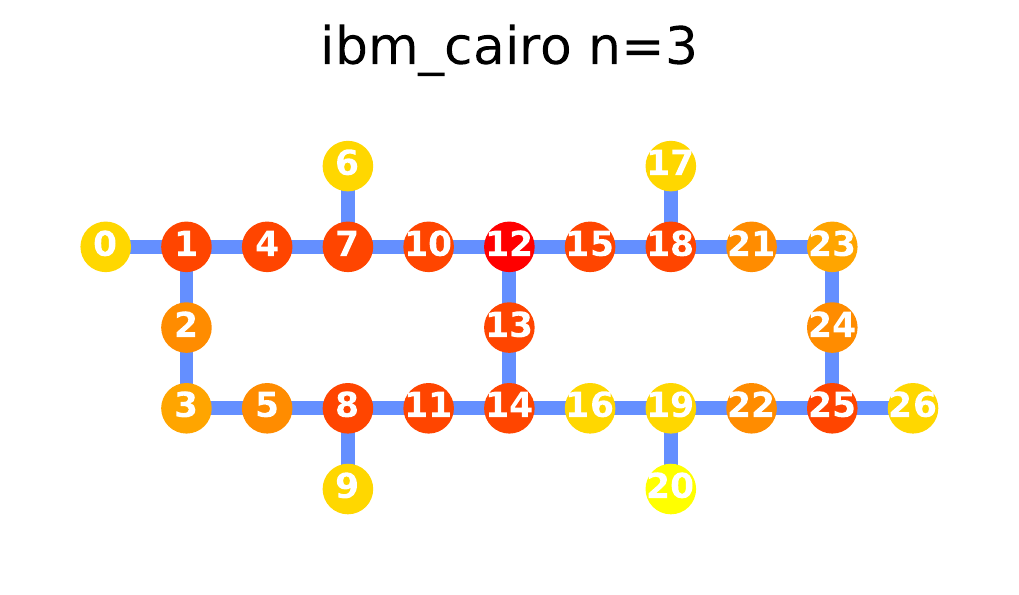}
    \includegraphics[width=0.24\textwidth]{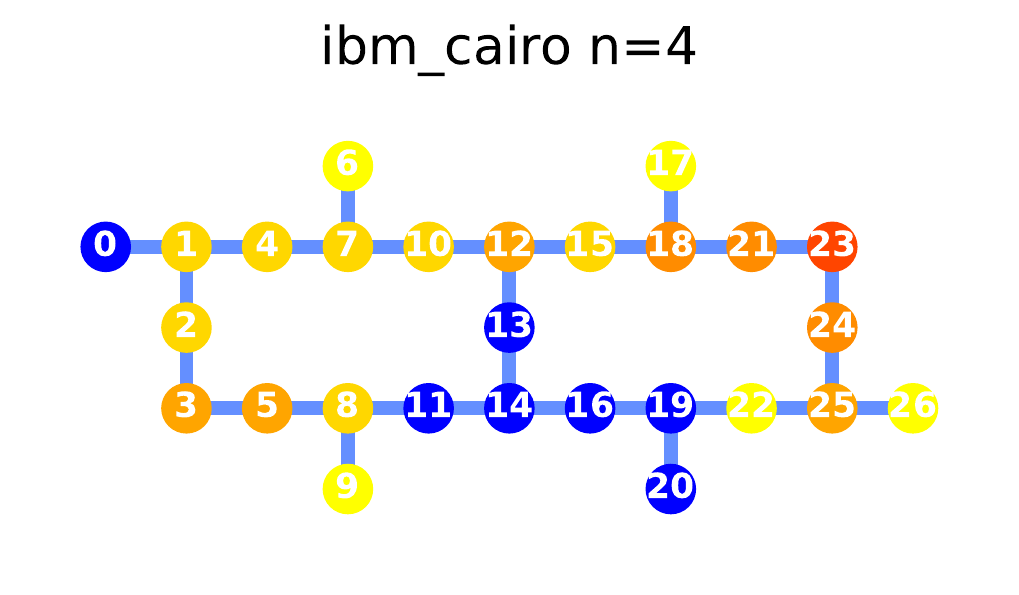}
    \includegraphics[width=0.24\textwidth]{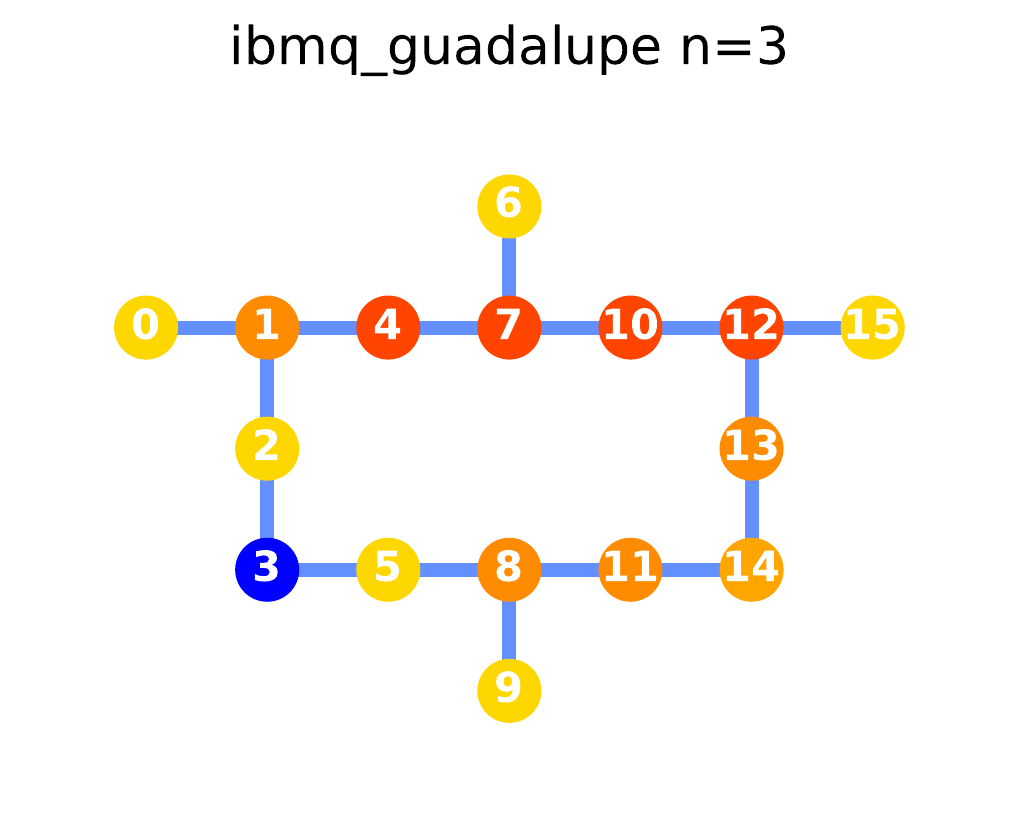}
    \includegraphics[width=0.24\textwidth]{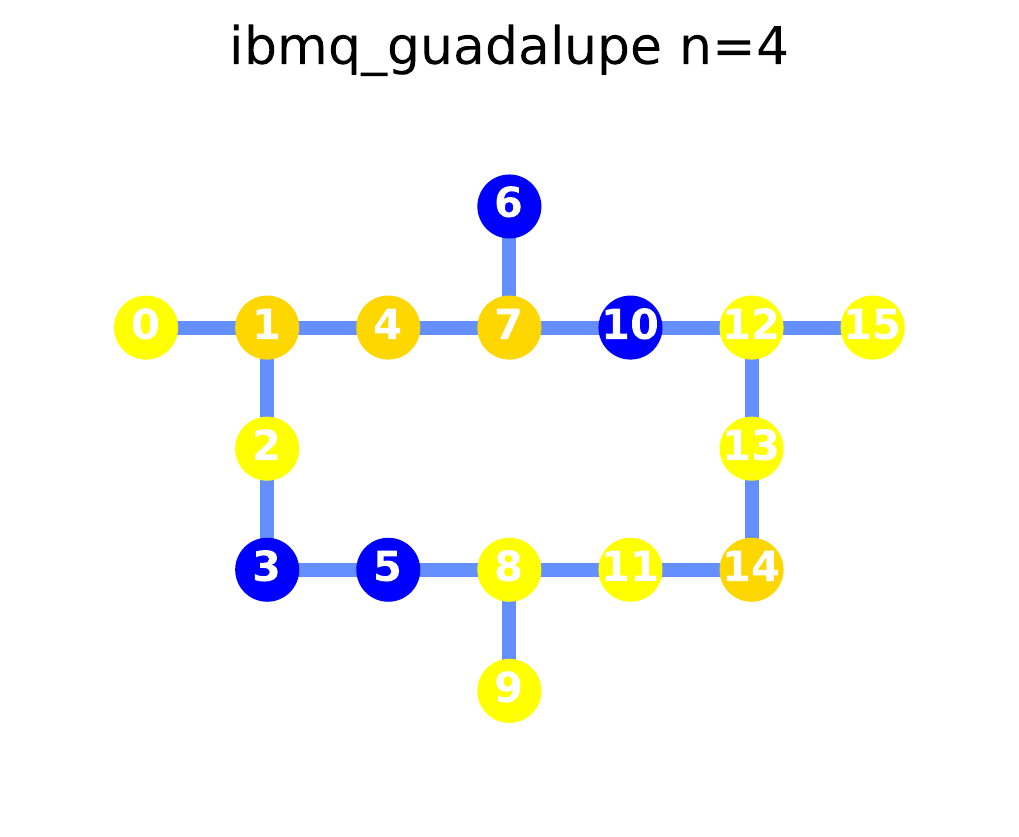}\\
    \includegraphics[width=0.24\textwidth]{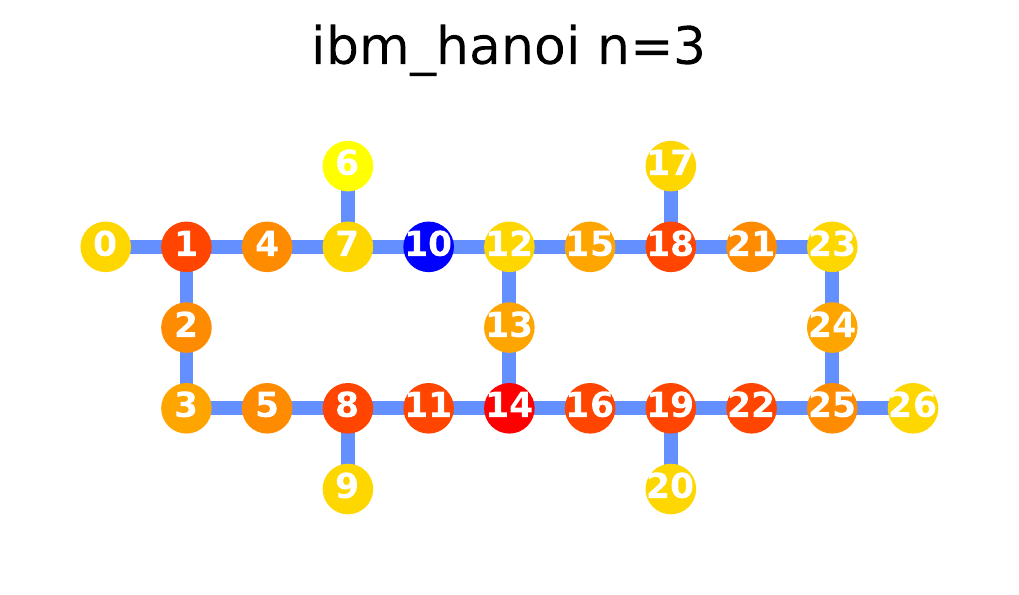}
    \includegraphics[width=0.24\textwidth]{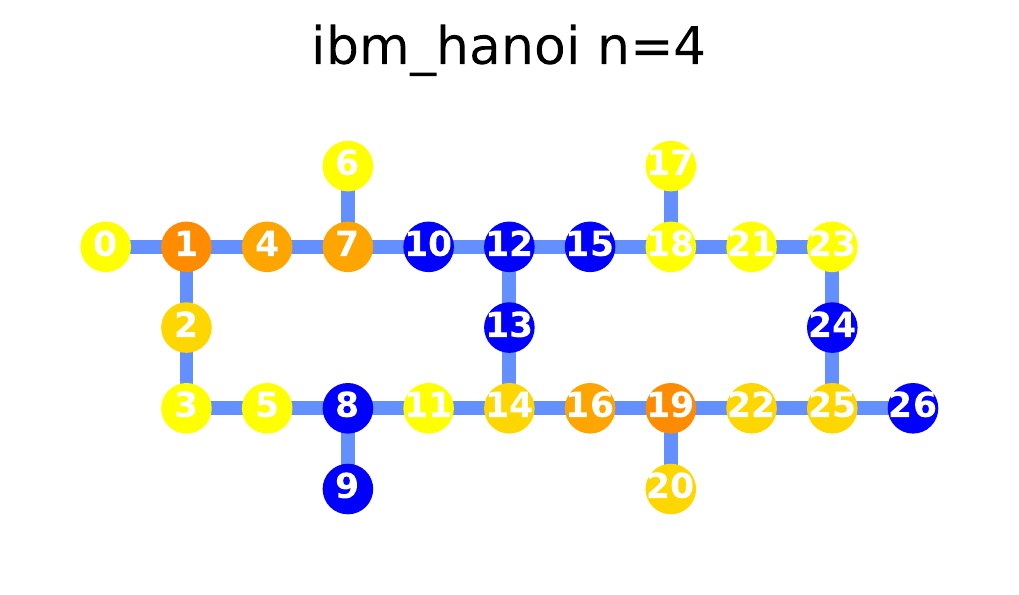}
    \includegraphics[width=0.24\textwidth]{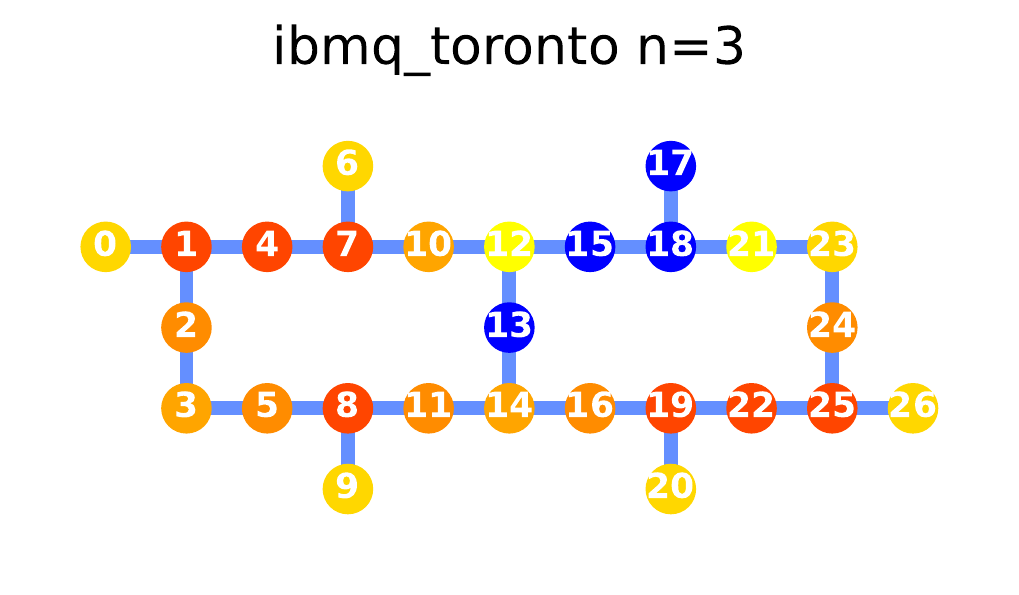}
    \includegraphics[width=0.24\textwidth]{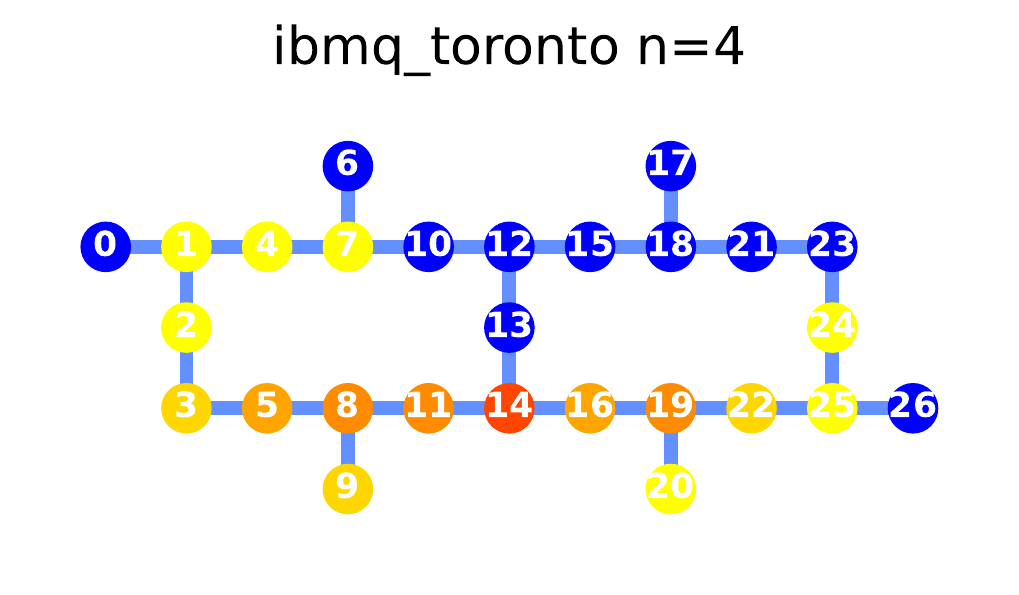}\\
    \includegraphics[width=0.5\textwidth]{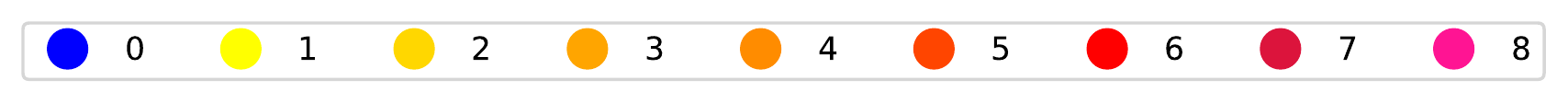}
    \caption{Heatmaps showing how many of the successful QV subgraphs each qubit was a member of across several IBM~Q backends for $n=3$ and $n=4$ sized QV circuits, corresponding to some of the data in Table~\ref{table:ibmq_connected_subgraph_results}. Due to the volume of circuits executed in order to obtain this data, these results are not fully self consistent because the noise profile of the backend can change over time, and these results were gathered over a time period of up to several weeks. }
    \label{fig:ibmq-heatmap}
\end{figure*}

Figure~\ref{fig:ibmq-heatmap} shows heatmaps of several IBM~Q backends in terms of QV protocol success counts across the entire chip. Notably, as with the error rates on the chip, the distribution of higher success rate qubits is not uniform across all qubits. While this is to be expected, this result shows the importance of backend connectivity and the error rates of particular gates; the QV value for a backend does not necessarily hold for all the qubits on the backend. 

Figure~\ref{fig:histogram_HOP_diff_IBMQ} shows the distribution of mean HOP values from the Qiskit transpiled QV circuits across all IBM~Q backends and connectivities, organized into three histograms corresponding to $n=3, 4, 5$. Two observations are noteworthy; first all three histograms seem to have bimodal characteristics, which could correspond to different processor generations. Second, although no $n=5$ QV protocol passed z-confidence of $0.99$, the histogram shows that some mean HOP values did cross the $\frac{2}{3}$ threshold; but the amount over $\frac{2}{3}$ was not significant.

\newcommand{\rot}[1]{\rotatebox{90}{#1}}
\begin{table*}[h]
    \scriptsize
    \hspace*{-1cm}
    \addtolength{\tabcolsep}{-2pt}
	\centering
	\begin{tabular}{@{}r rrrrrrrrrr rrrrrrrrr @{}}
		\toprule
		ibmq device	& \rot{lima}	& \rot{belem}	& \rot{quito}	& \rot{jakarta}	& \rot{bogota}	& \rot{manila}	& \rot{lagos}	& \rot{perth}	& \rot{casablanca}	& \rot{guadalupe} & \rot{sydney}	& \rot{toronto} & \rot{brooklyn} & \rot{hanoi} & \rot{cairo} & \rot{mumbai} & \rot{montreal} & \rot{auckland} & \rot{washington} \\
		\midrule[\heavyrulewidth]
 		\# qubits	& 5 & 5 & 5 & 7 & 5 & 5 & 7 & 7 & 7 & 16 & 27 & 27 & 65 & 27 & 27 & 27 & 27 & 27 & 127\\
		IBM $\log_2$ QV	& 3 & 4 & 4 & 4 & 5 & 5 & 5 & 5 & 5 & 5  & 5 & 5 & 5 & 6 & 6 & 7 & 7 & 6 & 6\\
		\midrule
		$n=3$ & 4/4 & 2/4 & 4/4 & 7/7 & 3/3 & 3/3 & 7/7 & 4/7 & 7/7 & 17/20 & 23/27 & 26/37 & 80/95 & 30/37 & 34/37 & 31/37 & 31/37 & 18/37  \\
		$n=4$ & 0/3 & 0/3 & 1/3 & 0/6 & 0/2 & 0/2 & 2/6 & 0/6 & 2/6 & 4/24  & 5/48 & 9/48 & 22/132 & 9/48 & 12/48 & 13/48 & 14/48 & 3/48 & 28/262*\\ 
		$n=5$ & 0/1 & 0/1 & 0/1 & 0/6 & 0/1 & 0/1 & 0/6 & 0/6 & 0/6 & 0/30  & 0/68 & 0/68 & 0/200 & 0/68 & 0/68 & 0/68 & 0/68 & 0/68\\ 
		\bottomrule
	\end{tabular}
	\addtolength{\tabcolsep}{1.45pt}
	\caption{Successful quantum volume experiments across the IBM~Q backends compared to the vendor provided $\log_2$~QV. Denominator is the number of connected subgraphs of size $n$ on the backend, numerator is the the number of those subgraphs that passed the quantum volume protocol test.\newline
	* On \texttt{ibm\_washington}, in part due to the significantly larger chip size than the other IBM~Q backends, not all circuits were able to be run; the true number of connected qubit subsets on \texttt{ibm\_washington} is 272, but we only tested 262 of those.}
\label{table:ibmq_connected_subgraph_results}
\end{table*}

\subsection{Custom Compilation using QV Passmanager: IBM~Q}
\label{sec:results_ibmq_connected_subg_custom_passmanager}
Table~\ref{table:ibmq_connected_subgraph_results_heavy_compilation} summarizes the QV results on \emph{a subset} of the IBM~Q backends when using the \texttt{custom QV passmanager} for circuit compilation. Compilation for \texttt{ibmq\_mumbai} and \texttt{ibm\_auckland} were successful, but failed to execute on the backends due to an internal error relating to the pulse instruction durations. The pulse gate duration and timing needs to be specified very precisely (see Figure~\ref{fig:QV_circuit_drawings}, circuit \textbf{(4)} where the QV passmanager specifies Delay gates in order to make the Pulse gates work correctly on the backend); it appears that the compilation to the \texttt{ibmq\_mumbai} and \texttt{ibm\_auckland} backends failed because of an error related to the circuit timing. Lastly, additional circuits compiled using this custom Passmanager were not executed on \texttt{ibm\_washington} due to the significant QPU time usage it would require. 

Importantly, this custom compiler increased the measured quantum volume on many of the backends compared to the default Qiskit transpiled circuits; going from $n=4$ when using the default Qiskit transpiler, to $n=5$. However, this custom passmanager did not work on all backends, required heavy computation time, and it did not consistently find the same QV values reported by the vendor, although not all qubit subsets were evaluated on all of the backends we tested. Therefore although it does improve circuit fidelity, the more custom compilation techniques are not feasible for a typical user. 

\begin{figure*}[h]
    \centering
    \includegraphics[width=0.45\textwidth]{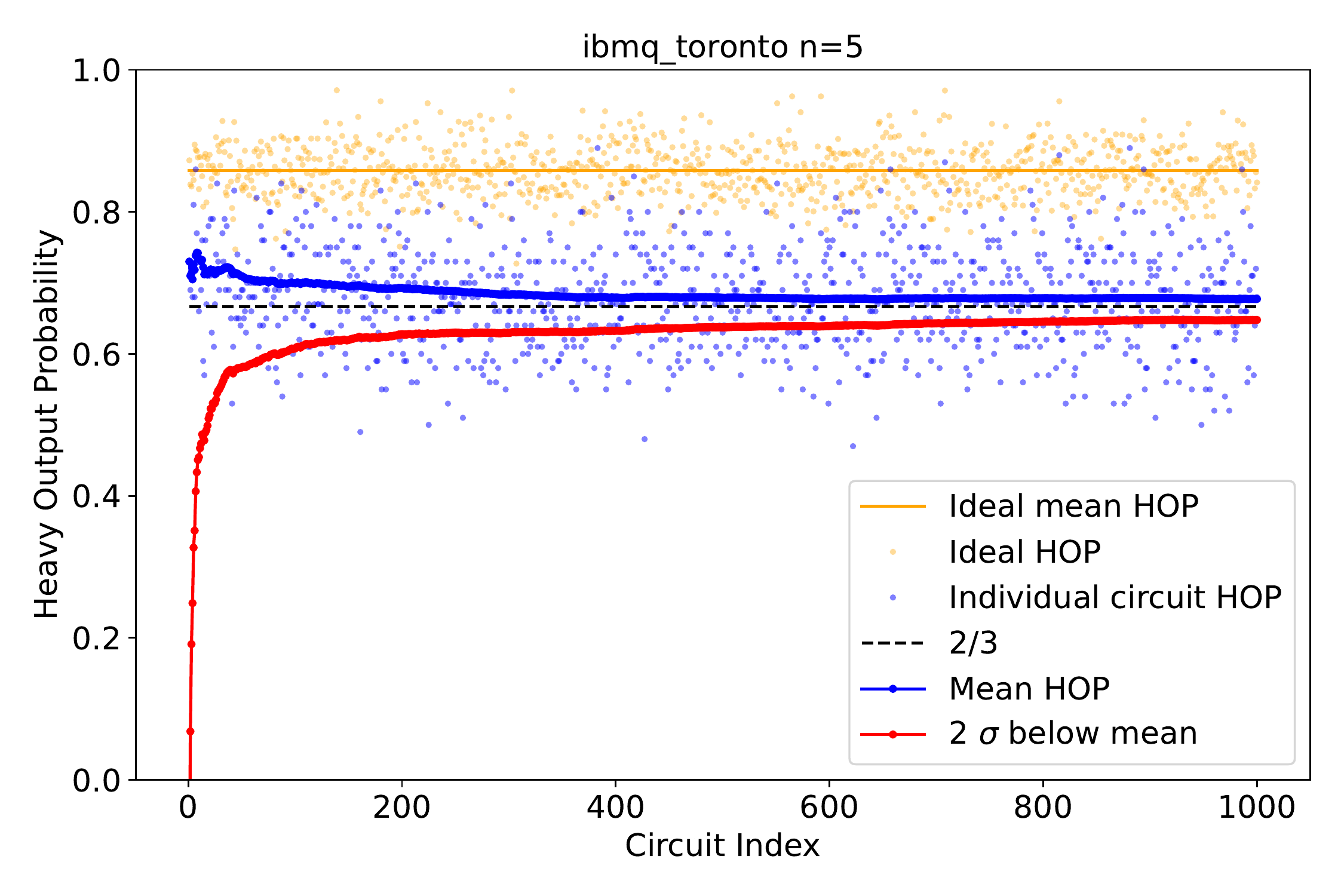}\hfill%
    \includegraphics[width=0.45\textwidth]{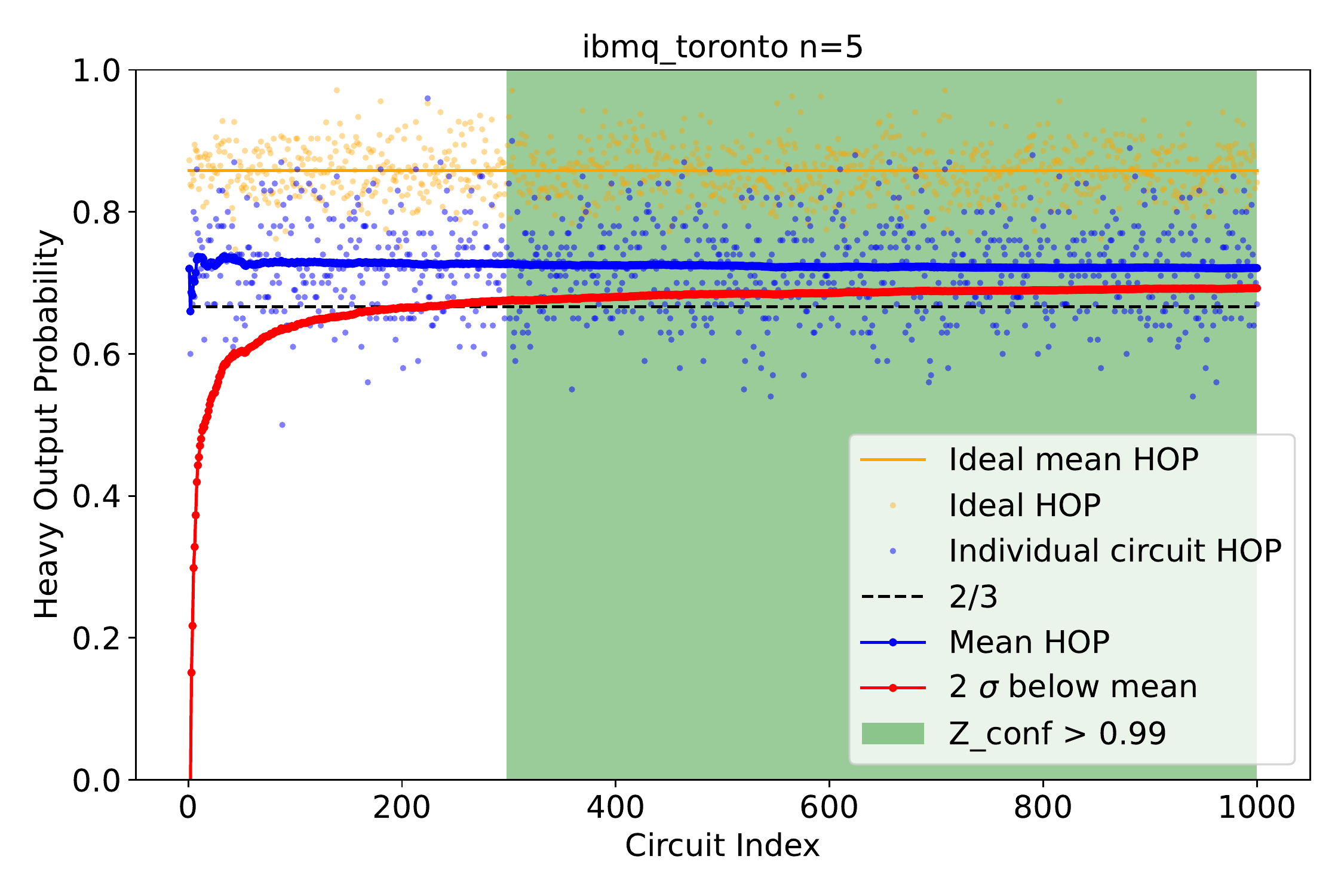}\\
    \includegraphics[width=0.45\textwidth]{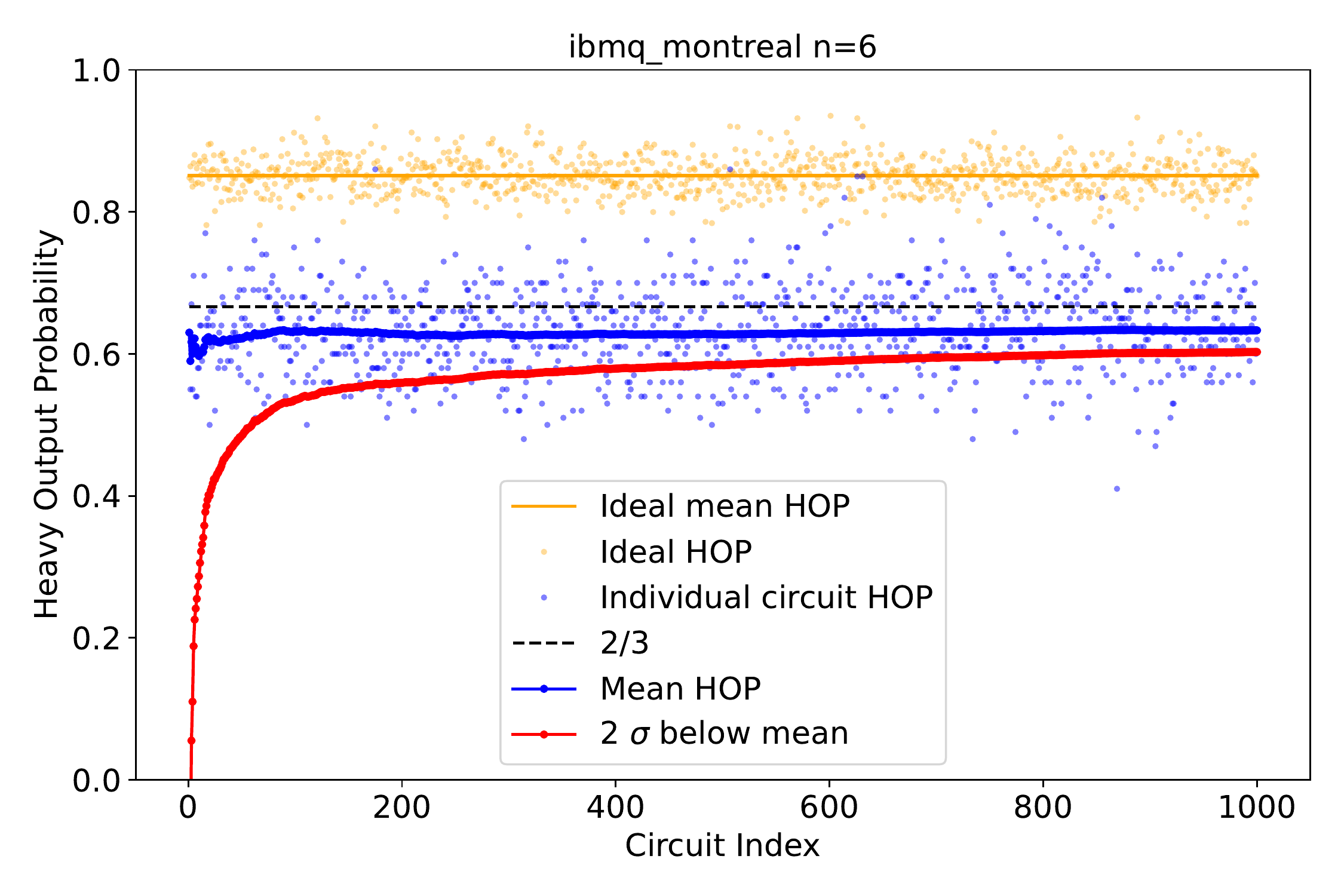}\hfill%
    \includegraphics[width=0.45\textwidth]{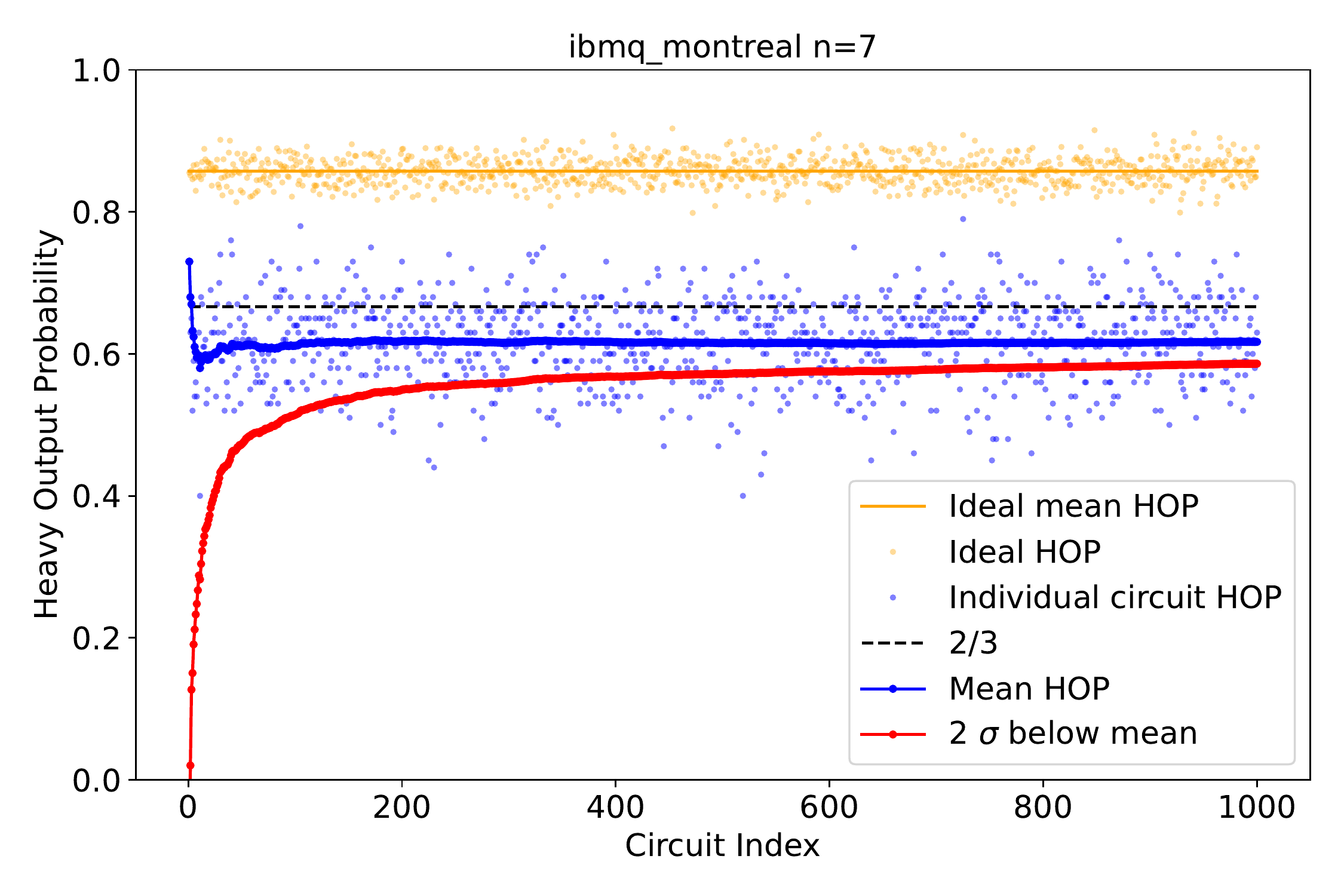}%
    \caption{IBM~Q custom QV passmanager compilation for $n=5$ on qubits \texttt{1, 2, 4, 7, 10} (top left), qubits \texttt{3, 5, 8, 11, 14} on \texttt{ibmq\_toronto} (top right). Then $n=6$ with qubits \texttt{14, 16, 19, 20, 22, 25} (bottom left) on \texttt{ibmq\_montreal} and $n=7$ with qubits \texttt{5, 8, 11, 14, 16, 19, 20} (bottom right) on \texttt{ibmq\_montreal}. These plots show a similar result to the default Qiskit transpiled circuits; the qubit subset choice can change the QV result very significantly. }
    \label{fig:heavy-compilation-ibmq}
\end{figure*}

Figure~\ref{fig:heavy-compilation-ibmq} shows a side-by-side of (different) qubit subset HOP results from \texttt{ibmq\_toronto} and \texttt{ibmq\_montreal}, showing which qubits are used on a chip can change the QV result. 

\begin{table*}[h]
	\centering
	\begin{tabular}{@{}r rrrrrrrrr @{}}
		\toprule
		ibmq device	& manila & bogota & guadalupe & lagos & toronto & hanoi & cairo & montreal & brooklyn\\ 
		\midrule[\heavyrulewidth]
		\# of qubits & 5 & 5 & 16 & 7 & 27 & 27 & 27 & 27 & 65\\
		IBM~Q log$_2$ QV & 5 & 5 & 5 & 5 & 5 & 6 & 6 & 7 & 5\\
		\midrule
		$n=3$ & 3/3 & 3/3 & 12/20 & 7/7 & 27/37 & 17/37 & 11/37 & 34/37 & 95/95\\ 
		$n=4$ & 2/2 & 0/2 & 4/24 & 4/4 & 3/16 & 2/20 & 0/4 & 20/30 & 7/13\\ 
		$n=5$ & 1/1 & 0/1 & 4/25 & 4/4 & 17/62 & 6/38 & 0/17 & 22/49 & 31/90\\ 
		$n=6$ &  &  &  &  &  & 0/37 & 0/48 & 0/36 & \\ 
		$n=7$ &  &  &  &  &  &  &  & 0/3 & \\ 
		\bottomrule
	\end{tabular}
	\caption{Successful quantum volume measures across some IBM~Q backends compared to the vendor provided $\log_2$~QV, when using the custom QV passmanager for compilation. Entries in the table have numerator equal to the number of subgraphs that passed the test, and denominator equal to the number of connected subgraphs that we could compile all $1,000$ QV circuits to in reasonable time. }
\label{table:ibmq_connected_subgraph_results_heavy_compilation}
\end{table*}

\subsection{Quantum Volume over time}
\label{sec:results_QV_over_time}
\begin{figure}[h]
    \centering
    \includegraphics[width=0.55\textwidth]{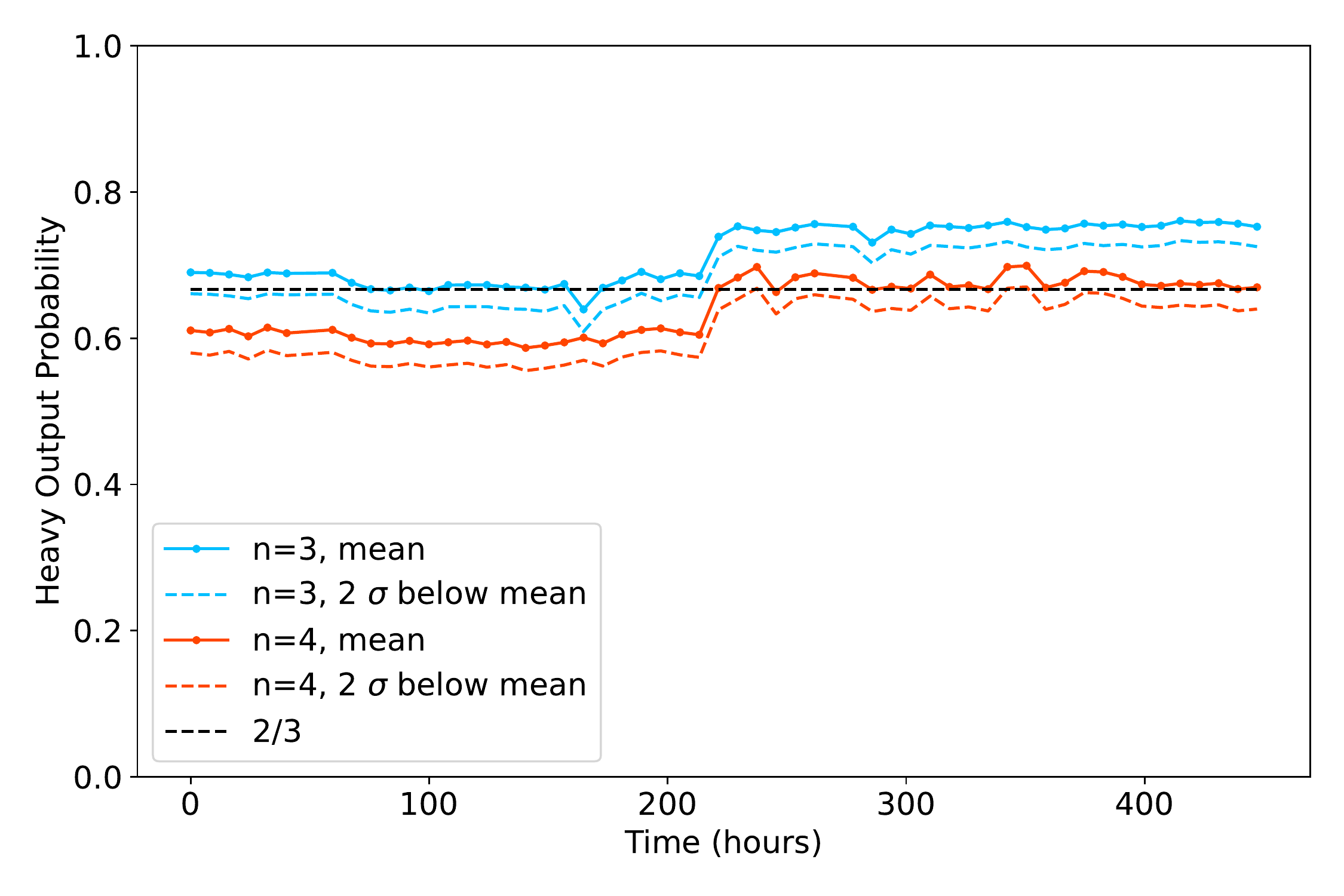}
    \caption{Fluctuations of quantum volume over 18 days:
    For repeated experiments over time on \texttt{ibmq\_auckland} we show the
    measured Mean QV HOP data points (each taken across all $1,000$ QV circuits of size $n$). The $2 \sigma$ below mean line is shown as dashed lines. Importantly, this shows that the measured QV drifts from 4 $(n<3)$ at the start to 8 $(n=3)$ at about $200$ hours. For a brief time at about $350$ hours we have QV 16, as the $2 \sigma$ below the mean ($n=4$) line just passes the $\frac{2}{3}$ threshold.
    Experiments for $n=3$ were run on qubits (10, 12, 13), experiments for $n=4$ on qubits (10, 12, 13, 14).}
    \label{fig:over_time}
\end{figure}

Another important aspect of a benchmark for quantum computers is that the noise profile on the device changes over time which means that benchmarks on a specific part of the hardware may not be consistent over time. Figure~\ref{fig:over_time} shows the mean HOP after having run all $1,000$ QV circuits for $n=3$ and $n=4$ on fixed qubit subsets of the ibm\_auckland device. The mean HOP achieved at each of these time points noticeably varies over the approximately $400$ hours of the data collection. This shows that at points this set of qubits fails to pass the $n=3$ QV test (because the mean HOP is below $\frac{2}{3}$, but at other points the mean HOP for $n=4$ goes above $\frac{2}{3}$. This shows that there is significant variation in the QV test over time. Therefore it is important to show the aggregate behavior of a NISQ device on the QV benchmark, for example what is shown in Figure~\ref{fig:ibmq-heatmap}. 

Although the $n=3$ and $n=4$ QV circuits were run with a small time delay in between, we would expect to see high correlation between the $n=3$ and $n=4$ lines in Figure~\ref{fig:over_time}. We observe this to be the case for the overall trends, however there are small discrepancies between the two lines; for example at approximately $180$ hours there is a dip in the blue $n=3$ line, but not a corresponding dip in the orange $n=4$ line.

\section{Discussion}
\label{sec:discussion}
Quantum Volume is designed to be a benchmark that can compare quantum backends to other quantum backends with different underlying hardware. What we found is that the particular QV protocol used (i.e. how many circuits are run), and how the QV circuit are compiled, massively impacts the measured quantum volume. For end users who will employ the simple compiler methods available in the quantum SDK's of the hardware vendors, the heavy compilation quantum volume results \emph{do not} reflect the expected backend fidelity they would experience when running jobs because they are not using those more advanced compiler options. 

The qubit subset enumeration QV results (for IBM-Q) reveals a lot more detail than the black-box method. These results give a more detailed analysis of the quantum device's capabilities; not only allowing more accurate comparisons across different backends, but also a more detailed picture of which regions of the device give the best results. Figure~\ref{fig:ibmq-heatmap} shows that the specific qubits used to execute circuits on the backend greatly impact the circuit fidelity. Using the approach of compiling the same circuits to different qubit subsets of a backend allows even greater evaluation of the performance of a backend; applying this methodology to application benchmarks~\cite{2021-application-benchmark, lubinski2021applicationoriented} is interesting future work. 

Although we were not able to exhaustively evaluate the the more advanced IBM~Q compilation features across all backends and qubit subsets, the more advanced compilation methods clearly increased the measured QV values (see Table~\ref{table:ibmq_connected_subgraph_results_heavy_compilation}).

The error rates and connectivity clearly translate to higher quantum volumes. The Quantinuum \texttt{H1-2} backend had the lowest overall error rate across all backends we tested, and it also had the highest quantum volume by a significant amount. 

We shows that importantly, although not surprisingly, the QV benchmark drifts over time even for fixed qubits. It is known that there is significant time dependence on NISQ noise profiles~\cite{dasgupta2021stability} and therefore it follows that benchmarks for these devices would also see time dependence. This is notable because it means that NISQ benchmarks also need to take into account the aggregate behavior of the device, not just a subset of qubits.

Other NISQ benchmarks, such as CLOPS for quantifying execution speed, should also continue to be evaluated in order to provide additional context for the performance of these backends. Another future research area is to quantify the correlation between NISQ benchmarks (such as QV) and error metrics such as \emph{aggregate error}~\cite{pelofske2021sampling, golden2022fairsampling}, cross-talk, and average qubit fidelity. On average it is clear that error rates, as well as connectivity and compilers, are the primary factors impacting NISQ device performance. However, exactly quantifying the error experienced by a circuit during execution can be difficult because it relies on time sensitive calibration data, as well as knowing the exact circuit that was executed on the backend. Additionally QV circuit execution can occur over an extended period of time. Therefore, determining the relationship between aggregate error and NISQ benchmarks is an interesting research avenue. 

Due to time and usage limitations (in particular the current trapped ion qubit technology costs more time per sample compared to superconducting qubit technology) we were unable to complete the full quantum volume tests on the Quantinuum H1-2 device. One of the limitations of the QV benchmark in this respect is that it requires many samples and many circuit executions, which is not always feasible on all platforms. One possible modification of the QV protocol would be to make use of data re-sampling in order to quantify HOP distributions using a smaller number of samples. This has been introduced in the literature~\cite{baldwin2021reexamining} and would make the QV protocol more accessible. 

Overall we find that quantum volume gives a good basis for comparing different NISQ backends if the settings for such a comparison are constant. There are many particular details which can impact the the measured quantum volume of a NISQ device. The most significant appears to be the compiler; on one hand, heavy optimization can yield better circuit fidelity, but on the other hand a poor choice of qubit layout can significantly hinder the circuit fidelity. Other important settings include how many circuits are used in the test. For instance Figure~\ref{fig:IBMQ_connected_subgraph} shows that the successful measurement of $n=4$ for \texttt{ibmq\_guadalupe} would not occur if we executed less than 400 circuits. Therefore, the quantum volume metric is only useful if there is a consistent basis for comparison (i.e. relatively consistent compiler, and consistent experimental settings). 

An important takeaway for users of NISQ computers when running near term applications is that it is prudent to try different qubit subsets on the hardware and to try different compilation strategies (for example noise adaptive compiler settings vs specifying some qubits to use) in order to get a good sense of what the device's full capabilities are. 

Lastly, the QV metric is designed specifically for gate model (i.e. universal) quantum computation devices; however there are restricted quantum computational devices in the NISQ-era including Quantum Annealers~\cite{Das2008, Kadowaki1998, Santoro2002, morita2008mathematical, boixo2014evidence, king2021scaling} and Boson Samplers~\cite{PhysRevLett.119.170501, Arrazola2021, Spagnolo2014, aaronson2011computational}. Although direct comparisons across these different quantum technologies are not always possible, benchmarks comparing the state of other NISQ-era technology could be useful.

\section{Acknowledgments}
\label{sec:acknowledgments}

We acknowledge the use of IBM Quantum services for this work. The views expressed are those of the authors, and do not reflect the official policy or position of IBM or the IBM Quantum team.

\noindent
This research used resources of the Oak Ridge Leadership Computing Facility, which is a DOE Office of Science User Facility supported under Contract DE-AC05-00OR22725.

\noindent
Research presented in this article was supported by the Laboratory Directed Research  and  Development  program  of  Los  Alamos  National  Laboratory under project number 20220656ER.

\noindent
We acknowledge the support of the LANL Information Science and Technology Institute.

\noindent
This research used resources provided by the Darwin testbed at Los Alamos National Laboratory (LANL) which is funded by the Computational Systems and Software Environments subprogram of LANL's Advanced Simulation and Computing program (NNSA/DOE). 

\noindent
The authors thank Georg Hahn for discussions on statistical methods to estimate the number of reasonable samples to use for QV benchmarking. 

\noindent
Figure~\ref{fig:flowchart} was created in Lucidchart, \url{www.lucidchart.com}

\setlength\bibitemsep{0pt}
\printbibliography

\appendix
\section{Gate definitions}
\label{sec:Appendix_gate_def}
Here we provide the exact gate definitions of all of the gate operations referenced for compilation to the different backends as well as the native gates on the different backends. 

\subsection{2-qubit gates}

We first define the parametrized 2-qubit gates used in this paper:
\begin{itemize}
    \item   XY is one of the native two qubit gates for the Rigetti systems, and it is parameterized by an angle $\theta$. 
\begin{align*}
XY(\theta) &= e^{i \theta/2 (XX+YY)/2}  \\ 
&= \begin{pmatrix} 1 & 0 & 0 & 0 \\ 0 & \cos(\theta/2) & i\sin(\theta/2) & 0 \\ 0 & i\sin(\theta/2) & \cos(\theta/2) & 0 \\ 0 & 0 & 0 & 1 \end{pmatrix}
\end{align*}
For fixed angle $\theta=\pi$, the $XY(\pi)$ gate is also known as $\mathit{iSWAP}$.
The iSWAP gate is used for illustration purposes to show a
Quil compiled circuit in Figure~\ref{fig:QV_circuit_drawings}. XY gates are native to the Rigetti hardware but currently not supported in Qiskit, though
the iSWAP gates in Figure~\ref{fig:QV_circuit_drawings} correspond to actual $\mathit{XY}(\pi)$ gates.

    \item   RXX is an angle parameterized gate used as the only two qubit gate when compiling the raw QV circuits in Qiskit to be submitted to the 11 qubit IonQ Harmony device through Amazon Braket. 
\begin{align*}
& RXX(\theta) \\ &= e^{-i \theta/2 (XX)} \\ 
&=\begin{pmatrix} cos(\frac{\theta}{2}) & 0 & 0 & -i sin(\frac{\theta}{2}) \\ 0 & cos(\frac{\theta}{2}) & -i sin(\frac{\theta}{2}) & 0 \\ 0 & -i sin(\frac{\theta}{2}) & cos(\frac{\theta}{2}) & 0 \\ -i sin(\frac{\theta}{2}) & 0 & 0 & cos(\frac{\theta}{2}) \end{pmatrix}
\end{align*}

    \item   CPHASE is one of the native two qubit gates for the Rigetti systems, and it is parameterized by the angle $\phi$. 
\begin{align*}
CPHASE(\phi) =  
\begin{pmatrix} 1 & 0 & 0 & 0 \\ 0 & 1 & 0 & 0 \\ 0 & 0 & 1 & 0 \\ 0 & 0 & 0 & e^{i \phi} \end{pmatrix}
\end{align*}

\end{itemize}

\noindent
Next, we define the more common and maybe better known unparametrized 2-qubit gates:
\begin{itemize}
    \item   CZ is the third two qubit gate for the Rigetti systems; it is not parameterized. 
\begin{gather*}
CZ = 
\begin{pmatrix} 1 & 0 & 0 & 0 \\ 0 & 1 & 0 & 0 \\ 0 & 0 & 1 & 0 \\ 0 & 0 & 0 & -1 \end{pmatrix}
\end{gather*}

    \item   The CNOT (CX) two qubit gate is a native two qubit gate on IBM~Q systems, and is used at various levels of the compilation procedures; for example the raw QV circuits are defined entirely in terms of U3 and CX gates. 
\begin{gather*}
CX = 
\begin{pmatrix} 1 & 0 & 0 & 0 \\ 0 & 1 & 0 & 0 \\ 0 & 0 & 0 & 1 \\ 0 & 0 & 1 & 0 \end{pmatrix}
\end{gather*}

    \item   The ECR gate is the two qubit gate used by the OQC Lucy backend. We compile the raw QV circuits to the OQC Lucy native gateset before submitting the circuits to the Amazon Braket system. 
\begin{gather*}
ECR_{q_0 , q_1} = 
\begin{pmatrix} 0 & 0 & 1 & i \\ 0 & 0 & i & 1 \\ 1 & -i & 0 & 0 \\ -i & 1 & 0 & 0 \end{pmatrix}
\end{gather*}

    \item   ZZ is the two qubit native gate for the Quantinuum H1-2 device. In the worfkflow for this paper the QV circuits are not compiled to the Quantinuum gateset on the user side, instead that compilation is done entirely on the server side.
\begin{gather*}
ZZ = 
\begin{pmatrix} 1 & 0 & 0 & 0 \\ 0 & i & 0 & 0 \\ 0 & 0 & i & 0 \\ 0 & 0 & 0 & 1 \end{pmatrix}
\end{gather*}

\end{itemize}

\subsection{1-qubit gates}

We again present parametrized gates first:

\begin{itemize}
    \item The rz gate is the most commonly used single qubit gate; it is used by the Quantinuum, IonQ, IBM~Q, OQC, and Rigetti devices. 
\begin{gather*}
rz(\lambda) = 
\begin{pmatrix} e^{-i \frac{\lambda}{2}} & 0 \\ 0 & e^{i \frac{\lambda}{2}} \end{pmatrix}
\end{gather*}

    \item   The ry gate is one of the single qubit gates we use when compiling the QV circuits to the IonQ gateset. 
\begin{gather*}
ry(\theta) = 
\begin{pmatrix} \cos{\frac{\theta}{2}} & - \sin{\frac{\theta}{2}} \\ \sin{\frac{\theta}{2}} & \cos{\frac{\theta}{2}} \end{pmatrix}
\end{gather*}

    \item   The rx gate is a single qubit gate used by both Rigetti and IonQ devices as a native gate. 
\begin{gather*}
rx(\theta) =
\begin{pmatrix} \cos{\frac{\theta}{2}} & - i \sin{\frac{\theta}{2}} \\ -i \sin{\frac{\theta}{2}} & \cos{\frac{\theta}{2}} \end{pmatrix}
\end{gather*}

    \item   $U_1$ is one of the native single qubit gates for the Quantinuum H1-2 device (the other single qubit gate is rz). As with the ZZ gate, we do not use this gate for compiling and submission to the H1-2 device as the compilation is handled server side. 
\begin{gather*}
U_1(\theta, \phi) = 
\begin{pmatrix} \cos{\frac{\theta}{2}} & - i e^{ -i \phi } \sin{\frac{\theta}{2}} \\ - i e^{ i \phi } \sin{\frac{\theta}{2}} & \cos{\frac{\theta}{2}} \end{pmatrix}
\end{gather*}

    \item   The U3 gate is used when the raw QV circuits are generated using Qiskit, as its parameters cover the whole space of single-qubit unitaries. Those raw QV circuits are defined entirely in terms of U3 and CX gates. Subsequent compilations transforms those circuits into other hardware specific gatesets. The compiled circuits returned from the OQC Lucy backend also make use of the U3 gate. 
\begin{gather*}
U3(\theta, \phi, \lambda) = 
\begin{pmatrix} \cos{\frac{\theta}{2}} & -e^{i \lambda} \sin{\frac{\theta}{2}} \\ e^{i \phi} \sin{\frac{\theta}{2}} & e^{i(\phi + \lambda)} \cos{\frac{\theta}{2}} \end{pmatrix}
\end{gather*}

\end{itemize}

\noindent
We conclude with the definitions for fixed single-qubit gates:
\begin{itemize}
    \item   sx is a native single qubit gate on IBM~Q and OQC devices.
\begin{gather*}
sx = \sqrt{X} = 
\begin{pmatrix} 1+i & 1-i \\ 1-i & 1+i \end{pmatrix}
\end{gather*}

    \item   The Pauli X gate is a native single qubit used on IBM~Q and OQC systems.
\begin{gather*}
x = 
\begin{pmatrix} 0 & 1 \\ 1 & 0 \end{pmatrix}
\end{gather*}

    \item   The Hadamard gate is used in description of the compiled QV circuits returned by the OQC device (written in OpenQASM). 
\begin{gather*}
H = 
\frac{1}{\sqrt{2}}\begin{pmatrix} 1 & 1 \\ 1 & -1 \end{pmatrix}
\end{gather*}

\end{itemize}

\end{document}